\PassOptionsToPackage{table}{xcolor}
\documentclass[sigconf,nonacm=true]{acmart}

\author{Arttu Paju}
\orcid{0000-0002-1398-2184}
\affiliation{
\institution{Tampere University}
\city{Tampere}
\country{Finland}
}
\email{arttu.paju@tuni.fi}

\author{Muhammad Owais Javed}
\orcid{0009-0003-9764-3001}
\affiliation{
\institution{Tampere University}
\city{Tampere}
\country{Finland}
}
\email{owais.javed@tuni.fi}

\author{Juha Nurmi}
\orcid{0000-0003-3071-9027}
\affiliation{
\institution{Tampere University}
\city{Tampere}
\country{Finland}
}
\email{juha.nurmi@tuni.fi}

\author{Juha Savim\"{a}ki}
\orcid{0000-0002-5252-7774}
\affiliation{
\institution{Tampere University, Unikie Oy}
\city{Tampere}
\country{Finland}
}
\email{juha.savimaki@tuni.fi}

\author{Brian McGillion}
\orcid{0000-0003-2683-5295}
\affiliation{
\institution{Technology Innovation Institute (TII)}
\city{Abu Dhabi}
\country{UAE}
}
\email{brian@ssrc.tii.ae}

\author{Billy Bob Brumley}
\orcid{0000-0001-9160-0463}
\affiliation{
\institution{Tampere University}
\city{Tampere}
\country{Finland}
}
\email{billy.brumley@tuni.fi}

\usepackage{totcount}
\newtotcounter{citenum}
\def\oldcite{}
\let\oldcite=\bibcite%
\def\bibcite{\stepcounter{citenum}\oldcite}

\usepackage{multirow}

\newcounter{rqcounter}
\newtotcounter{totalcontainerscounter}
\newtotcounter{totalopensourcecontainerscounter}
\newtotcounter{totalframeworkscounter}
\newtotcounter{totalopensourceframeworkscounter}

\makeatletter
\newcommand{\rqcounterautorefname}{RQ\@gobble}
\makeatother

\newcommand{\bulletg}{\textcolor{lightgray}{\bullet}}

\newcommand{\Paragraph}[2][.]{\subsubsection*{#2#1}}

\newcommand{\KEYWORDS}{%
Trusted Execution Environment;
TEE\null;
Confidential Computing;
Privacy and Confidentiality;
Usability;
Application Security}

\title[{SoK}: A Systematic Review of {TEE} Usage]{{SoK}: A Systematic Review of {TEE} Usage for Developing Trusted Applications}

\begin{abstract}
Trusted Execution Environments (TEEs) are a feature of modern central processing units (CPUs) that aim to provide a high assurance,
isolated environment in which to run workloads that demand both confidentiality and integrity.
Hardware and software components in the CPU isolate workloads,
commonly referred to as Trusted Applications (TAs), from the main operating system (OS).
\\
This article aims to analyse the TEE ecosystem, determine its usability, and suggest improvements where necessary to make adoption easier.
\\
To better understand TEE usage, we gathered academic and practical examples from a total of 223 references.
We summarise the literature and provide a publication timeline, along with insights into the evolution of TEE research and deployment.
We categorise TAs into major groups and analyse the tools available to developers.
\\
Lastly, we evaluate trusted container projects, test performance,
and identify the requirements for migrating applications inside them.
 \end{abstract}
\ccsdesc[500]{Security and privacy~Trusted computing}
\ccsdesc[300]{Security and privacy~Software security engineering}
\ccsdesc[300]{Security and privacy~Domain-specific security and privacy architectures}
\ccsdesc[300]{Software and its engineering~Application specific development environments}
\keywords{\KEYWORDS{}}
\copyrightyear{2023}
\acmYear{2023}
\setcopyright{rightsretained}
\acmConference[ARES 2023]{The 18th International Conference on Availability,
Reliability and Security}{August 29-September 1, 2023}{Benevento, Italy}
\acmBooktitle{The 18th International Conference on Availability, Reliability and
Security (ARES 2023), August 29-September 1, 2023, Benevento,
Italy}\acmDOI{10.1145/3600160.3600169}
\acmISBN{979-8-4007-0772-8/23/08}

\begin{document}

\maketitle

\section{Introduction}\label{sec:intro}

Often, sensitive data is processed on general-purpose operating systems (OSs) which are prone to compromise
due to the large number of complex features and services they support.
Typically, when an OS is compromised, the applications and their data are also compromised \citep{fischer2019rpc}.
For instance, if an adversary takes control of an Internet of Things (IoT) device or a cloud instance,
the adversary can also access the processes running there \citep{fischer2019rpc,DBLP:conf/dais/SegarraDLAPS19}.

To help mitigate these risks, modern central processing units (CPUs) support a mode of operation that isolates the applications which manage
sensitive data from the rest of the system. These isolated environments generally only support enough functionality
to enable the processing of this sensitive data. This reduced functionality leads to less code, hence, a smaller
Trusted Computing Base (TCB), which in turn enables us to derive trust in those components. Thus, these modes are
generally referred to as Trusted Execution Environments (TEEs).

A TEE is a component of the CPU that comprises both hardware and software features with
the aim of ensuring the confidentiality and integrity of the code and data loaded inside.
The code that runs inside the TEE is often referred to as a Trusted Application (TA), although it does not have
to be a full application in the traditional sense; it may comprise only the parts of a larger application that
process sensitive data.

\begin{figure}[ht]
    \newcommand{\CREVIEW}{\cellcolor[HTML]{011959}}
    \newcommand{\CFRAMEWORK}{\cellcolor[HTML]{3c6d55}}
    \newcommand{\CCONTAINER}{\cellcolor[HTML]{d29343}}
    \newcommand{\CAPPLICATION}{\cellcolor[HTML]{faccfa}}
    \renewcommand*{\arraystretch}{1.5}
    \resizebox{1.0\linewidth}{!}{%
    \begin{tabular}{*{30}{c}}
\CREVIEW{}&\multicolumn{3}{l}{\textbf{Review}}
&&
&&
&&
&&
&&
&&
&\CREVIEW{}
&\citep{DBLP:journals/corr/abs-2109-01923}
&&
\\
\CFRAMEWORK{}&\multicolumn{3}{l}{\textbf{Framework}}
&&
&&
&&
&&
&&
&\CREVIEW{}
&\citep{DBLP:journals/corr/abs-2105-00378}
&\CFRAMEWORK{}
&\citep{Teekap2022}
&&
\\
\CCONTAINER{}&\multicolumn{3}{l}{\textbf{Container}}
&&
&&
&&
&&
&&
&\CFRAMEWORK{}
&\citep{secGear}
&\CFRAMEWORK{}
&\citep{TeaclaveTZ2022}
&&
\\
\CAPPLICATION{}&\multicolumn{3}{l}{\textbf{Application}}
&&
&&
&&
&&
&&
&\CFRAMEWORK{}
&\citep{linux-sgx}
&\CFRAMEWORK{}
&\citep{Kinibi520a21}
&&
\\
&
&&
&&
&&
&&
&&
&\CREVIEW{}
&\citep{DBLP:journals/corr/abs-1910-04957}
&\CFRAMEWORK{}
&\citep{EdgelessRT2022}
&\CCONTAINER{}
&\citep{Mystikos2022}
&&
\\
&
&&
&&
&&
&&
&&
&\CFRAMEWORK{}
&\citep{mTower}
&\CFRAMEWORK{}
&\citep{DBLP:conf/eurosys/LeeKSAS20}
&\CCONTAINER{}
&\citep{Enarx2022}
&&
\\
&
&&
&&
&&
&&
&&
&\CFRAMEWORK{}
&\citep{SamsungKnoxTizenSDK21}
&\CCONTAINER{}
&\citep{TeaclaveIncubating2022}
&\CCONTAINER{}
&\citep{Ego2022}
&&
\\
&
&&
&&
&&
&&
&&
&\CFRAMEWORK{}
&\citep{DBLP:conf/ndss/BrasserGJSS19}
&\CCONTAINER{}
&\citep{DBLP:journals/corr/abs-2005-05996}
&\CCONTAINER{}
&\citep{Decentriq2022}
&&
\\
&
&&
&&
&&
&&
&&
&\CFRAMEWORK{}
&\citep{Ccf2022}
&\CCONTAINER{}
&\citep{DBLP:conf/asplos/ShenTCCWXXY20}
&\CCONTAINER{}
&\citep{DBLP:conf/ndss/AhmadKSSFL21}
&&
\\
&
&&
&&
&&
&&
&&
&\CCONTAINER{}
&\citep{DBLP:journals/corr/abs-1908-11143}
&\CAPPLICATION{}
&\citep{mobilecoin}
&\CCONTAINER{}
&\citep{DBLP:conf/icde/MenetreyPFS21}
&&
\\
&
&&
&&
&&
&&
&&
&\CCONTAINER{}
&\citep{DBLP:conf/usenix/GhosnLB19}
&\CAPPLICATION{}
&\citep{SafeTraceGit}
&\CCONTAINER{}
&\citep{DBLP:conf/dsn/0004WCWLCWSCT21}
&&
\\
&
&&
&&
&&
&&
&&
&\CCONTAINER{}
&\citep{DBLP:conf/middleware/GoltzscheNKK19}
&\CAPPLICATION{}
&\citep{Keybuster}
&\CAPPLICATION{}
&\citep{enarx-shim}
&&
\\
&
&&
&&
&&
&&
&&
&\CAPPLICATION{}
&\citep{wasm-micro-runtime}
&\CAPPLICATION{}
&\citep{DBLP:journals/wpc/AhnLK20}
&\CAPPLICATION{}
&\citep{DOKMAI2021983}
&&
\\
&
&&
&&
&&
&&
&&
&\CAPPLICATION{}
&\citep{steamlink-sdk}
&\CAPPLICATION{}
&\citep{DBLP:journals/scn/WangZY20}
&\CAPPLICATION{}
&\citep{DBLP:journals/pvldb/SunWL021}
&&
\\
&
&&
&&
&&
&&
&&
&\CAPPLICATION{}
&\citep{kmgk}
&\CAPPLICATION{}
&\citep{DBLP:journals/corr/abs-2010-08855}
&\CAPPLICATION{}
&\citep{DBLP:journals/jpdc/DaiWWLZJ21}
&&
\\
&
&&
&&
&&
&&
&&
&\CAPPLICATION{}
&\citep{FuzzingOPTee19}
&\CAPPLICATION{}
&\citep{DBLP:journals/access/LeeJCKPL20}
&\CAPPLICATION{}
&\citep{DBLP:journals/iotj/ValadaresSPG21}
&&
\\
&
&&
&&
&&
&&
&&
&\CAPPLICATION{}
&\citep{DBLP:conf/uss/MateticWSKKC19}
&\CAPPLICATION{}
&\citep{DBLP:conf/uss/SchwarzR20}
&\CAPPLICATION{}
&\citep{DBLP:journals/iacr/LiWLWWLD21}
&&
\\
&
&&
&&
&&
&&
&&
&\CAPPLICATION{}
&\citep{DBLP:conf/usenix/ParkZLL19}
&\CAPPLICATION{}
&\citep{DBLP:conf/srds/MullerBCFGS20}
&\CAPPLICATION{}
&\citep{DBLP:journals/compsec/JeonK21a}
&&
\\
&
&&
&&
&&
&\CREVIEW{}
&\citep{Tamrakar2017}
&&
&\CAPPLICATION{}
&\citep{DBLP:conf/systor/TrachOGBF19}
&\CAPPLICATION{}
&\citep{DBLP:conf/sp/ZhuH0WCZWZYZM20}
&\CAPPLICATION{}
&\citep{DBLP:journals/chinaf/HuaYGXCZ21}
&&
\\
&
&&
&&
&&
&\CFRAMEWORK{}
&\citep{TeaclaveSGX2022}
&\CFRAMEWORK{}
&\citep{SamsungKnox22}
&\CAPPLICATION{}
&\citep{DBLP:conf/systor/BrennerK19}
&\CAPPLICATION{}
&\citep{DBLP:conf/nsdi/HuntJMSHRW20}
&\CAPPLICATION{}
&\citep{DBLP:journals/access/OhNJCP21}
&&
\\
&
&&
&&
&&
&\CFRAMEWORK{}
&\citep{SamsungTeegrisSDK17}
&\CFRAMEWORK{}
&\citep{OpenEnclave2022}
&\CAPPLICATION{}
&\citep{DBLP:conf/sosp/LindNEKSP19}
&\CAPPLICATION{}
&\citep{DBLP:conf/ndss/PaccagnellaDH0F20}
&\CAPPLICATION{}
&\citep{DBLP:conf/uss/BahmaniBDJKSS21}
&&
\\
&
&&
&&
&&
&\CCONTAINER{}
&\citep{DBLP:conf/usenix/TsaiPV17}
&\CFRAMEWORK{}
&\citep{Asylo2022}
&\CAPPLICATION{}
&\citep{DBLP:conf/ndss/WeiserWBMMS19}
&\CAPPLICATION{}
&\citep{DBLP:conf/msn/SuYLZBZ20}
&\CAPPLICATION{}
&\citep{DBLP:conf/systor/MirandaEPP21}
&&
\\
&
&&
&&
&&
&\CAPPLICATION{}
&\citep{DBLP:journals/cluster/ChangJCXCA17}
&\CCONTAINER{}
&\citep{DBLP:journals/tocs/HuntZXPW18}
&\CAPPLICATION{}
&\citep{DBLP:conf/middleware/SilvaMCNRR19}
&\CAPPLICATION{}
&\citep{DBLP:conf/mobisys/MoSKDLCH20}
&\CAPPLICATION{}
&\citep{DBLP:conf/osdi/LiZZ0XA021}
&\CREVIEW{}
&\citep{DBLP:journals/corr/abs-2205-12742}
\\
&
&&
&&
&\CFRAMEWORK{}
&\citep{linux-sgx2022}
&\CAPPLICATION{}
&\citep{DBLP:conf/sosp/FerraiuoloBHP17}
&\CCONTAINER{}
&\citep{Anjuna2022}
&\CAPPLICATION{}
&\citep{DBLP:conf/iclr/TramerB19}
&\CAPPLICATION{}
&\citep{DBLP:conf/mobisys/MirzamohammadiL20}
&\CAPPLICATION{}
&\citep{DBLP:conf/osdi/FengLDYJXZ021}
&\CREVIEW{}
&\citep{DBLP:conf/uss/ShakevskyRW22}
\\
&
&&
&&
&\CFRAMEWORK{}
&\citep{TrustyTEE20}
&\CAPPLICATION{}
&\citep{DBLP:conf/nsdi/KimHHKH17}
&\CAPPLICATION{}
&\citep{TizenFX}
&\CAPPLICATION{}
&\citep{DBLP:conf/fc/WustMSMKC19}
&\CAPPLICATION{}
&\citep{DBLP:conf/middleware/QuocGAKBF20}
&\CAPPLICATION{}
&\citep{DBLP:conf/mobisys/MoHKMPK21}
&\CREVIEW{}
&\citep{DBLP:conf/seed/AkramAPL22}
\\
&
&&
&&
&\CFRAMEWORK{}
&\citep{DBLP:conf/uss/CostanLD16}
&\CAPPLICATION{}
&\citep{DBLP:conf/eurosec/GoltzscheWMRPK17}
&\CAPPLICATION{}
&\citep{DBLP:conf/www/KrawieckaKPMA18}
&\CAPPLICATION{}
&\citep{DBLP:conf/eurosys/KimPWJH19}
&\CAPPLICATION{}
&\citep{DBLP:conf/icc/LiLM020}
&\CAPPLICATION{}
&\citep{DBLP:conf/micro/KangXJWKYKLJ021}
&\CCONTAINER{}
&\citep{DBLP:journals/tissec/CuiSSSY22}
\\
&
&&
&&
&\CCONTAINER{}
&\citep{FortanixEDP2022}
&\CAPPLICATION{}
&\citep{DBLP:conf/esorics/ChandraKLKKT17}
&\CAPPLICATION{}
&\citep{DBLP:conf/uss/MateticSMJC18}
&\CAPPLICATION{}
&\citep{DBLP:conf/eurosec/AmjadKM19}
&\CAPPLICATION{}
&\citep{DBLP:conf/dsn/FuhryHKK20}
&\CAPPLICATION{}
&\citep{DBLP:conf/icnidc/XuZ0Z21}
&\CCONTAINER{}
&\citep{DBLP:conf/sp/ZhaoLZL22}
\\
&
&&
&&
&\CCONTAINER{}
&\citep{DBLP:conf/osdi/ArnautovTGKMPLM16}
&\CAPPLICATION{}
&\citep{DBLP:conf/ccs/Subramanyan0LDS17}
&\CAPPLICATION{}
&\citep{DBLP:conf/srds/ArnautovBFFGKOM18}
&\CAPPLICATION{}
&\citep{DBLP:conf/dsn/DjokoLL19}
&\CAPPLICATION{}
&\citep{DBLP:conf/dsd/SardarQF20}
&\CAPPLICATION{}
&\citep{DBLP:conf/ic2e/TruongGGW21}
&\CAPPLICATION{}
&\citep{mmledger}
\\
&
&&
&\CFRAMEWORK{}
&\citep{DBLP:conf/trustcom/McGillionDNA15}
&\CAPPLICATION{}
&\citep{DBLP:conf/uss/SaltaformaggioB16}
&\CAPPLICATION{}
&\citep{DBLP:conf/ccs/ShaonKLK17}
&\CAPPLICATION{}
&\citep{DBLP:conf/osdi/VolosVB18}
&\CAPPLICATION{}
&\citep{DBLP:conf/dais/SegarraDLAPS19}
&\CAPPLICATION{}
&\citep{DBLP:conf/codaspy/HuCJCFML20}
&\CAPPLICATION{}
&\citep{DBLP:conf/eurosp/MondalMRG21}
&\CAPPLICATION{}
&\citep{DBLP:journals/corr/abs-2202-07165}
\\
&
&&
&\CAPPLICATION{}
&\citep{DBLP:conf/sp/SchusterCFGPMR15}
&\CAPPLICATION{}
&\citep{DBLP:conf/uss/OhrimenkoSFMNVC16}
&\CAPPLICATION{}
&\citep{DBLP:conf/ccs/KarandeBLK17}
&\CAPPLICATION{}
&\citep{DBLP:conf/nsdi/PoddarLPR18}
&\CAPPLICATION{}
&\citep{DBLP:conf/ccs/ZhaoZQFF19}
&\CAPPLICATION{}
&\citep{DBLP:conf/codaspy/DharPKC20}
&\CAPPLICATION{}
&\citep{DBLP:conf/cf/ZhangWCHM21}
&\CAPPLICATION{}
&\citep{DBLP:conf/uss/ChenZ22}
\\
\CFRAMEWORK{}
&\citep{QSEEInitialRelease}
&&
&\CAPPLICATION{}
&\citep{DBLP:conf/ndss/JangKKKK15}
&\CAPPLICATION{}
&\citep{DBLP:conf/middleware/PiresPFF16}
&\CAPPLICATION{}
&\citep{DBLP:conf/ccs/FischVBG17}
&\CAPPLICATION{}
&\citep{DBLP:conf/icdcs/GoltzschePMBBFK18}
&\CAPPLICATION{}
&\citep{DBLP:conf/ccs/DuanWYZW019}
&\CAPPLICATION{}
&\citep{DBLP:conf/closer/RochaVPGPW20}
&\CAPPLICATION{}
&\citep{DBLP:conf/acsac/GaoDC21}
&\CAPPLICATION{}
&\citep{DBLP:conf/sp/PatatSF22}
\\
\CFRAMEWORK{}
&\citep{DBLP:conf/trust/NamilukoPS13}
&\CREVIEW{}
&\citep{DBLP:conf/mobilecloud/ArfaouiGT14}
&\CAPPLICATION{}
&\citep{DBLP:conf/mobisys/LiLCX15}
&\CAPPLICATION{}
&\citep{DBLP:conf/middleware/BaumanL16}
&\CAPPLICATION{}
&\citep{DBLP:conf/IEEEares/ShepherdAM17}
&\CAPPLICATION{}
&\citep{DBLP:conf/eurosys/AublinKOMPLKFEP18}
&\CAPPLICATION{}
&\citep{DBLP:conf/ccs/ChenZL19}
&\CAPPLICATION{}
&\citep{DBLP:conf/ccs/ParkAL20}
&\CAPPLICATION{}
&\citep{AtlasMScThesis}
&\CAPPLICATION{}
&\citep{ApplePasskeys}
\\
\CAPPLICATION{}
&\citep{DBLP:conf/openidentity/Rijswijk-DeijP13}
&\CFRAMEWORK{}
&\citep{Optee2022}
&\CAPPLICATION{}
&\citep{DBLP:conf/ccs/SunSWJ15}
&\CAPPLICATION{}
&\citep{DBLP:conf/ccs/ZhangCCJS16}
&\CAPPLICATION{}
&\citep{ContactDiscoveryService}
&\CAPPLICATION{}
&\citep{DBLP:conf/acsac/TranLKBS18}
&\CAPPLICATION{}
&\citep{DBLP:conf/asplos/JangTKSH19}
&\CAPPLICATION{}
&\citep{DBLP:conf/acsac/SuzakiTGM20}
&\CAPPLICATION{}
&\citep{9566165}
&\CAPPLICATION{}
&\citep{9512478}
\\
\multicolumn{2}{l}{\textbf{2013}}
&\multicolumn{2}{l}{\textbf{2014}}
&\multicolumn{2}{l}{\textbf{2015}}
&\multicolumn{2}{l}{\textbf{2016}}
&\multicolumn{2}{l}{\textbf{2017}}
&\multicolumn{2}{l}{\textbf{2018}}
&\multicolumn{2}{l}{\textbf{2019}}
&\multicolumn{2}{l}{\textbf{2020}}
&\multicolumn{2}{l}{\textbf{2021}}
&\multicolumn{2}{l}{\textbf{2022}}
     \end{tabular}%
    }
    \Description[TEE literature ordered and categorised]
    {We order the literature by the publication year and categorise them under the categories of review, framework, container, and application.}
    \caption{TEE literature falls under the categories of review, framework, container, and application.}\label{fig:refs}
\end{figure}

The end user of an application or a system is becoming increasingly aware of the need for security, however,
they lack the technical knowledge to make informed decisions. As such, the onus is on the developers and maintainers
of software to make the correct choices for the user in the most transparent manner possible. For this reason, we take the
software developer's perspective and review TEE software development kits (SDKs) and trusted containers (tcons) in order
to determine their usability and, consequently, the likelihood of their adoption by applications.
Our research questions (RQs) are:\\
\textbf{\autoref{rq:apps}}.\refstepcounter{rqcounter}\label{rq:apps}
\textbf{Which use case classification describes TAs?}\\
\textbf{\autoref{rq:frameworks}}.\refstepcounter{rqcounter}\label{rq:frameworks}
\textbf{Which SDKs are available for TA development?}\\
\textbf{\autoref{rq:containers}}.\refstepcounter{rqcounter}\label{rq:containers}
\textbf{What types of tcons are available?}\\
\textbf{\autoref{rq:test}}.\refstepcounter{rqcounter}\label{rq:test}
\textbf{What are the usability implications of porting existing applications to tcons?}

TEE implementations are available from a variety of hardware vendors,
including AMD Secure Encrypted Virtualization (SEV), Intel Software Guard Extensions (SGX), Intel Trusted Domain Extensions (TDX),
ARM TrustZone, and RISC-V Keystone \citep{embeddedbits}. In addition to TEEs, there are a number of solutions
that utilise trusted co-processors:
AMD Platform Security Processor (PSP), Google Titan M, and Apple Secure Enclave Processor (SEP)
provide many of the same benefits, however, they are discrete from the main CPU\null.

Cryptographic primitives are utilised extensively to ensure the confidentiality and integrity of the TEE/TA throughout
its lifecycle. Attestation of the TEE assures that it is in a known good state before code is loaded; signed binaries
ensure that only approved code is loaded; encrypted and integrity-checked data protects it from being read or modified
by untrusted parties. All of this is bound to the CPU which protects it from the main OS, potential attackers, and also
from the user. As surprising as it may sound, the legitimate user of a system may be considered a potential adversary in
certain cases, for example in the case of Digital Rights Management (DRM) or selected banking operations where there is
the potential for financial gain by subverting the system's normal operation.

This separation of TEE and Rich Execution Environment (REE)
allows for more rigorous verification to be applied to the
security-critical parts of a system \citep{DBLP:conf/mobilecloud/ArfaouiGT14}.

\Paragraph{The scope of this review is to understand practical TEE deployments}
\autoref{fig:refs} illustrates our references, categorised and ordered by year.
In cases of multiple references for the same topic, we list the most peer-reviewed one.

Based on our methodology, we systematically collected 208 of 223 references for this article from 11 March 2022 to 7 June 2022,
and only added 15 references after this date.
On 13 February 2023, we updated preprint papers with published versions for references that were officially published after our systematic collection of references,
but we did not search for additional articles published after that date.
Therefore, it is probable that we are missing references published after May 2022.

Our TEE reference search methodology encompasses both academic and applied activity.
We discover that the publications fit well under the categories of review, framework, container, and application.
The selection of these categories is based on the references' natural fit within them.

We notice that publication velocity increased after 2015.
Between 2013 and 2015, there were few publications, but the rate of publication began rising in 2016.
We believe this is likely due to the introduction of Intel SGX in 2015, which sparked a surge in interest in TEE-related research.
Most ARM TrustZone deployments, which are the most common commercially available TEEs, are proprietary and
designed for embedded use cases which renders them inaccessible to most researchers.
19 of the \total{totalcontainerscounter} containers listed in \autoref{tab:containers} support Intel SGX hardware.
Moreover, 70 of the 103 applications listed in \autoref{tab:apps} support SGX\null.
These facts support the argument that SGX technology is the cause of an increase in publications after 2015.

The number of publications in 2018 remained almost unchanged from the previous year.
In 2019, the number of publications nearly doubled and continued to rise in 2020 and 2021.
There are more publications in 2021 than ever before, and we speculate this trend will continue in the foreseeable future.

In \autoref{fig:refs}, 103 of the 154 academic publications demonstrate applications.
43 articles cover frameworks and containers.
8 publications are reviews, including surveys and systematisation of knowledge.

This provides the motivation for our article: systematisation of knowledge is required because there are numerous articles but few reviews.
These reviews have a limited perspective, typically cover only one piece of hardware, and do not seek real-world usage examples.
Our knowledge systematisation assists software developers and encompasses heterogeneous hardware.

This preliminary categorisation of TEE papers shows how academic literature focuses on application demonstrations
and helps form a basis for our research questions.
It is the initial beginning of our extensive systematisation of knowledge.

\begin{table}[htb]
    \Description[Most cited TEE-related articles]{We present 11 most cited academic TEE-related articles.}
    \caption{Most cited TEE-related articles.}\label{tab:recommended_articles}
    \resizebox{1.0\linewidth}{!}{%
\begin{tabular}{p{0.9\linewidth}rrl}
    \toprule
    \textbf{Name} &
    \textbf{Citations} &
    \textbf{Stars} &
    \textbf{Reference} \\
    \midrule
    VC3: Trustworthy Data Analytics in the Cloud Using SGX & 313 & - & \citep{DBLP:conf/sp/SchusterCFGPMR15} \\
    Town Crier: An Authenticated Data Feed for Smart Contracts & 281 & 125 & \citep{DBLP:conf/ccs/ZhangCCJS16,TownCrierGit} \\
    Hypervision Across Worlds: Real-time Kernel Protection from the ARM TrustZone Secure World & 160 & - & \citep{DBLP:conf/ccs/AzabNSCBGMS14} \\
    SCONE\null: Secure Linux Containers With Intel SGX & 123 & - & \citep{DBLP:conf/osdi/ArnautovTGKMPLM16,Scone2022} \\
    Keystone: An Open Framework for Architecting Trusted Execution Environments & 101 & 347 & \citep{DBLP:conf/eurosys/LeeKSAS20,Keystone2022} \\
    Graphene-SGX\null: A Practical Library OS for Unmodified Applications on SGX & 72 & 327 & \citep{DBLP:conf/usenix/TsaiPV17,Gramine2022} \\
    SGX-Log: Securing System Logs With SGX & 68 & 15 & \citep{DBLP:conf/ccs/KarandeBLK17,SgxLogGit} \\
    Komodo: Using Verification to Disentangle Secure-Enclave Hardware from Software & 67 & 92 & \citep{DBLP:conf/sosp/FerraiuoloBHP17,Komodo} \\
    AdAttester: Secure Online Mobile Advertisement Attestation Using TrustZone & 58 & - & \citep{DBLP:conf/mobisys/LiLCX15} \\
    Sanctum: Minimal Hardware Extensions for Strong Software Isolation & 52 & 49 + 22 & \citep{DBLP:conf/uss/CostanLD16,SanctumOldGit,SanctumNewGit} \\
    Teechain: A Secure Payment Network with Asynchronous Blockchain Access & 52 & 47 & \citep{DBLP:conf/sosp/LindNEKSP19,TeechainGit} \\
    \bottomrule
  \end{tabular}
     }
\end{table}

\begin{table}[htb]
    \Description[Most starred TEE-related repositories]{We present 10 most starred TEE-related git repositories.}
    \caption{Most starred TEE-related repositories.}\label{tab:recommended_repos}
    \resizebox{1.0\linewidth}{!}{%
\begin{tabular}{lrl}
    \toprule
    \textbf{Name} &
    \textbf{Stars} &
    \textbf{Reference} \\
    \midrule
    WebAssembly Micro Runtime & 3,425 & \citep{wasm-micro-runtime} \\
    MobileCoin & 1,102 & \citep{mobilecoin} \\
    Intel Software Guard Extensions for Linux* OS & 1,090 & \citep{linux-sgx2022} \\
    Occlum & 1,067 & \citep{Occlum2022,DBLP:conf/asplos/ShenTCCWXXY20} \\
    Teaclave SGX SDK & 1,065 & \citep{TeaclaveSGX2022} \\
    Enarx: Confidential Computing with WebAssembly & 1,048 & \citep{Enarx2022} \\
    Asylo & 925 & \citep{Asylo2022} \\
    Open Enclave SDK\null& 866 & \citep{OpenEnclave2022} \\
    Teaclave: A Universal Secure Computing Platform & 646 & \citep{TeaclaveIncubating2022} \\
    The Confidential Consortium Framework & 640 & \citep{Ccf2022} \\
    \bottomrule
  \end{tabular}
     }
\end{table}

Since our survey comprises of 223 references, we wish to highlight some of the most essential ones.
In \autoref{tab:recommended_articles} and \autoref{tab:recommended_repos},
we recommend 11 TEE-related articles and 10 repositories, respectively.
We base our suggestions on the number of citations for publications and the number of stars for repositories,
which we collected between 23 January 2023 and 25 January 2023.
We collected the number of citations from the \textit{IEEE Xplore}\footnote{\url{https://ieeexplore.ieee.org/}},
\textit{Springer Link}\footnote{\url{https://link.springer.com/}}, and
\textit{ACM Digital Library}\footnote{\url{https://dl.acm.org/}} databases.
This methodology has limitations when it comes to technical and research papers published elsewhere:
the collection method does not take these publications into account.
Similarly, \textit{GitHub}\footnote{\url{https://github.com/}} alone is utilised to determine the number of stars for each repository.

The chosen articles and repositories give a great overview of some of the most important real-world TEE use cases.

\subsection{Related work}\label{sec:related}

As there is prior work on systematising TEE knowledge, we began studying publications and resources that organise TEE utilisation.
These data sources cover the following topics.

\Paragraph{Software~development~kits~({SDKs})}
Each CPU vendor has its own TEE\null.
To assist TA development, there are numerous SDKs to aid the software developer \citep{DBLP:journals/corr/abs-2109-01923}.
Intel SGX SDK \citep{linux-sgx2022}, OP-TEE \citep{Optee2022}, etc.\ intend to make the development easier.

\Paragraph{Trusted~containers~(tcons)}
To execute an application within a TEE, a developer must apply framework-specific modifications to the original application, which can be a time-consuming operation.
Trusted containers solve this usability issue by allowing direct execution of unmodified binary code within a TEE,
or by performing automated transformations on source code prior to loading it into a TEE executable \citep{DBLP:conf/osdi/ArnautovTGKMPLM16}.
Certain tcons support multiple hardware backends, eliminating the need for a software
developer to make hardware design selections at the code level \citep{DBLP:journals/corr/abs-2109-01923}.
We utilise the existing work on tcons by \citet{DBLP:journals/corr/abs-2109-01923} in our categorisations in \autoref{tab:frameworks}
and \autoref{tab:containers}.
Their work provides a comprehensive analysis of 15 existing tcon solutions' designs and implementations, highlighting the most common security pitfalls.
We are not evaluating containers in terms of security,
but rather analysing the software wrapper stack and hardware support of \total{totalcontainerscounter} containers.
Additionally, we check which containers are open source and active as of 2022.
We conclude by comparing the active tcons, benchmarking the performance of various tcons, and discussing the usability of the tcons from our perspective.

\Paragraph{Applications~of~TEEs}
\citet{Tamrakar2017} covers several applications of TEEs, including attestation mechanisms and access control.
Our categorisation of TEE utilisation in \autoref{tab:apps} is not based on said work,
yet we included the applications presented therein.
We also used the study of attestation mechanisms for TEEs by \citet{DBLP:conf/dais/MenetreyGKPFSR22} for systemising knowledge of
TEE attestation applications.
\citet{DBLP:journals/corr/abs-2105-00378} explore how TEEs can be used in conjunction with security technologies
such as homomorphic encryption and differential privacy for efficient \emph{software-hardware-security} codesign.
They propose that security techniques must be combined in order to overcome the inherent limitations of existing technologies.

\Paragraph{Curated~lists~of~TEE~publications}
\citet{SgxPapersGit} maintains a curated list of SGX papers while \citet{TzPapersGit} maintains a similar list for TrustZone publications.
Whereas the former aims to list all peer-reviewed publications regarding SGX,
the latter focuses on attacks against TrustZone-based TEEs and is primarily composed of technical reports, blog postings, and hacking conference presentations.

\Paragraph{TEE~hardware~security}
\citet{DBLP:journals/corr/abs-1910-04957} systemise knowledge of hardware security support for TEEs.
\citet{DBLP:journals/corr/abs-2205-12742} present a systematisation of knowledge pertaining to how
various hardware-based TEE solutions meet the security goals of verifiable launch, run-time isolation, trusted I/O, and secure storage.
This survey is valuable for understanding how present TEE solution designs achieve their security goals
and how existing knowledge can be applied to the development of future TEE solutions.

\Paragraph{Attacks~against~TEEs}
There are also other surveys on TEEs not directly relevant to our work.
For example, presenting how TEEs reduce the attack surface but do not eliminate it.
Numerous attacks have been launched against TEE protection mechanisms and TA implementations \citep{DBLP:journals/csur/FeiYDX21}.
Researchers and practitioners target security flaws and propose solutions for real-world applications, for example,
\citet{DBLP:conf/sp/Cerdeira0FP20} and \citet{DBLP:journals/sensors/Koutroumpouchos21} present a security analysis of popular TrustZone-assisted TEE systems.
\citet{DBLP:conf/seed/AkramAPL22} present a systematisation of knowledge pertaining existing TEEs,
highlighting common mechanisms of security guarantees, and offering comparative analyses of different TEE proposals.
They also bring up the current limitations of TEEs for high-performance computing systems.

\subsection{TEE use cases}\label{sec:uses}

A TEE technology provides extra protection for various sensitive applications.
The following are the most prevalent usage scenarios \citep{Tamrakar2017}:

\Paragraph{Digital~rights~management}
Copyright holders frequently use TEEs to prevent consumers from copying video or audio \citep{DBLP:conf/sp/PatatSF22}.
TEEs protect digitally encoded media on connected devices, including smartphones, tablets, and high-definition televisions \citep{DBLP:conf/ccs/Asokan19,DBLP:journals/ieeesp/EkbergKA14}.
Along with the fact that the TEE and the device's display are connected via a protected hardware channel, this prevents the device's owner from reading stored secrets.

\Paragraph{Online~payments}
Mobile wallets, peer-to-peer payments, cryptocurrency wallets, and the use of
a mobile device as a point-of-sale terminal -- all have well-defined security requirements.
Blockchain systems use lightweight clients,
which outsource the computational and storage load over full blockchain nodes \citep{DBLP:conf/uss/MateticWSKKC19}.
It is possible to use TEEs to protect the privacy of the light clients without compromising the performance of the assisting full nodes \citep{DBLP:conf/uss/MateticWSKKC19}.
TEEs can be used as trusted backend systems to provide
the required security to facilitate financial transactions.
This may necessitate the entry of a PIN, password, or biometric identifier by the user.

\Paragraph{Authentication}
TEEs are commonly used to implement biometric identity methods (facial recognition, fingerprint sensor, and voice authorisation).
For instance, Android OS can save
fingerprint biometrics in the TEE because it is inaccessible and encrypted from the ordinary OS environment \citep{Biometric-Authentication}.
Often, biometric identifications are convenient to use and more difficult to steal than PINs and passwords.
TEEs can be utilised to protect the biometric identification method. However, increasingly, biometric data is being stored and verified
directly on the sensors and only an attestation is shared with the TEE\null.
Similarly to biometric identification information, cryptographic private keys can also be stored in the TEE\null.
Combining the biometric identification information and the private keys allows passwordless authentication standards
such as Apple's passkeys \citep{ApplePasskeys}.

\Paragraph{Trusted~cloud}
Typically, when a cloud (the server or the backend) is compromised, the adversary gains access to the cloud's processes and data.
TEEs provide protection against compromised infrastructure:
the adversary is unable to access selective parts of the TA, which safeguards sensitive code and data.

\Paragraph{Privacy-preserving~data~analysis}
Machine learning has become an essential part of data processing in several application domains, such as healthcare, stock prediction,
and artificial intelligence.
Sometimes these applications process sensitive data,
and to protect said data, a TEE-based solution
can be used to maintain the integrity of the machine learning process and prevent attacks \citep{DBLP:journals/isci/ChenLLXLL20}.

\Paragraph{Runtime~integrity}
TEEs can be used for runtime integrity, such as real-time kernel protection. If an attacker targets kernel binaries,
the security monitoring service can shut down if it is isolated in a secure environment \citep{DBLP:conf/ccs/AzabNSCBGMS14}.

\Paragraph{Secure~modular~programming}
As it decouples functionalities into small, self-contained modules,
modular programming is an efficient way to build software architectures for software assets that encourages reuse.
In this instance, each module contains everything necessary to perform its intended function,
and the TEE permits the execution of the module while protecting it from the vulnerabilities of other modules.
\section{Methodology}\label{sec:methodology}

\subsection{Collecting references for the review}

We began our search for scientific literature with
\textit{Google Scholar}\footnote{\url{https://scholar.google.com/}},
\textit{arXiv open-access archive}\footnote{\url{https://arxiv.org/}},
\textit{the DBLP computer science bibliography}\footnote{\url{https://dblp.org/}},
\textit{Andor}\footnote{\url{https://andor.tuni.fi/}},
\textit{ACM Digital Library}, and
\textit{IEEE Xplore}
using TEE-related search terms, such as
``TEE'', ``Trusted Execution Environment'', ``OP-TEE'', ``TrustZone'', ``(Intel) SGX'', ``AMD SEV'', ``confidential computing'', etc.
While this paints an overall picture of TEE-related scholarly work,
it does not cover more applied aspects, such as toolkits and deployments.

To address this gap,
we then mined real source code using
the Sourcegraph\footnote{\url{https://sourcegraph.com/}}
search engine, to find examples of practical TEE utilisation.
Sourcegraph covers \textit{GitLab}\footnote{\url{https://gitlab.com/}}, \textit{GitHub},
and \textit{BitBucket}\footnote{\url{https://bitbucket.org/}}, as well as other public software source repositories.
\autoref{tab:search} details our search terms regarding Sourcegraph,
with examples\footnote{\url{https://sourcegraph.com/search?q=context:global+Op-TEE&patternType=literal}}.
The most difficult aspect of the mining process was locating appropriate TEE applications, development frameworks, and container repositories.
Typically, a keyword search yields thousands of repositories. These repositories contain OSs and kernels, as well as forks and projects with work-in-progress status.
Furthermore, we specified ``code'' as the search type, then sorted and filtered the results to identify the most relevant ones, then finally, manually examined the results.

We based our selection of important phrases on the constants, variables, and functions utilised in the source code of each TEE-based application.
The alternative method for picking specific search phrases was to consult the documentation of various TEE-based frameworks and containers,
such as the GlobalPlatform API \citep{GlobalPlatformAPI}.
It reveals applications and other frameworks, containers, and repositories.
However, this required combing through each repository manually to obtain the desired results.

\begin{table}[ht]
    \Description[Search of real-world TEE applications]{Using the Sourcegraph search engine, we compile real-world applications of TEEs with various search terms.}
    \caption{Using the Sourcegraph search engine, we compile real-world applications of TEEs with the provided search terms.}\label{tab:search}
    \resizebox{1.0\linewidth}{!}{%
\begin{tabular}{p{0.5\linewidth}lll}
  \toprule
  \textbf{Search~terms} & \textbf{Applications} & \textbf{Containers} & \textbf{Frameworks}\ \\
  \midrule
  {SGX\_CREATE\_ENCLAVE\_\-EX\_PCL\_\-BIT\_IDX} & 1\citep{ContactDiscoveryService} & 1\citep{Twine2021} & 2\citep{TeaclaveSGX2022,OpenEnclave2022} \\
  {TEEC\_InvokeCommand} & 5\citep{TizenFX,darknetz,Keybuster,kmgk,steamlink-sdk} & 0 & 1\citep{mTower}\\
  {SGX\_CREATE\_ENCLAVE\_\-EX\_\-SWITCHLESS} & 3\citep{ShieldStore,sengsgx,linux-sgx-mage} & 0 & 2\citep{linux-sgx,secGear} \\
  {TEEC\_MEMREF\_TEMP\_\-OUTPUT} & 3\citep{optee_fuzzer,RetroScope,PPFLgit} & 0 & 0 \\
  {sgx\_enclave\_id\_t} & 2\citep{mobilecoin,wasm-micro-runtime} & 1\citep{Occlum2022} & 1\citep{Asylo2022} \\
  {TEEC\_RegisterSharedMemory} & 0 & 0 & 1\citep{optee_client,trustonic-tee-user-space} \\
  {enarx} & 2\citep{mmledger,enarx-shim} & 0 & 0 \\
  \bottomrule
\end{tabular}
     }
\end{table}

\subsection{Dimensions for knowledge systematisation}

Based on related work and our observations while gathering and reviewing the publications, we organise the TEE literature and practical work.

\begin{figure}
\centering
\includegraphics[width=\linewidth]{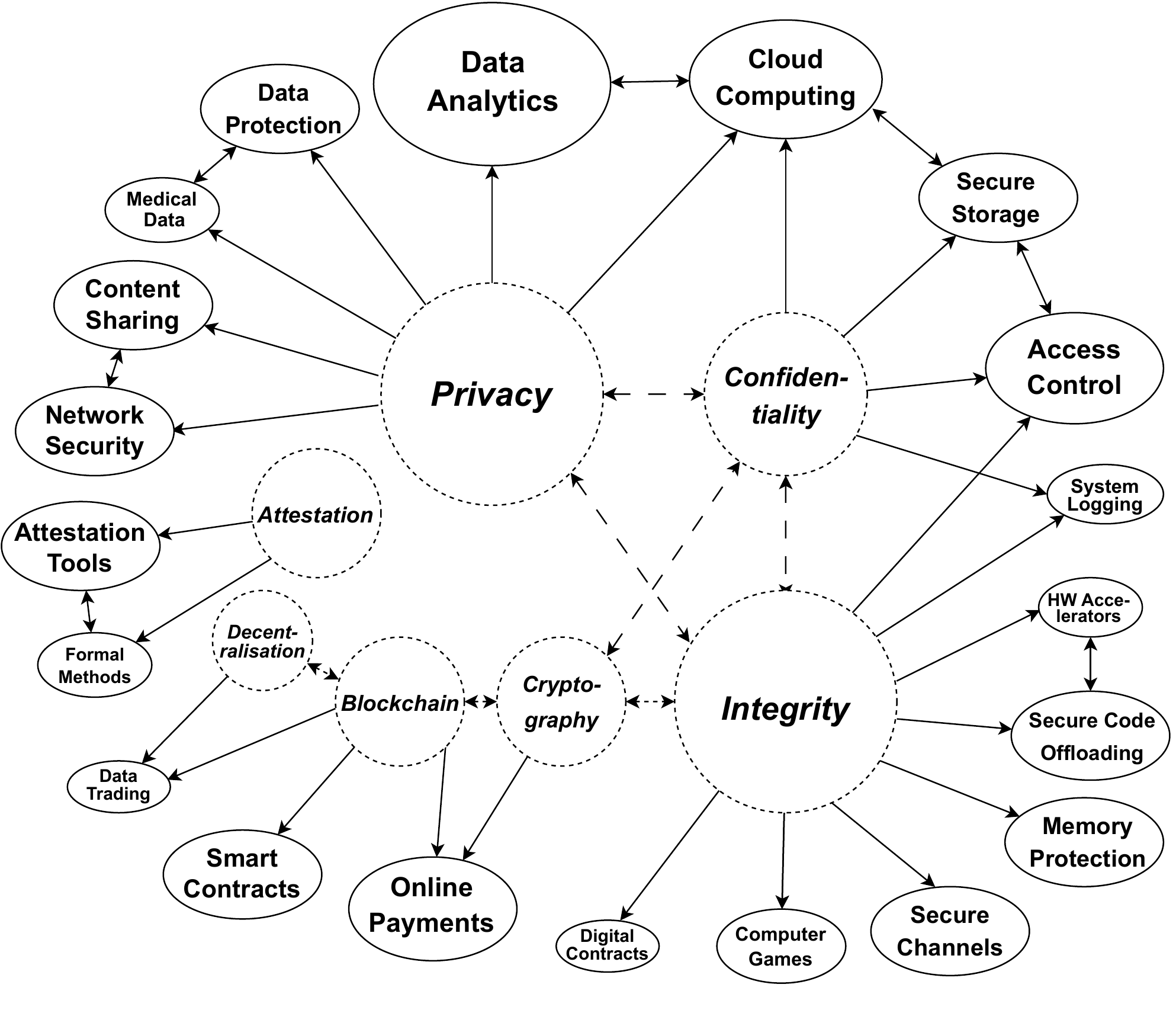}
\Description[TEE application category relationships]
{We categorise example TEE applications into 21 different use cases and highlight the most significant relationships between the categories and the related security properties and mechanisms.}
\caption{We define which classification aggregates the usage examples after reviewing the TEE example applications.
Only the most significant relationships between the categories and the security properties and mechanisms are depicted.}\label{fig:methodology}
\end{figure}

\Paragraph{In~\autoref{sec:applications}~we~address~\autoref*{rq:apps}}
Our goal is to assist the reader in comprehending TEEs, how they are utilised, and when, how, and why they could be used.
To accomplish this, we tag the applications with 92 distinct keywords,
which we merge into 21 primary categories and seven distinct security properties and mechanisms based on initial similarities.
We discover that the primary 21 use cases for TEEs in application development are
\textit{data analytics},
\textit{cloud computing},
\textit{access control},
\textit{data protection},
\textit{online payments},
\textit{memory protection},
\textit{attestation tools},
\textit{secure storage},
\textit{network security},
\textit{secure channels},
\textit{content sharing},
\textit{secure code offloading},
\textit{smart contracts},
\textit{computer games},
\textit{hardware accelerators},
\textit{formal methods},
\textit{medical data},
\textit{secure system logging},
\textit{web search},
\textit{data trading},
and \textit{digital contracts}.
Additionally, we discover that the main security properties and mechanisms related to the use cases are
\textit{privacy},
\textit{integrity},
\textit{confidentiality},
\textit{cryptography},
\textit{attestation},
\textit{blockchain},
and \textit{decentralisation}.
This is the classification we utilise while reviewing existing TA demonstrations and practical implementations.
\autoref{fig:methodology} illustrates our categorisation of the key use cases and the related security properties and mechanisms.
Only the strongest relationships are shown in the figure.
The size of a category, security property, or mechanism approximately corresponds to its prevalence in existing implementations.
\autoref{tab:apps} shows which applications are related to each primary category.

\Paragraph{In~\autoref{sec:frameworks}~we~address~\autoref*{rq:frameworks}}
We compare the software frameworks targeting developers.
It is difficult to compare TEE software development tools due to a lack of similar work and public information about their features.
Hence, we compile \autoref{tab:frameworks} detailing the available tools, their supported programming languages, their software licences,
the hardware architecture they support, and whether or not they are actively developed as of 2022.

\Paragraph{In~\autoref{sec:containers}~we~address~\autoref*{rq:containers}}
We organise the tcons for the developers.
Again, it is difficult to directly compare tcon tools due to the absence of shared and unique characteristics.
In addition, some containers are not actively developed, while others, such as the Enarx container \citep{Enarx2022}, are updated every month with new features.
In response, we compile \autoref{tab:containers}, which details the available tools, interfaces, software licences,
activity of the container project as of 2022, and hardware supported by each tcon.

\Paragraph{In~\autoref{sec:benchmark}~we~address~\autoref*{rq:test}}
We compare the actively developed containers.
Additionally, we present \autoref{fig:rust-tcp-proxy}, which exemplifies a required tcon-specific modification to existing code,
demonstrating how challenging it can be to use tcons.
Finally, we benchmark the performance of an existing test application using various tcons and present the findings in \autoref{tab:benchmark}.

\subsection{Limitations and bias}
\Paragraph{Lack~of~documentation~of~closed-source~systems}
Companies that own proprietary solutions utilising TEEs typically withhold information about their systems from the public.
Therefore, it is difficult to find detailed information regarding closed-source solutions that employ TEEs.
Because of this, the data we gathered might be biased towards open-source software and might not show the whole picture of reality.
For example, there may be several more closed-source applications and development frameworks for ARM TrustZone than we present in this paper.

\Paragraph{Lack~of~citation~data}
Certain venues or sources do not disclose the number of citations.
This imposes restrictions on which technical and research papers can be listed in \autoref{tab:recommended_articles}.

\Paragraph{Date~of~initial~release}
It is often difficult to discover when a specific application, framework, or container was first released.
Due to this, the publication year information in \autoref{fig:refs} may not be entirely accurate.

\Paragraph{Manual~keyword~search}
The likelihood of omitting relevant repositories is the most significant shortcoming of a manual search.
Although using Sourcegraph as a repository search engine simplifies the search process, it also generates a large number of irrelevant results.
There is a chance of missing other applications, development frameworks, and containers that employ different keywords not on our list.
\section{Application scenarios for TEE}\label{sec:applications}

\textit{\autoref*{rq:apps}: Which use case classification describes TAs?}

The TEE isolates and protects the TA code and data in terms of confidentiality and integrity.
While we may be unaware, there are a large number of gadgets around us, most notably smartphones, set-top boxes, videogame consoles,
and Smart TVs, that utilise a TEE\null. The number of gadgets utilising a TEE that are designed for many different purposes results in a wide range of use cases.
These use cases vary from everyday user applications to backend services, such as mobile financial services and cloud services \citep{Tamrakar2017}.
To address \autoref*{rq:apps}, \autoref{tab:apps} combines TEE application scenarios based on our categorisation.

\begin{table*}[!htb]
\Description[TEE applications categorised]
{We classify TEE applications into 21 different use cases, show supported hardware architectures, show if they are deployed to real users, and if they are published under an open-source licence.}
\caption{We classified TEE application scenarios into 21 groups.}\label{tab:apps}
\newcounter{totalappscounter}
\centering
\resizebox*{!}{0.95\textheight}{%
\setlength{\tabcolsep}{1pt}
\begin{tabular}{llc*{30}{c}}
\toprule
\newcounter{dataanalyticscounter}%
\newcounter{cloudcomputingcounter}%
\newcounter{accesscontrolcounter}%
\newcounter{dataprotectioncounter}%
\newcounter{onlinepaymentscounter}%
\newcounter{memoryprotectioncounter}%
\newcounter{attestationtoolcounter}%
\newcounter{securecodeoffloadingcounter}%
\newcounter{securestoragecounter}%
\newcounter{networksecuritycounter}%
\newcounter{securechannelscounter}%
\newcounter{contentsharingcounter}%
\newcounter{smartcontractscounter}%
\newcounter{computergamescounter}%
\newcounter{hardwareacceleratorscounter}%
\newcounter{formalmethodscounter}%
\newcounter{medicaldatacounter}%
\newcounter{securesystemloggingcounter}%
\newcounter{websearchcounter}%
\newcounter{datatradingcounter}%
\newcounter{digitalcontractscounter}%
&&&
\rotatebox[origin=r]{270}{\textbf{Data analytics}}&
\rotatebox[origin=r]{270}{\textbf{Cloud computing}}&
\rotatebox[origin=r]{270}{\textbf{Access control}}&
\rotatebox[origin=r]{270}{\textbf{Data protection}}&
\rotatebox[origin=r]{270}{\textbf{Online payments}}&
\rotatebox[origin=r]{270}{\textbf{Memory protection}}&
\rotatebox[origin=r]{270}{\textbf{Attestation tool}}&
\rotatebox[origin=r]{270}{\textbf{Secure code offloading}}&
\rotatebox[origin=r]{270}{\textbf{Secure storage}}&
\rotatebox[origin=r]{270}{\textbf{Network security}}&
\rotatebox[origin=r]{270}{\textbf{Secure channels}}&
\rotatebox[origin=r]{270}{\textbf{Content sharing}}&
\rotatebox[origin=r]{270}{\textbf{Smart contracts}}&
\rotatebox[origin=r]{270}{\textbf{Computer games}}&
\rotatebox[origin=r]{270}{\textbf{Hardware accelerators}}&
\rotatebox[origin=r]{270}{\textbf{Formal methods}}&
\rotatebox[origin=r]{270}{\textbf{Medical data}}&
\rotatebox[origin=r]{270}{\textbf{Secure system logging}}&
\rotatebox[origin=r]{270}{\textbf{Web search}}&
\rotatebox[origin=r]{270}{\textbf{Data trading}}&
\rotatebox[origin=r]{270}{\textbf{Digital contracts}}&
\phantom{z}&
\rotatebox[origin=r]{270}{\textbf{Intel SGX}}&
\rotatebox[origin=r]{270}{\textbf{ARM TrustZone}}&
\rotatebox[origin=r]{270}{\textbf{RISC-V}}&
\rotatebox[origin=r]{270}{\textbf{GPU TEEs}}&
\phantom{z}&
\rotatebox[origin=r]{270}{\textbf{Open source}}&
\rotatebox[origin=r]{270}{\textbf{Real deployments}}\\\midrule
\multirow{5}{*}{\rotatebox[origin=c]{270}{\textbf{$\leq$ 2015 \phantom{$\leq$}}}}
&{AdAttester}: Secure Online Mobile Advertisement Attestation Using {TrustZone}&\addtocounter{totalappscounter}{1}\citep{DBLP:conf/mobisys/LiLCX15}&
$\circ$&
$\circ$&
$\circ$&
$\circ$&
$\circ$&
$\circ$&
\stepcounter{attestationtoolcounter}$\bullet$&
$\circ$&
$\circ$&
$\circ$&
$\circ$&
$\circ$&
$\circ$&
$\circ$&
$\circ$&
$\circ$&
$\circ$&
$\circ$&
$\circ$&
$\circ$&
$\circ$&
&
$\circ$&
$\bullet$&
$\circ$&
$\circ$&
&
$\circ$&
$\circ$&
\\
&{SeCReT}: Secure Channel between Rich Execution Environment and {TEE}&\addtocounter{totalappscounter}{1}\citep{DBLP:conf/ndss/JangKKKK15}&
$\circ$&
$\circ$&
$\circ$&
$\circ$&
$\circ$&
\stepcounter{memoryprotectioncounter}$\bullet$&
$\circ$&
$\circ$&
$\circ$&
$\circ$&
\stepcounter{securechannelscounter}$\bullet$&
$\circ$&
$\circ$&
$\circ$&
$\circ$&
$\circ$&
$\circ$&
$\circ$&
$\circ$&
$\circ$&
$\circ$&
&
$\circ$&
$\bullet$&
$\circ$&
$\circ$&
&
$\circ$&
$\circ$&
\\
&{TrustOTP}: Transforming Smartphones into Secure One-Time Password Tokens&\addtocounter{totalappscounter}{1}\citep{DBLP:conf/ccs/SunSWJ15}&
$\circ$&
$\circ$&
\stepcounter{accesscontrolcounter}$\bullet$&
$\circ$&
$\circ$&
$\circ$&
$\circ$&
$\circ$&
$\circ$&
$\circ$&
$\circ$&
$\circ$&
$\circ$&
$\circ$&
$\circ$&
$\circ$&
$\circ$&
$\circ$&
$\circ$&
$\circ$&
$\circ$&
&
$\circ$&
$\bullet$&
$\circ$&
$\circ$&
&
$\circ$&
$\bullet$&
\\
&Using TEEs in Two-factor Authentication: comparing approaches&\addtocounter{totalappscounter}{1}\citep{DBLP:conf/openidentity/Rijswijk-DeijP13}&
$\circ$&
$\circ$&
\stepcounter{accesscontrolcounter}$\bullet$&
$\circ$&
$\circ$&
$\circ$&
$\circ$&
$\circ$&
$\circ$&
$\circ$&
$\circ$&
$\circ$&
$\circ$&
$\circ$&
$\circ$&
$\circ$&
$\circ$&
$\circ$&
$\circ$&
$\circ$&
$\circ$&
&
$\circ$&
$\bullet$&
$\circ$&
$\circ$&
&
$\circ$&
$\circ$&
\\
&{VC3}: Trustworthy Data Analytics in the Cloud Using {SGX}&\addtocounter{totalappscounter}{1}\citep{DBLP:conf/sp/SchusterCFGPMR15}&
\stepcounter{dataanalyticscounter}$\bullet$&
\stepcounter{cloudcomputingcounter}$\bullet$&
$\circ$&
$\circ$&
$\circ$&
$\circ$&
$\circ$&
$\circ$&
$\circ$&
$\circ$&
$\circ$&
$\circ$&
$\circ$&
$\circ$&
$\circ$&
$\circ$&
$\circ$&
$\circ$&
$\circ$&
$\circ$&
$\circ$&
&
$\bullet$&
$\circ$&
$\circ$&
$\circ$&
&
$\circ$&
$\circ$&
\\
\midrule
\multirow{5}{*}{\rotatebox[origin=c]{270}{\textbf{2016}}}
&A Case for Protecting Computer Games With {SGX}&\addtocounter{totalappscounter}{1}\citep{DBLP:conf/middleware/BaumanL16}&
$\circ$&
$\circ$&
$\circ$&
$\circ$&
$\circ$&
$\circ$&
$\circ$&
$\circ$&
$\circ$&
$\circ$&
$\circ$&
$\circ$&
$\circ$&
\stepcounter{computergamescounter}$\bullet$&
$\circ$&
$\circ$&
$\circ$&
$\circ$&
$\circ$&
$\circ$&
$\circ$&
&
$\bullet$&
$\circ$&
$\circ$&
$\circ$&
&
$\circ$&
$\circ$&
\\
&Oblivious Multi-Party Machine Learning on Trusted Processors&\addtocounter{totalappscounter}{1}\citep{DBLP:conf/uss/OhrimenkoSFMNVC16}&
\stepcounter{dataanalyticscounter}$\bullet$&
$\circ$&
$\circ$&
$\circ$&
$\circ$&
$\circ$&
$\circ$&
$\circ$&
$\circ$&
$\circ$&
$\circ$&
$\circ$&
$\circ$&
$\circ$&
$\circ$&
$\circ$&
$\circ$&
$\circ$&
$\circ$&
$\circ$&
$\circ$&
&
$\bullet$&
$\circ$&
$\circ$&
$\circ$&
&
$\circ$&
$\circ$&
\\
&Screen after Previous Screens: Spatial-Temporal Recreation of {Android} App Displays from Memory Images&\addtocounter{totalappscounter}{1}\citep{DBLP:conf/uss/SaltaformaggioB16,RetroScope}&
$\circ$&
$\circ$&
$\circ$&
$\circ$&
$\circ$&
\stepcounter{memoryprotectioncounter}$\bullet$&
$\circ$&
$\circ$&
$\circ$&
$\circ$&
$\circ$&
$\circ$&
$\circ$&
$\circ$&
$\circ$&
$\circ$&
$\circ$&
$\circ$&
$\circ$&
$\circ$&
$\circ$&
&
$\circ$&
$\bullet$&
$\circ$&
$\circ$&
&
$\bullet$&
$\bullet$&
\\
&Secure Content-Based Routing Using {Intel} Software Guard Extensions&\addtocounter{totalappscounter}{1}\citep{DBLP:conf/middleware/PiresPFF16,ScbrGit}&
$\circ$&
$\circ$&
$\circ$&
$\circ$&
$\circ$&
$\circ$&
$\circ$&
$\circ$&
$\circ$&
\stepcounter{networksecuritycounter}$\bullet$&
$\circ$&
\stepcounter{contentsharingcounter}$\bullet$&
$\circ$&
$\circ$&
$\circ$&
$\circ$&
$\circ$&
$\circ$&
$\circ$&
$\circ$&
$\circ$&
&
$\bullet$&
$\circ$&
$\circ$&
$\circ$&
&
$\bullet$&
$\bullet$&
\\
&Town Crier: An Authenticated Data Feed for Smart Contracts&\addtocounter{totalappscounter}{1}\citep{DBLP:conf/ccs/ZhangCCJS16,TownCrierGit}&
$\circ$&
$\circ$&
\stepcounter{accesscontrolcounter}$\bullet$&
$\circ$&
$\circ$&
$\circ$&
$\circ$&
$\circ$&
$\circ$&
$\circ$&
$\circ$&
$\circ$&
\stepcounter{smartcontractscounter}$\bullet$&
$\circ$&
$\circ$&
$\circ$&
$\circ$&
$\circ$&
$\circ$&
$\circ$&
$\circ$&
&
$\bullet$&
$\circ$&
$\circ$&
$\circ$&
&
$\bullet$&
$\bullet$&
\\
\midrule
\multirow{11}{*}{\rotatebox[origin=c]{270}{\textbf{2017}}}
&A Formal Foundation for Secure Remote Execution of Enclaves&\addtocounter{totalappscounter}{1}\citep{DBLP:conf/ccs/Subramanyan0LDS17}&
$\circ$&
$\circ$&
$\circ$&
$\circ$&
$\circ$&
$\circ$&
$\circ$&
$\circ$&
$\circ$&
$\circ$&
$\circ$&
$\circ$&
$\circ$&
$\circ$&
$\circ$&
\stepcounter{formalmethodscounter}$\bullet$&
$\circ$&
$\circ$&
$\circ$&
$\circ$&
$\circ$&
&
$\bullet$&
$\circ$&
$\bullet$&
$\circ$&
&
$\circ$&
$\circ$&
\\
&Enhancing Security and Privacy of {Tor's} Ecosystem by Using TEEs&\addtocounter{totalappscounter}{1}\citep{DBLP:conf/nsdi/KimHHKH17,SgxTorGit}&
$\circ$&
$\circ$&
$\circ$&
$\circ$&
$\circ$&
$\circ$&
$\circ$&
$\circ$&
$\circ$&
\stepcounter{networksecuritycounter}$\bullet$&
$\circ$&
$\circ$&
$\circ$&
$\circ$&
$\circ$&
$\circ$&
$\circ$&
$\circ$&
$\circ$&
$\circ$&
$\circ$&
&
$\bullet$&
$\circ$&
$\circ$&
$\circ$&
&
$\bullet$&
$\bullet$&
\\
&Establishing Mutually Trusted Channels for Remote Sensing Devices with TEEs&\addtocounter{totalappscounter}{1}\citep{DBLP:conf/IEEEares/ShepherdAM17}&
$\circ$&
$\circ$&
$\circ$&
$\circ$&
$\circ$&
$\circ$&
$\circ$&
$\circ$&
$\circ$&
$\circ$&
\stepcounter{securechannelscounter}$\bullet$&
$\circ$&
$\circ$&
$\circ$&
$\circ$&
$\circ$&
$\circ$&
$\circ$&
$\circ$&
$\circ$&
$\circ$&
&
$\bullet$&
$\circ$&
$\circ$&
$\circ$&
&
$\circ$&
$\circ$&
\\
&{IRON}: Functional Encryption using {Intel} {SGX}&\addtocounter{totalappscounter}{1}\citep{DBLP:conf/ccs/FischVBG17}&
$\circ$&
$\circ$&
\stepcounter{accesscontrolcounter}$\bullet$&
$\circ$&
$\circ$&
$\circ$&
$\circ$&
$\circ$&
$\circ$&
$\circ$&
$\circ$&
$\circ$&
$\circ$&
$\circ$&
$\circ$&
\stepcounter{formalmethodscounter}$\bullet$&
$\circ$&
$\circ$&
$\circ$&
$\circ$&
$\circ$&
&
$\bullet$&
$\circ$&
$\circ$&
$\circ$&
&
$\circ$&
$\circ$&
\\
&{Komodo}: Using verification to disentangle secure-enclave hardware from software&\addtocounter{totalappscounter}{1}\citep{DBLP:conf/sosp/FerraiuoloBHP17,Komodo}&
$\circ$&
$\circ$&
$\circ$&
$\circ$&
$\circ$&
$\circ$&
$\circ$&
$\circ$&
$\circ$&
$\circ$&
$\circ$&
$\circ$&
$\circ$&
$\circ$&
$\circ$&
$\circ$&
$\circ$&
$\circ$&
$\circ$&
$\circ$&
$\circ$&
&
$\bullet$&
$\bullet$&
$\circ$&
$\circ$&
&
$\bullet$&
$\bullet$&
\\
&{MIPE}: a practical memory integrity protection method in a {TEE}&\addtocounter{totalappscounter}{1}\citep{DBLP:journals/cluster/ChangJCXCA17}&
$\circ$&
$\circ$&
$\circ$&
$\circ$&
$\circ$&
\stepcounter{memoryprotectioncounter}$\bullet$&
$\circ$&
$\circ$&
$\circ$&
$\circ$&
$\circ$&
$\circ$&
$\circ$&
$\circ$&
$\circ$&
$\circ$&
$\circ$&
$\circ$&
$\circ$&
$\circ$&
$\circ$&
&
$\circ$&
$\bullet$&
$\circ$&
$\circ$&
&
$\circ$&
$\circ$&
\\
&{Private Contact Discovery Service}&\addtocounter{totalappscounter}{1}\citep{ContactDiscoveryService}&
$\circ$&
$\circ$&
$\circ$&
\stepcounter{dataprotectioncounter}$\bullet$&
$\circ$&
$\circ$&
$\circ$&
$\circ$&
$\circ$&
$\circ$&
$\circ$&
$\circ$&
$\circ$&
$\circ$&
$\circ$&
$\circ$&
$\circ$&
$\circ$&
$\circ$&
$\circ$&
$\circ$&
&
$\bullet$&
$\circ$&
$\circ$&
$\circ$&
&
$\bullet$&
$\bullet$&
\\
&Securing Data Analytics on {SGX} with Randomization&\addtocounter{totalappscounter}{1}\citep{DBLP:conf/esorics/ChandraKLKKT17,secure-analytics-sgx}&
\stepcounter{dataanalyticscounter}$\bullet$&
$\circ$&
$\circ$&
$\circ$&
$\circ$&
$\circ$&
$\circ$&
$\circ$&
$\circ$&
$\circ$&
$\circ$&
$\circ$&
$\circ$&
$\circ$&
$\circ$&
$\circ$&
$\circ$&
$\circ$&
$\circ$&
$\circ$&
$\circ$&
&
$\bullet$&
$\circ$&
$\circ$&
$\circ$&
&
$\bullet$&
$\bullet$&
\\
&SGX-BigMatrix: {A} Practical Encrypted Data Analytic Framework With Trusted Processors&\addtocounter{totalappscounter}{1}\citep{DBLP:conf/ccs/ShaonKLK17}&
\stepcounter{dataanalyticscounter}$\bullet$&
\stepcounter{cloudcomputingcounter}$\bullet$&
$\circ$&
$\circ$&
$\circ$&
$\circ$&
$\circ$&
$\circ$&
$\circ$&
$\circ$&
$\circ$&
$\circ$&
$\circ$&
$\circ$&
$\circ$&
$\circ$&
$\circ$&
$\circ$&
$\circ$&
$\circ$&
$\circ$&
&
$\bullet$&
$\circ$&
$\circ$&
$\circ$&
&
$\circ$&
$\circ$&
\\
&{SGX-Log}: Securing System Logs With {SGX}&\addtocounter{totalappscounter}{1}\citep{DBLP:conf/ccs/KarandeBLK17,SgxLogGit}&
$\circ$&
$\circ$&
$\circ$&
$\circ$&
$\circ$&
$\circ$&
$\circ$&
$\circ$&
$\circ$&
$\circ$&
$\circ$&
$\circ$&
$\circ$&
$\circ$&
$\circ$&
$\circ$&
$\circ$&
\stepcounter{securesystemloggingcounter}$\bullet$&
$\circ$&
$\circ$&
$\circ$&
&
$\bullet$&
$\circ$&
$\circ$&
$\circ$&
&
$\bullet$&
$\bullet$&
\\
&{TrustJS}: Trusted Client-side Execution of {JavaScript}&\addtocounter{totalappscounter}{1}\citep{DBLP:conf/eurosec/GoltzscheWMRPK17}&
$\circ$&
$\circ$&
$\circ$&
$\circ$&
$\circ$&
$\circ$&
$\circ$&
\stepcounter{securecodeoffloadingcounter}$\bullet$&
$\circ$&
\stepcounter{networksecuritycounter}$\bullet$&
$\circ$&
$\circ$&
$\circ$&
$\circ$&
$\circ$&
$\circ$&
$\circ$&
$\circ$&
$\circ$&
$\circ$&
$\circ$&
&
$\bullet$&
$\circ$&
$\circ$&
$\circ$&
&
$\circ$&
$\circ$&
\\
\midrule
\multirow{9}{*}{\rotatebox[origin=c]{270}{\textbf{2018}}}
&{CYCLOSA}: Decentralizing Private Web Search through {SGX}-Based Browser Extensions&\addtocounter{totalappscounter}{1}\citep{DBLP:conf/icdcs/GoltzschePMBBFK18}&
$\circ$&
$\circ$&
$\circ$&
\stepcounter{dataprotectioncounter}$\bullet$&
$\circ$&
$\circ$&
$\circ$&
$\circ$&
$\circ$&
$\circ$&
$\circ$&
$\circ$&
$\circ$&
$\circ$&
$\circ$&
$\circ$&
$\circ$&
$\circ$&
\stepcounter{websearchcounter}$\bullet$&
$\circ$&
$\circ$&
&
$\bullet$&
$\circ$&
$\circ$&
$\circ$&
&
$\circ$&
$\circ$&
\\
&{DelegaTEE}: Brokered Delegation Using TEEs&\addtocounter{totalappscounter}{1}\citep{DBLP:conf/uss/MateticSMJC18}&
$\circ$&
$\circ$&
\stepcounter{accesscontrolcounter}$\bullet$&
$\circ$&
$\circ$&
$\circ$&
$\circ$&
$\circ$&
$\circ$&
$\circ$&
$\circ$&
$\circ$&
$\circ$&
$\circ$&
$\circ$&
$\circ$&
$\circ$&
$\circ$&
$\circ$&
$\circ$&
$\circ$&
&
$\bullet$&
$\circ$&
$\circ$&
$\circ$&
&
$\circ$&
$\circ$&
\\
&{Graviton}: TEEs on {GPUs}&\addtocounter{totalappscounter}{1}\citep{DBLP:conf/osdi/VolosVB18}&
$\circ$&
$\circ$&
$\circ$&
$\circ$&
$\circ$&
$\circ$&
$\circ$&
\stepcounter{securecodeoffloadingcounter}$\bullet$&
$\circ$&
$\circ$&
$\circ$&
$\circ$&
$\circ$&
$\circ$&
\stepcounter{hardwareacceleratorscounter}$\bullet$&
$\circ$&
$\circ$&
$\circ$&
$\circ$&
$\circ$&
$\circ$&
&
$\circ$&
$\circ$&
$\circ$&
$\bullet$&
&
$\circ$&
$\circ$&
\\
&{LibSEAL}: revealing service integrity violations using trusted execution&\addtocounter{totalappscounter}{1}\citep{DBLP:conf/eurosys/AublinKOMPLKFEP18,LibSealGit}&
$\circ$&
$\circ$&
$\circ$&
\stepcounter{dataprotectioncounter}$\bullet$&
$\circ$&
$\circ$&
$\circ$&
$\circ$&
$\circ$&
$\circ$&
$\circ$&
$\circ$&
$\circ$&
$\circ$&
$\circ$&
$\circ$&
$\circ$&
$\circ$&
$\circ$&
$\circ$&
$\circ$&
&
$\bullet$&
$\circ$&
$\circ$&
$\circ$&
&
$\bullet$&
$\bullet$&
\\
&{Obscuro}: A {Bitcoin} Mixer using TEEs&\addtocounter{totalappscounter}{1}\citep{DBLP:conf/acsac/TranLKBS18,ObscuroGit}&
$\circ$&
$\circ$&
$\circ$&
$\circ$&
\stepcounter{onlinepaymentscounter}$\bullet$&
$\circ$&
$\circ$&
$\circ$&
$\circ$&
$\circ$&
$\circ$&
$\circ$&
$\circ$&
$\circ$&
$\circ$&
$\circ$&
$\circ$&
$\circ$&
$\circ$&
$\circ$&
$\circ$&
&
$\bullet$&
$\circ$&
$\circ$&
$\circ$&
&
$\bullet$&
$\bullet$&
\\
&{PubSub-SGX}: Exploiting TEEs for Privacy-Preserving Publish/Subscribe Systems&\addtocounter{totalappscounter}{1}\citep{DBLP:conf/srds/ArnautovBFFGKOM18,PubSubGit}&
$\circ$&
$\circ$&
$\circ$&
$\circ$&
$\circ$&
$\circ$&
$\circ$&
$\circ$&
$\circ$&
\stepcounter{networksecuritycounter}$\bullet$&
$\circ$&
\stepcounter{contentsharingcounter}$\bullet$&
$\circ$&
$\circ$&
$\circ$&
$\circ$&
$\circ$&
$\circ$&
$\circ$&
$\circ$&
$\circ$&
&
$\bullet$&
$\circ$&
$\circ$&
$\circ$&
&
$\bullet$&
$\bullet$&
\\
&{SafeBricks}: Shielding Network Functions in the Cloud&\addtocounter{totalappscounter}{1}\citep{DBLP:conf/nsdi/PoddarLPR18,SafeBricksGit}&
$\circ$&
\stepcounter{cloudcomputingcounter}$\bullet$&
$\circ$&
$\circ$&
$\circ$&
$\circ$&
$\circ$&
$\circ$&
$\circ$&
\stepcounter{networksecuritycounter}$\bullet$&
\stepcounter{securechannelscounter}$\bullet$&
$\circ$&
$\circ$&
$\circ$&
$\circ$&
$\circ$&
$\circ$&
$\circ$&
$\circ$&
$\circ$&
$\circ$&
&
$\bullet$&
$\circ$&
$\circ$&
$\circ$&
&
$\bullet$&
$\bullet$&
\\
&{SafeKeeper}: Protecting Web Passwords using TEEs&\addtocounter{totalappscounter}{1}\citep{DBLP:conf/www/KrawieckaKPMA18,SafeKeeperGit}&
$\circ$&
$\circ$&
\stepcounter{accesscontrolcounter}$\bullet$&
$\circ$&
$\circ$&
$\circ$&
$\circ$&
$\circ$&
$\circ$&
$\circ$&
$\circ$&
$\circ$&
$\circ$&
$\circ$&
$\circ$&
$\circ$&
$\circ$&
$\circ$&
$\circ$&
$\circ$&
$\circ$&
&
$\bullet$&
$\circ$&
$\circ$&
$\circ$&
&
$\bullet$&
$\bullet$&
\\
&{TizenFX}&\addtocounter{totalappscounter}{1}\citep{TizenFX}&
$\circ$&
$\circ$&
\stepcounter{accesscontrolcounter}$\bullet$&
$\circ$&
$\circ$&
$\circ$&
$\circ$&
$\circ$&
$\circ$&
$\circ$&
$\circ$&
$\circ$&
$\circ$&
$\circ$&
$\circ$&
$\circ$&
$\circ$&
$\circ$&
$\circ$&
$\circ$&
$\circ$&
&
$\circ$&
$\bullet$&
$\circ$&
$\circ$&
&
$\bullet$&
$\bullet$&
\\
\midrule
\multirow{21}{*}{\rotatebox[origin=c]{270}{\textbf{2019}}}
&{BITE}: {Bitcoin} Lightweight Client Privacy using Trusted Execution&\addtocounter{totalappscounter}{1}\citep{DBLP:conf/uss/MateticWSKKC19}&
$\circ$&
$\circ$&
$\circ$&
$\circ$&
\stepcounter{onlinepaymentscounter}$\bullet$&
$\circ$&
$\circ$&
\stepcounter{securecodeoffloadingcounter}$\bullet$&
$\circ$&
$\circ$&
$\circ$&
$\circ$&
$\circ$&
$\circ$&
$\circ$&
$\circ$&
$\circ$&
$\circ$&
$\circ$&
$\circ$&
$\circ$&
&
$\bullet$&
$\circ$&
$\circ$&
$\circ$&
&
$\circ$&
$\circ$&
\\
&{Clemmys}: towards secure remote execution in {FaaS}&\addtocounter{totalappscounter}{1}\citep{DBLP:conf/systor/TrachOGBF19}&
$\circ$&
\stepcounter{cloudcomputingcounter}$\bullet$&
$\circ$&
$\circ$&
$\circ$&
$\circ$&
$\circ$&
$\circ$&
$\circ$&
$\circ$&
$\circ$&
$\circ$&
$\circ$&
$\circ$&
$\circ$&
$\circ$&
$\circ$&
$\circ$&
$\circ$&
$\circ$&
$\circ$&
&
$\bullet$&
$\circ$&
$\circ$&
$\circ$&
&
$\circ$&
$\circ$&
\\
&Forward and Backward Private Searchable Encryption with {SGX}&\addtocounter{totalappscounter}{1}\citep{DBLP:conf/eurosec/AmjadKM19}&
$\circ$&
\stepcounter{cloudcomputingcounter}$\bullet$&
$\circ$&
$\circ$&
$\circ$&
$\circ$&
$\circ$&
$\circ$&
\stepcounter{securestoragecounter}$\bullet$&
$\circ$&
$\circ$&
$\circ$&
$\circ$&
$\circ$&
$\circ$&
$\circ$&
$\circ$&
$\circ$&
$\circ$&
$\circ$&
$\circ$&
&
$\bullet$&
$\circ$&
$\circ$&
$\circ$&
&
$\circ$&
$\circ$&
\\
&{Fuzzing OP-TEE with AFL}&\addtocounter{totalappscounter}{1}\citep{FuzzingOPTee19,optee_fuzzer}&
$\circ$&
$\circ$&
$\circ$&
$\circ$&
$\circ$&
$\circ$&
$\circ$&
$\circ$&
$\circ$&
$\circ$&
$\circ$&
$\circ$&
$\circ$&
$\circ$&
$\circ$&
$\circ$&
$\circ$&
$\circ$&
$\circ$&
$\circ$&
$\circ$&
&
$\circ$&
$\bullet$&
$\circ$&
$\circ$&
&
$\bullet$&
$\bullet$&
\\
&Heterogeneous Isolated Execution for Commodity {GPUs}&\addtocounter{totalappscounter}{1}\citep{DBLP:conf/asplos/JangTKSH19}&
$\circ$&
$\circ$&
$\circ$&
$\circ$&
$\circ$&
$\circ$&
$\circ$&
\stepcounter{securecodeoffloadingcounter}$\bullet$&
$\circ$&
$\circ$&
$\circ$&
$\circ$&
$\circ$&
$\circ$&
\stepcounter{hardwareacceleratorscounter}$\bullet$&
$\circ$&
$\circ$&
$\circ$&
$\circ$&
$\circ$&
$\circ$&
&
$\circ$&
$\circ$&
$\circ$&
$\bullet$&
&
$\circ$&
$\bullet$&
\\
&{LightBox}: Full-stack Protected Stateful Middlebox at Lightning Speed&\addtocounter{totalappscounter}{1}\citep{DBLP:conf/ccs/DuanWYZW019,LightBoxGit}&
$\circ$&
$\circ$&
$\circ$&
$\circ$&
$\circ$&
$\circ$&
$\circ$&
$\circ$&
$\circ$&
\stepcounter{networksecuritycounter}$\bullet$&
\stepcounter{securechannelscounter}$\bullet$&
$\circ$&
$\circ$&
$\circ$&
$\circ$&
$\circ$&
$\circ$&
$\circ$&
$\circ$&
$\circ$&
$\circ$&
&
$\bullet$&
$\circ$&
$\circ$&
$\circ$&
&
$\bullet$&
$\bullet$&
\\
&{NeXUS}: Practical and Secure Access Control on Untrusted Storage Platforms using Client-Side {SGX}&\addtocounter{totalappscounter}{1}\citep{DBLP:conf/dsn/DjokoLL19,NexusGit}&
$\circ$&
$\circ$&
\stepcounter{accesscontrolcounter}$\bullet$&
$\circ$&
$\circ$&
$\circ$&
$\circ$&
$\circ$&
\stepcounter{securestoragecounter}$\bullet$&
$\circ$&
$\circ$&
$\circ$&
$\circ$&
$\circ$&
$\circ$&
$\circ$&
$\circ$&
$\circ$&
$\circ$&
$\circ$&
$\circ$&
&
$\bullet$&
$\circ$&
$\circ$&
$\circ$&
&
$\bullet$&
$\bullet$&
\\
&{OPERA}: Open Remote Attestation for {Intel's} Secure Enclaves&\addtocounter{totalappscounter}{1}\citep{DBLP:conf/ccs/ChenZL19}&
$\circ$&
$\circ$&
$\circ$&
$\circ$&
$\circ$&
$\circ$&
\stepcounter{attestationtoolcounter}$\bullet$&
$\circ$&
$\circ$&
$\circ$&
$\circ$&
$\circ$&
$\circ$&
$\circ$&
$\circ$&
$\circ$&
$\circ$&
$\circ$&
$\circ$&
$\circ$&
$\circ$&
&
$\bullet$&
$\circ$&
$\circ$&
$\circ$&
&
$\circ$&
$\circ$&
\\
&{OP-TEE based keymaster and gatekeeper HIDL HAL}&\addtocounter{totalappscounter}{1}\citep{kmgk}&
$\circ$&
$\circ$&
\stepcounter{accesscontrolcounter}$\bullet$&
$\circ$&
$\circ$&
$\circ$&
$\circ$&
$\circ$&
$\circ$&
$\circ$&
$\circ$&
$\circ$&
$\circ$&
$\circ$&
$\circ$&
$\circ$&
$\circ$&
$\circ$&
$\circ$&
$\circ$&
$\circ$&
&
$\circ$&
$\bullet$&
$\circ$&
$\circ$&
&
$\bullet$&
$\bullet$&
\\
&{PrivaTube}: Privacy-Preserving Edge-Assisted Video Streaming&\addtocounter{totalappscounter}{1}\citep{DBLP:conf/middleware/SilvaMCNRR19}&
$\circ$&
$\circ$&
$\circ$&
\stepcounter{dataprotectioncounter}$\bullet$&
$\circ$&
$\circ$&
$\circ$&
$\circ$&
$\circ$&
\stepcounter{networksecuritycounter}$\bullet$&
$\circ$&
\stepcounter{contentsharingcounter}$\bullet$&
$\circ$&
$\circ$&
$\circ$&
$\circ$&
$\circ$&
$\circ$&
$\circ$&
$\circ$&
$\circ$&
&
$\bullet$&
$\circ$&
$\circ$&
$\circ$&
&
$\circ$&
$\circ$&
\\
&{SDK for the Valve Steam Link}&\addtocounter{totalappscounter}{1}\citep{steamlink-sdk}&
$\circ$&
$\circ$&
$\circ$&
$\circ$&
$\circ$&
$\circ$&
$\circ$&
$\circ$&
$\circ$&
$\circ$&
$\circ$&
\stepcounter{contentsharingcounter}$\bullet$&
$\circ$&
\stepcounter{computergamescounter}$\bullet$&
$\circ$&
$\circ$&
$\circ$&
$\circ$&
$\circ$&
$\circ$&
$\circ$&
&
$\circ$&
$\bullet$&
$\circ$&
$\circ$&
&
$\bullet$&
$\bullet$&
\\
&{SecTEE}: {A} Software-based Approach to Secure Enclave Architecture Using {TEE}&\addtocounter{totalappscounter}{1}\citep{DBLP:conf/ccs/ZhaoZQFF19}&
$\circ$&
$\circ$&
$\circ$&
$\circ$&
$\circ$&
$\circ$&
$\circ$&
$\circ$&
$\circ$&
$\circ$&
$\circ$&
$\circ$&
$\circ$&
$\circ$&
$\circ$&
$\circ$&
$\circ$&
$\circ$&
$\circ$&
$\circ$&
$\circ$&
&
$\circ$&
$\bullet$&
$\circ$&
$\circ$&
&
$\circ$&
$\circ$&
\\
&{ShieldStore}: Shielded In-memory Key-value Storage with {SGX}&\addtocounter{totalappscounter}{1}\citep{DBLP:conf/eurosys/KimPWJH19,ShieldStore}&
$\circ$&
\stepcounter{cloudcomputingcounter}$\bullet$&
$\circ$&
$\circ$&
$\circ$&
$\circ$&
$\circ$&
$\circ$&
$\circ$&
$\circ$&
$\circ$&
$\circ$&
$\circ$&
$\circ$&
$\circ$&
$\circ$&
$\circ$&
$\circ$&
$\circ$&
$\circ$&
$\circ$&
&
$\bullet$&
$\circ$&
$\circ$&
$\circ$&
&
$\bullet$&
$\bullet$&
\\
&{Slalom}: Fast, Verifiable and Private Execution of Neural Networks in Trusted Hardware&\addtocounter{totalappscounter}{1}\citep{DBLP:conf/iclr/TramerB19,SlalomGit}&
\stepcounter{dataanalyticscounter}$\bullet$&
$\circ$&
$\circ$&
$\circ$&
$\circ$&
$\circ$&
$\circ$&
$\circ$&
$\circ$&
$\circ$&
$\circ$&
$\circ$&
$\circ$&
$\circ$&
$\circ$&
$\circ$&
$\circ$&
$\circ$&
$\circ$&
$\circ$&
$\circ$&
&
$\bullet$&
$\circ$&
$\circ$&
$\circ$&
&
$\bullet$&
$\bullet$&
\\
&{StreamBox-TZ}: Secure Stream Analytics at the Edge with {TrustZone}&\addtocounter{totalappscounter}{1}\citep{DBLP:conf/usenix/ParkZLL19}&
\stepcounter{dataanalyticscounter}$\bullet$&
\stepcounter{cloudcomputingcounter}$\bullet$&
$\circ$&
$\circ$&
$\circ$&
$\circ$&
$\circ$&
$\circ$&
$\circ$&
$\circ$&
$\circ$&
$\circ$&
$\circ$&
$\circ$&
$\circ$&
$\circ$&
$\circ$&
$\circ$&
$\circ$&
$\circ$&
$\circ$&
&
$\circ$&
$\bullet$&
$\circ$&
$\circ$&
&
$\circ$&
$\circ$&
\\
&{Teechain}: a secure payment network with asynchronous blockchain access&\addtocounter{totalappscounter}{1}\citep{DBLP:conf/sosp/LindNEKSP19,DBLP:journals/corr/LindEPS16,TeechainGit}&
$\circ$&
$\circ$&
$\circ$&
$\circ$&
\stepcounter{onlinepaymentscounter}$\bullet$&
$\circ$&
$\circ$&
$\circ$&
$\circ$&
$\circ$&
$\circ$&
$\circ$&
$\circ$&
$\circ$&
$\circ$&
$\circ$&
$\circ$&
$\circ$&
$\circ$&
$\circ$&
$\circ$&
&
$\bullet$&
$\circ$&
$\circ$&
$\circ$&
&
$\bullet$&
$\bullet$&
\\
&{TIMBER-V}: Tag-Isolated Memory Bringing Fine-grained Enclaves to {RISC-V}&\addtocounter{totalappscounter}{1}\citep{DBLP:conf/ndss/WeiserWBMMS19,TimberVGit}&
$\circ$&
$\circ$&
$\circ$&
$\circ$&
$\circ$&
\stepcounter{memoryprotectioncounter}$\bullet$&
$\circ$&
$\circ$&
$\circ$&
$\circ$&
$\circ$&
$\circ$&
$\circ$&
$\circ$&
$\circ$&
$\circ$&
$\circ$&
$\circ$&
$\circ$&
$\circ$&
$\circ$&
&
$\circ$&
$\circ$&
$\bullet$&
$\circ$&
&
$\bullet$&
$\bullet$&
\\
&Trust more, serverless&\addtocounter{totalappscounter}{1}\citep{DBLP:conf/systor/BrennerK19}&
$\circ$&
\stepcounter{cloudcomputingcounter}$\bullet$&
$\circ$&
$\circ$&
$\circ$&
$\circ$&
$\circ$&
$\circ$&
$\circ$&
$\circ$&
$\circ$&
$\circ$&
$\circ$&
$\circ$&
$\circ$&
$\circ$&
$\circ$&
$\circ$&
$\circ$&
$\circ$&
$\circ$&
&
$\bullet$&
$\circ$&
$\circ$&
$\circ$&
&
$\circ$&
$\circ$&
\\
&Using TEEs for Secure Stream Processing of Medical Data&\addtocounter{totalappscounter}{1}\citep{DBLP:conf/dais/SegarraDLAPS19}&
$\circ$&
$\circ$&
$\circ$&
\stepcounter{dataprotectioncounter}$\bullet$&
$\circ$&
$\circ$&
$\circ$&
$\circ$&
$\circ$&
$\circ$&
$\circ$&
$\circ$&
$\circ$&
$\circ$&
$\circ$&
$\circ$&
\stepcounter{medicaldatacounter}$\bullet$&
$\circ$&
$\circ$&
$\circ$&
$\circ$&
&
$\bullet$&
$\circ$&
$\circ$&
$\circ$&
&
$\circ$&
$\circ$&
\\
&{WebAssembly Micro Runtime (WAMR)}&\addtocounter{totalappscounter}{1}\citep{wasm-micro-runtime}&
$\circ$&
$\circ$&
$\circ$&
$\circ$&
$\circ$&
$\circ$&
$\circ$&
$\circ$&
$\circ$&
$\circ$&
\stepcounter{securechannelscounter}$\bullet$&
$\circ$&
$\circ$&
$\circ$&
$\circ$&
$\circ$&
$\circ$&
$\circ$&
$\circ$&
$\circ$&
$\circ$&
&
$\bullet$&
$\circ$&
$\circ$&
$\circ$&
&
$\bullet$&
$\bullet$&
\\
&{ZLiTE}: Lightweight Clients for Shielded {Zcash} Transactions Using Trusted Execution&\addtocounter{totalappscounter}{1}\citep{DBLP:conf/fc/WustMSMKC19}&
$\circ$&
$\circ$&
$\circ$&
$\circ$&
\stepcounter{onlinepaymentscounter}$\bullet$&
$\circ$&
$\circ$&
$\circ$&
$\circ$&
$\circ$&
$\circ$&
$\circ$&
$\circ$&
$\circ$&
$\circ$&
$\circ$&
$\circ$&
$\circ$&
$\circ$&
$\circ$&
$\circ$&
&
$\bullet$&
$\circ$&
$\circ$&
$\circ$&
&
$\circ$&
$\circ$&
\\
\midrule
\multirow{24}{*}{\rotatebox[origin=c]{270}{\textbf{2020}}}
&{BDTF}: A Blockchain-Based Data Trading Framework with {TEE}&\addtocounter{totalappscounter}{1}\citep{DBLP:conf/msn/SuYLZBZ20}&
$\circ$&
$\circ$&
$\circ$&
$\circ$&
$\circ$&
$\circ$&
$\circ$&
$\circ$&
$\circ$&
$\circ$&
$\circ$&
$\circ$&
$\circ$&
$\circ$&
$\circ$&
$\circ$&
$\circ$&
$\circ$&
$\circ$&
\stepcounter{datatradingcounter}$\bullet$&
$\circ$&
&
$\bullet$&
$\circ$&
$\circ$&
$\circ$&
&
$\circ$&
$\circ$&
\\
&{BlackMirror}: Preventing Wallhacks in {3D} Online {FPS} Games&\addtocounter{totalappscounter}{1}\citep{DBLP:conf/ccs/ParkAL20}&
$\circ$&
$\circ$&
$\circ$&
$\circ$&
$\circ$&
$\circ$&
$\circ$&
$\circ$&
$\circ$&
$\circ$&
$\circ$&
$\circ$&
$\circ$&
\stepcounter{computergamescounter}$\bullet$&
$\circ$&
$\circ$&
$\circ$&
$\circ$&
$\circ$&
$\circ$&
$\circ$&
&
$\bullet$&
$\circ$&
$\circ$&
$\circ$&
&
$\circ$&
$\circ$&
\\
&{Custos}: Practical Tamper-Evident Auditing of Operating Systems Using Trusted Execution&\addtocounter{totalappscounter}{1}\citep{DBLP:conf/ndss/PaccagnellaDH0F20}&
$\circ$&
$\circ$&
$\circ$&
$\circ$&
$\circ$&
$\circ$&
$\circ$&
$\circ$&
$\circ$&
$\circ$&
$\circ$&
$\circ$&
$\circ$&
$\circ$&
$\circ$&
$\circ$&
$\circ$&
\stepcounter{securesystemloggingcounter}$\bullet$&
$\circ$&
$\circ$&
$\circ$&
&
$\bullet$&
$\circ$&
$\circ$&
$\circ$&
&
$\circ$&
$\circ$&
\\
&{CVShield}: Guarding Sensor Data in Connected Vehicle with {TEE}&\addtocounter{totalappscounter}{1}\citep{DBLP:conf/codaspy/HuCJCFML20}&
$\circ$&
$\circ$&
$\circ$&
\stepcounter{dataprotectioncounter}$\bullet$&
$\circ$&
$\circ$&
$\circ$&
$\circ$&
$\circ$&
$\circ$&
$\circ$&
$\circ$&
$\circ$&
$\circ$&
$\circ$&
$\circ$&
$\circ$&
$\circ$&
$\circ$&
$\circ$&
$\circ$&
&
$\circ$&
$\bullet$&
$\circ$&
$\circ$&
&
$\circ$&
$\circ$&
\\
&{DarkneTZ}: towards model privacy at the edge using TEEs&\addtocounter{totalappscounter}{1}\citep{DBLP:conf/mobisys/MoSKDLCH20,darknetz}&
\stepcounter{dataanalyticscounter}$\bullet$&
$\circ$&
$\circ$&
$\circ$&
$\circ$&
$\circ$&
$\circ$&
$\circ$&
$\circ$&
$\circ$&
$\circ$&
$\circ$&
$\circ$&
$\circ$&
$\circ$&
$\circ$&
$\circ$&
$\circ$&
$\circ$&
$\circ$&
$\circ$&
&
$\bullet$&
$\circ$&
$\circ$&
$\circ$&
&
$\bullet$&
$\bullet$&
\\
&Design and Implementation of Hardware-Based Remote Attestation for a Secure {Internet} of {Things}&\addtocounter{totalappscounter}{1}\citep{DBLP:journals/wpc/AhnLK20}&
$\circ$&
$\circ$&
$\circ$&
$\circ$&
$\circ$&
$\circ$&
\stepcounter{attestationtoolcounter}$\bullet$&
$\circ$&
$\circ$&
$\circ$&
$\circ$&
$\circ$&
$\circ$&
$\circ$&
$\circ$&
$\circ$&
$\circ$&
$\circ$&
$\circ$&
$\circ$&
$\circ$&
&
$\circ$&
$\bullet$&
$\circ$&
$\circ$&
&
$\circ$&
$\circ$&
\\
&Enabling Rack-scale Confidential Computing using Heterogeneous {TEE}&\addtocounter{totalappscounter}{1}\citep{DBLP:conf/sp/ZhuH0WCZWZYZM20}&
$\circ$&
$\circ$&
$\circ$&
$\circ$&
$\circ$&
$\circ$&
$\circ$&
\stepcounter{securecodeoffloadingcounter}$\bullet$&
$\circ$&
$\circ$&
$\circ$&
$\circ$&
$\circ$&
$\circ$&
\stepcounter{hardwareacceleratorscounter}$\bullet$&
$\circ$&
$\circ$&
$\circ$&
$\circ$&
$\circ$&
$\circ$&
&
$\circ$&
$\circ$&
$\circ$&
$\bullet$&
&
$\circ$&
$\bullet$&
\\
&Fine-Grained Access Control-Enabled Logging Method on {ARM} {TrustZone}&\addtocounter{totalappscounter}{1}\citep{DBLP:journals/access/LeeJCKPL20}&
$\circ$&
$\circ$&
\stepcounter{accesscontrolcounter}$\bullet$&
$\circ$&
$\circ$&
$\circ$&
$\circ$&
$\circ$&
\stepcounter{securestoragecounter}$\bullet$&
$\circ$&
$\circ$&
$\circ$&
$\circ$&
$\circ$&
$\circ$&
$\circ$&
$\circ$&
$\circ$&
$\circ$&
$\circ$&
$\circ$&
&
$\circ$&
$\bullet$&
$\circ$&
$\circ$&
&
$\circ$&
$\circ$&
\\
&{GOAT:} {GPU} Outsourcing of Deep Learning Training With Async.\ Probabilistic Integrity Verification&\addtocounter{totalappscounter}{1}\citep{DBLP:journals/corr/abs-2010-08855}&
\stepcounter{dataanalyticscounter}$\bullet$&
$\circ$&
$\circ$&
$\circ$&
$\circ$&
$\circ$&
$\circ$&
$\circ$&
$\circ$&
$\circ$&
$\circ$&
$\circ$&
$\circ$&
$\circ$&
$\circ$&
$\circ$&
$\circ$&
$\circ$&
$\circ$&
$\circ$&
$\circ$&
&
$\bullet$&
$\circ$&
$\circ$&
$\circ$&
&
$\circ$&
$\circ$&
\\
&{Keybuster}&\addtocounter{totalappscounter}{1}\citep{Keybuster,DBLP:conf/uss/ShakevskyRW22}&
$\circ$&
$\circ$&
$\circ$&
$\circ$&
$\circ$&
\stepcounter{memoryprotectioncounter}$\bullet$&
$\circ$&
$\circ$&
$\circ$&
$\circ$&
$\circ$&
$\circ$&
$\circ$&
$\circ$&
$\circ$&
$\circ$&
$\circ$&
$\circ$&
$\circ$&
$\circ$&
$\circ$&
&
$\circ$&
$\bullet$&
$\circ$&
$\circ$&
&
$\bullet$&
$\bullet$&
\\
&{MobileCoin: Private payments for mobile devices}&\addtocounter{totalappscounter}{1}\citep{mobilecoin,DBLP:journals/corr/abs-2111-12364}&
$\circ$&
$\circ$&
$\circ$&
$\circ$&
\stepcounter{onlinepaymentscounter}$\bullet$&
$\circ$&
$\circ$&
$\circ$&
$\circ$&
$\circ$&
$\circ$&
$\circ$&
$\circ$&
$\circ$&
$\circ$&
$\circ$&
$\circ$&
$\circ$&
$\circ$&
$\circ$&
$\circ$&
&
$\bullet$&
$\circ$&
$\circ$&
$\circ$&
&
$\bullet$&
$\bullet$&
\\
&Privacy-preserving Payment Channel Networks using {TEE}&\addtocounter{totalappscounter}{1}\citep{DBLP:conf/icc/LiLM020}&
$\circ$&
$\circ$&
$\circ$&
$\circ$&
\stepcounter{onlinepaymentscounter}$\bullet$&
$\circ$&
$\circ$&
$\circ$&
$\circ$&
$\circ$&
$\circ$&
$\circ$&
$\circ$&
$\circ$&
$\circ$&
$\circ$&
$\circ$&
$\circ$&
$\circ$&
$\circ$&
$\circ$&
&
$\bullet$&
$\circ$&
$\circ$&
$\circ$&
&
$\circ$&
$\circ$&
\\
&{ProximiTEE}: Hardened {SGX} Attestation by Proximity Verification&\addtocounter{totalappscounter}{1}\citep{DBLP:conf/codaspy/DharPKC20}&
$\circ$&
$\circ$&
$\circ$&
$\circ$&
$\circ$&
$\circ$&
\stepcounter{attestationtoolcounter}$\bullet$&
$\circ$&
$\circ$&
$\circ$&
$\circ$&
$\circ$&
$\circ$&
$\circ$&
$\circ$&
$\circ$&
$\circ$&
$\circ$&
$\circ$&
$\circ$&
$\circ$&
&
$\bullet$&
$\circ$&
$\circ$&
$\circ$&
&
$\circ$&
$\circ$&
\\
&Reboot-Oriented {IoT}: Life Cycle Management in {TEE} for Disposable {IoT} devices&\addtocounter{totalappscounter}{1}\citep{DBLP:conf/acsac/SuzakiTGM20}&
$\circ$&
$\circ$&
$\circ$&
$\circ$&
$\circ$&
\stepcounter{memoryprotectioncounter}$\bullet$&
$\circ$&
$\circ$&
$\circ$&
$\circ$&
$\circ$&
$\circ$&
$\circ$&
$\circ$&
$\circ$&
$\circ$&
$\circ$&
$\circ$&
$\circ$&
$\circ$&
$\circ$&
&
$\circ$&
$\bullet$&
$\circ$&
$\circ$&
&
$\circ$&
$\circ$&
\\
&{SafeTrace: COVID-19 Self-reporting with Privacy}&\addtocounter{totalappscounter}{1}\citep{SafeTraceGit}&
\stepcounter{dataanalyticscounter}$\bullet$&
$\circ$&
$\circ$&
\stepcounter{dataprotectioncounter}$\bullet$&
$\circ$&
$\circ$&
$\circ$&
$\circ$&
$\circ$&
$\circ$&
$\circ$&
$\circ$&
$\circ$&
$\circ$&
$\circ$&
$\circ$&
\stepcounter{medicaldatacounter}$\bullet$&
$\circ$&
$\circ$&
$\circ$&
$\circ$&
&
$\bullet$&
$\circ$&
$\circ$&
$\circ$&
&
$\bullet$&
$\bullet$&
\\
&Secure Cloud Storage with Client-side Encryption using a {TEE}&\addtocounter{totalappscounter}{1}\citep{DBLP:conf/closer/RochaVPGPW20}&
$\circ$&
\stepcounter{cloudcomputingcounter}$\bullet$&
$\circ$&
$\circ$&
$\circ$&
$\circ$&
$\circ$&
$\circ$&
\stepcounter{securestoragecounter}$\bullet$&
$\circ$&
$\circ$&
$\circ$&
$\circ$&
$\circ$&
$\circ$&
$\circ$&
$\circ$&
$\circ$&
$\circ$&
$\circ$&
$\circ$&
&
$\bullet$&
$\circ$&
$\circ$&
$\circ$&
&
$\circ$&
$\circ$&
\\
&{secureTF}: {A} Secure {TensorFlow} Framework&\addtocounter{totalappscounter}{1}\citep{DBLP:conf/middleware/QuocGAKBF20}&
\stepcounter{dataanalyticscounter}$\bullet$&
\stepcounter{cloudcomputingcounter}$\bullet$&
$\circ$&
$\circ$&
$\circ$&
$\circ$&
$\circ$&
$\circ$&
$\circ$&
$\circ$&
$\circ$&
$\circ$&
$\circ$&
$\circ$&
$\circ$&
$\circ$&
$\circ$&
$\circ$&
$\circ$&
$\circ$&
$\circ$&
&
$\bullet$&
$\circ$&
$\circ$&
$\circ$&
&
$\circ$&
$\circ$&
\\
&{SeGShare}: Secure Group File Sharing in the Cloud using Enclaves&\addtocounter{totalappscounter}{1}\citep{DBLP:conf/dsn/FuhryHKK20}&
$\circ$&
$\circ$&
$\circ$&
$\circ$&
$\circ$&
$\circ$&
$\circ$&
$\circ$&
$\circ$&
$\circ$&
$\circ$&
\stepcounter{contentsharingcounter}$\bullet$&
$\circ$&
$\circ$&
$\circ$&
$\circ$&
$\circ$&
$\circ$&
$\circ$&
$\circ$&
$\circ$&
&
$\bullet$&
$\circ$&
$\circ$&
$\circ$&
&
$\circ$&
$\circ$&
\\
&{SENG}, the {SGX}-Enforcing Network Gateway: Authorizing Communication from Shielded Clients&\addtocounter{totalappscounter}{1}\citep{DBLP:conf/uss/SchwarzR20,sengsgx}&
$\circ$&
$\circ$&
\stepcounter{accesscontrolcounter}$\bullet$&
$\circ$&
$\circ$&
$\circ$&
$\circ$&
$\circ$&
$\circ$&
$\circ$&
$\circ$&
$\circ$&
$\circ$&
$\circ$&
$\circ$&
$\circ$&
$\circ$&
$\circ$&
$\circ$&
$\circ$&
$\circ$&
&
$\bullet$&
$\circ$&
$\circ$&
$\circ$&
&
$\bullet$&
$\bullet$&
\\
&Tabellion: secure legal contracts on mobile devices&\addtocounter{totalappscounter}{1}\citep{DBLP:conf/mobisys/MirzamohammadiL20}&
$\circ$&
$\circ$&
$\circ$&
$\circ$&
$\circ$&
$\circ$&
$\circ$&
$\circ$&
$\circ$&
$\circ$&
$\circ$&
$\circ$&
$\circ$&
$\circ$&
$\circ$&
$\circ$&
$\circ$&
$\circ$&
$\circ$&
$\circ$&
\stepcounter{digitalcontractscounter}$\bullet$&
&
$\bullet$&
$\bullet$&
$\circ$&
$\circ$&
&
$\circ$&
$\bullet$&
\\
&{Telekine}: Secure Computing with Cloud {GPUs}&\addtocounter{totalappscounter}{1}\citep{DBLP:conf/nsdi/HuntJMSHRW20}&
\stepcounter{dataanalyticscounter}$\bullet$&
\stepcounter{cloudcomputingcounter}$\bullet$&
$\circ$&
$\circ$&
$\circ$&
$\circ$&
$\circ$&
\stepcounter{securecodeoffloadingcounter}$\bullet$&
$\circ$&
$\circ$&
$\circ$&
$\circ$&
$\circ$&
$\circ$&
\stepcounter{hardwareacceleratorscounter}$\bullet$&
$\circ$&
$\circ$&
$\circ$&
$\circ$&
$\circ$&
$\circ$&
&
$\circ$&
$\circ$&
$\circ$&
$\bullet$&
&
$\circ$&
$\circ$&
\\
&Towards Formalization of Enhanced Privacy {ID} ({EPID})-based Remote Attestation in {Intel} {SGX}&\addtocounter{totalappscounter}{1}\citep{DBLP:conf/dsd/SardarQF20}&
$\circ$&
$\circ$&
$\circ$&
$\circ$&
$\circ$&
$\circ$&
\stepcounter{attestationtoolcounter}$\bullet$&
$\circ$&
$\circ$&
$\circ$&
$\circ$&
$\circ$&
$\circ$&
$\circ$&
$\circ$&
\stepcounter{formalmethodscounter}$\bullet$&
$\circ$&
$\circ$&
$\circ$&
$\circ$&
$\circ$&
&
$\bullet$&
$\circ$&
$\circ$&
$\circ$&
&
$\circ$&
$\circ$&
\\
&{TZ4Fabric}: Executing Smart Contracts with {ARM} {TrustZone}&\addtocounter{totalappscounter}{1}\citep{DBLP:conf/srds/MullerBCFGS20,TZ4FabricGit}&
$\circ$&
$\circ$&
$\circ$&
$\circ$&
$\circ$&
$\circ$&
$\circ$&
$\circ$&
$\circ$&
$\circ$&
$\circ$&
$\circ$&
\stepcounter{smartcontractscounter}$\bullet$&
$\circ$&
$\circ$&
$\circ$&
$\circ$&
$\circ$&
$\circ$&
$\circ$&
$\circ$&
&
$\circ$&
$\bullet$&
$\circ$&
$\circ$&
&
$\bullet$&
$\bullet$&
\\
&{TZ-MRAS}: A Remote Attestation Scheme for the Mobile Terminal Based on {ARM} {TrustZone}&\addtocounter{totalappscounter}{1}\citep{DBLP:journals/scn/WangZY20}&
$\circ$&
$\circ$&
$\circ$&
$\circ$&
$\circ$&
$\circ$&
\stepcounter{attestationtoolcounter}$\bullet$&
$\circ$&
$\circ$&
$\circ$&
$\circ$&
$\circ$&
$\circ$&
$\circ$&
$\circ$&
$\circ$&
$\circ$&
$\circ$&
$\circ$&
$\circ$&
$\circ$&
&
$\circ$&
$\bullet$&
$\circ$&
$\circ$&
&
$\circ$&
$\circ$&
\\
\midrule
\multirow{22}{*}{\rotatebox[origin=c]{270}{\textbf{2021}}}
&Atlas: Automated Scale-out of Trust-Oblivious Systems to TEEs&\addtocounter{totalappscounter}{1}\citep{AtlasMScThesis,AtlasRuntime}&
$\circ$&
$\circ$&
$\circ$&
$\circ$&
$\circ$&
$\circ$&
$\circ$&
\stepcounter{securecodeoffloadingcounter}$\bullet$&
$\circ$&
$\circ$&
$\circ$&
$\circ$&
$\circ$&
$\circ$&
$\circ$&
$\circ$&
$\circ$&
$\circ$&
$\circ$&
$\circ$&
$\circ$&
&
$\bullet$&
$\circ$&
$\circ$&
$\circ$&
&
$\bullet$&
$\bullet$&
\\
&Bringing Decentralized Search to Decentralized Services&\addtocounter{totalappscounter}{1}\citep{DBLP:conf/osdi/LiZZ0XA021}&
$\circ$&
$\circ$&
$\circ$&
$\circ$&
$\circ$&
$\circ$&
$\circ$&
$\circ$&
$\circ$&
$\circ$&
$\circ$&
$\circ$&
$\circ$&
$\circ$&
$\circ$&
$\circ$&
$\circ$&
$\circ$&
\stepcounter{websearchcounter}$\bullet$&
$\circ$&
$\circ$&
&
$\bullet$&
$\circ$&
$\circ$&
$\circ$&
&
$\bullet$&
$\bullet$&
\\
&Building Enclave-Native Storage Engines for Practical Encrypted Databases&\addtocounter{totalappscounter}{1}\citep{DBLP:journals/pvldb/SunWL021}&
$\circ$&
\stepcounter{cloudcomputingcounter}$\bullet$&
$\circ$&
$\circ$&
$\circ$&
$\circ$&
$\circ$&
$\circ$&
\stepcounter{securestoragecounter}$\bullet$&
$\circ$&
$\circ$&
$\circ$&
$\circ$&
$\circ$&
$\circ$&
$\circ$&
$\circ$&
$\circ$&
$\circ$&
$\circ$&
$\circ$&
&
$\bullet$&
$\circ$&
$\circ$&
$\circ$&
&
$\circ$&
$\circ$&
\\
&{CURE}: A Security Architecture with CUstomizable and Resilient Enclaves&\addtocounter{totalappscounter}{1}\citep{DBLP:conf/uss/BahmaniBDJKSS21}&
$\circ$&
$\circ$&
$\circ$&
$\circ$&
$\circ$&
$\circ$&
$\circ$&
$\circ$&
$\circ$&
$\circ$&
$\circ$&
$\circ$&
$\circ$&
$\circ$&
$\circ$&
$\circ$&
$\circ$&
$\circ$&
$\circ$&
$\circ$&
$\circ$&
&
$\circ$&
$\circ$&
$\bullet$&
$\circ$&
&
$\circ$&
$\bullet$&
\\
&Distributed Learning in {TEE}: A Case Study of Federated Learning in {SGX}&\addtocounter{totalappscounter}{1}\citep{DBLP:conf/icnidc/XuZ0Z21}&
\stepcounter{dataanalyticscounter}$\bullet$&
$\circ$&
$\circ$&
$\circ$&
$\circ$&
$\circ$&
$\circ$&
$\circ$&
$\circ$&
$\circ$&
$\circ$&
$\circ$&
$\circ$&
$\circ$&
$\circ$&
$\circ$&
$\circ$&
$\circ$&
$\circ$&
$\circ$&
$\circ$&
&
$\bullet$&
$\circ$&
$\circ$&
$\circ$&
&
$\circ$&
$\circ$&
\\
&{Enarx Shim SGX}&\addtocounter{totalappscounter}{1}\citep{enarx-shim}&
$\circ$&
$\circ$&
$\circ$&
$\circ$&
$\circ$&
$\circ$&
$\circ$&
$\circ$&
$\circ$&
$\circ$&
$\circ$&
$\circ$&
$\circ$&
$\circ$&
$\circ$&
$\circ$&
$\circ$&
$\circ$&
$\circ$&
$\circ$&
$\circ$&
&
$\bullet$&
$\circ$&
$\circ$&
$\circ$&
&
$\bullet$&
$\bullet$&
\\
&Formal Verification of a {TEE}-Based Architecture for {IoT} Applications&\addtocounter{totalappscounter}{1}\citep{DBLP:journals/iotj/ValadaresSPG21}&
$\circ$&
$\circ$&
$\circ$&
\stepcounter{dataprotectioncounter}$\bullet$&
$\circ$&
$\circ$&
$\circ$&
$\circ$&
$\circ$&
$\circ$&
$\circ$&
$\circ$&
$\circ$&
$\circ$&
$\circ$&
$\circ$&
$\circ$&
$\circ$&
$\circ$&
$\circ$&
$\circ$&
&
$\bullet$&
$\circ$&
$\circ$&
$\circ$&
&
$\circ$&
$\circ$&
\\
&{IceClave}: A {TEE} for In-Storage Computing&\addtocounter{totalappscounter}{1}\citep{DBLP:conf/micro/KangXJWKYKLJ021}&
$\circ$&
$\circ$&
$\circ$&
$\circ$&
$\circ$&
$\circ$&
$\circ$&
$\circ$&
\stepcounter{securestoragecounter}$\bullet$&
$\circ$&
$\circ$&
$\circ$&
$\circ$&
$\circ$&
$\circ$&
$\circ$&
$\circ$&
$\circ$&
$\circ$&
$\circ$&
$\circ$&
&
$\circ$&
$\bullet$&
$\circ$&
$\circ$&
&
$\circ$&
$\circ$&
\\
&{IvyCross}: A Trustworthy and Privacy-preserving Framework for Blockchain Interoperability&\addtocounter{totalappscounter}{1}\citep{DBLP:journals/iacr/LiWLWWLD21}&
$\circ$&
$\circ$&
$\circ$&
$\circ$&
$\circ$&
$\circ$&
$\circ$&
$\circ$&
$\circ$&
$\circ$&
$\circ$&
$\circ$&
\stepcounter{smartcontractscounter}$\bullet$&
$\circ$&
$\circ$&
$\circ$&
$\circ$&
$\circ$&
$\circ$&
$\circ$&
$\circ$&
&
$\bullet$&
$\circ$&
$\circ$&
$\circ$&
&
$\circ$&
$\circ$&
\\
&{MeetGo}: A {TEE} for Remote Applications on {FPGA}&\addtocounter{totalappscounter}{1}\citep{DBLP:journals/access/OhNJCP21}&
\stepcounter{dataanalyticscounter}$\bullet$&
$\circ$&
$\circ$&
$\circ$&
\stepcounter{onlinepaymentscounter}$\bullet$&
$\circ$&
$\circ$&
$\circ$&
$\circ$&
$\circ$&
\stepcounter{securechannelscounter}$\bullet$&
$\circ$&
$\circ$&
$\circ$&
$\circ$&
$\circ$&
$\circ$&
$\circ$&
$\circ$&
$\circ$&
$\circ$&
&
$\bullet$&
$\circ$&
$\circ$&
$\circ$&
&
$\circ$&
$\circ$&
\\
&Memory-Efficient Deep Learning Inference in TEEs&\addtocounter{totalappscounter}{1}\citep{DBLP:conf/ic2e/TruongGGW21}&
\stepcounter{dataanalyticscounter}$\bullet$&
$\circ$&
$\circ$&
$\circ$&
$\circ$&
$\circ$&
$\circ$&
$\circ$&
$\circ$&
$\circ$&
$\circ$&
$\circ$&
$\circ$&
$\circ$&
$\circ$&
$\circ$&
$\circ$&
$\circ$&
$\circ$&
$\circ$&
$\circ$&
&
$\bullet$&
$\circ$&
$\circ$&
$\circ$&
&
$\circ$&
$\circ$&
\\
&Poster: {FLATEE}: Federated Learning Across TEEs&\addtocounter{totalappscounter}{1}\citep{DBLP:conf/eurosp/MondalMRG21}&
\stepcounter{dataanalyticscounter}$\bullet$&
$\circ$&
$\circ$&
$\circ$&
$\circ$&
$\circ$&
$\circ$&
$\circ$&
$\circ$&
$\circ$&
$\circ$&
$\circ$&
$\circ$&
$\circ$&
$\circ$&
$\circ$&
$\circ$&
$\circ$&
$\circ$&
$\circ$&
$\circ$&
&
$\bullet$&
$\circ$&
$\circ$&
$\circ$&
&
$\circ$&
$\circ$&
\\
&{PPFL}: privacy-preserving federated learning with TEEs&\addtocounter{totalappscounter}{1}\citep{DBLP:conf/mobisys/MoHKMPK21,PPFLgit}&
\stepcounter{dataanalyticscounter}$\bullet$&
\stepcounter{cloudcomputingcounter}$\bullet$&
$\circ$&
$\circ$&
$\circ$&
$\circ$&
$\circ$&
$\circ$&
$\circ$&
$\circ$&
$\circ$&
$\circ$&
$\circ$&
$\circ$&
$\circ$&
$\circ$&
$\circ$&
$\circ$&
$\circ$&
$\circ$&
$\circ$&
&
$\bullet$&
$\circ$&
$\circ$&
$\circ$&
&
$\bullet$&
$\bullet$&
\\
&Privacy-preserving genotype imputation in a {TEE}&\addtocounter{totalappscounter}{1}\citep{DOKMAI2021983,SmacGit}&
$\circ$&
$\circ$&
$\circ$&
\stepcounter{dataprotectioncounter}$\bullet$&
$\circ$&
$\circ$&
$\circ$&
$\circ$&
$\circ$&
$\circ$&
$\circ$&
$\circ$&
$\circ$&
$\circ$&
$\circ$&
$\circ$&
\stepcounter{medicaldatacounter}$\bullet$&
$\circ$&
$\circ$&
$\circ$&
$\circ$&
&
$\bullet$&
$\circ$&
$\circ$&
$\circ$&
&
$\circ$&
$\bullet$&
\\
&{S2Dedup}: {SGX}-enabled secure deduplication&\addtocounter{totalappscounter}{1}\citep{DBLP:conf/systor/MirandaEPP21,S2DedupGit}&
$\circ$&
$\circ$&
$\circ$&
\stepcounter{dataprotectioncounter}$\bullet$&
$\circ$&
$\circ$&
$\circ$&
$\circ$&
$\circ$&
$\circ$&
$\circ$&
\stepcounter{contentsharingcounter}$\bullet$&
$\circ$&
$\circ$&
$\circ$&
$\circ$&
$\circ$&
$\circ$&
$\circ$&
$\circ$&
$\circ$&
&
$\bullet$&
$\circ$&
$\circ$&
$\circ$&
&
$\bullet$&
$\bullet$&
\\
&Scalable Memory Protection in the {PENGLAI} Enclave&\addtocounter{totalappscounter}{1}\citep{DBLP:conf/osdi/FengLDYJXZ021,PenglaiGit}&
$\circ$&
$\circ$&
$\circ$&
$\circ$&
$\circ$&
\stepcounter{memoryprotectioncounter}$\bullet$&
$\circ$&
$\circ$&
$\circ$&
$\circ$&
$\circ$&
$\circ$&
$\circ$&
$\circ$&
$\circ$&
$\circ$&
$\circ$&
$\circ$&
$\circ$&
$\circ$&
$\circ$&
&
$\circ$&
$\circ$&
$\bullet$&
$\circ$&
&
$\bullet$&
$\bullet$&
\\
&{ShuffleFL}: gradient-preserving federated learning using {TEE}&\addtocounter{totalappscounter}{1}\citep{DBLP:conf/cf/ZhangWCHM21}&
\stepcounter{dataanalyticscounter}$\bullet$&
$\circ$&
$\circ$&
$\circ$&
$\circ$&
$\circ$&
$\circ$&
$\circ$&
$\circ$&
$\circ$&
$\circ$&
$\circ$&
$\circ$&
$\circ$&
$\circ$&
$\circ$&
$\circ$&
$\circ$&
$\circ$&
$\circ$&
$\circ$&
&
$\bullet$&
$\circ$&
$\circ$&
$\circ$&
&
$\circ$&
$\circ$&
\\
&{TEEKAP}: Self-Expiring Data Capsule using {TEE}&\addtocounter{totalappscounter}{1}\citep{DBLP:conf/acsac/GaoDC21,Teekap2022}&
$\circ$&
$\circ$&
\stepcounter{accesscontrolcounter}$\bullet$&
$\circ$&
$\circ$&
$\circ$&
$\circ$&
$\circ$&
$\circ$&
$\circ$&
$\circ$&
$\circ$&
$\circ$&
$\circ$&
$\circ$&
$\circ$&
$\circ$&
$\circ$&
$\circ$&
$\circ$&
$\circ$&
&
$\bullet$&
$\circ$&
$\circ$&
$\circ$&
&
$\circ$&
$\circ$&
\\
&{Tora}: A Trusted Blockchain Oracle Based on a Decentralized {TEE} Network&\addtocounter{totalappscounter}{1}\citep{9566165,ToraRepo}&
$\circ$&
$\circ$&
$\circ$&
$\circ$&
$\circ$&
$\circ$&
$\circ$&
$\circ$&
$\circ$&
$\circ$&
$\circ$&
$\circ$&
\stepcounter{smartcontractscounter}$\bullet$&
$\circ$&
$\circ$&
$\circ$&
$\circ$&
$\circ$&
$\circ$&
$\circ$&
$\circ$&
&
$\bullet$&
$\circ$&
$\circ$&
$\circ$&
&
$\bullet$&
$\bullet$&
\\
&{TrustZone}-based secure lightweight wallet for hyperledger fabric&\addtocounter{totalappscounter}{1}\citep{DBLP:journals/jpdc/DaiWWLZJ21}&
$\circ$&
$\circ$&
$\circ$&
$\circ$&
\stepcounter{onlinepaymentscounter}$\bullet$&
$\circ$&
$\circ$&
$\circ$&
$\circ$&
$\circ$&
$\circ$&
$\circ$&
$\circ$&
$\circ$&
$\circ$&
$\circ$&
$\circ$&
$\circ$&
$\circ$&
$\circ$&
$\circ$&
&
$\circ$&
$\bullet$&
$\circ$&
$\circ$&
&
$\circ$&
$\circ$&
\\
&{TZ-Container}: protecting container from untrusted {OS} with {ARM} {TrustZone}&\addtocounter{totalappscounter}{1}\citep{DBLP:journals/chinaf/HuaYGXCZ21}&
$\circ$&
\stepcounter{cloudcomputingcounter}$\bullet$&
$\circ$&
$\circ$&
$\circ$&
$\circ$&
$\circ$&
$\circ$&
\stepcounter{securestoragecounter}$\bullet$&
$\circ$&
$\circ$&
$\circ$&
$\circ$&
$\circ$&
$\circ$&
$\circ$&
$\circ$&
$\circ$&
$\circ$&
$\circ$&
$\circ$&
&
$\circ$&
$\bullet$&
$\circ$&
$\circ$&
&
$\circ$&
$\circ$&
\\
&{TZMon}: Improving mobile game security with {ARM} {TrustZone}&\addtocounter{totalappscounter}{1}\citep{DBLP:journals/compsec/JeonK21a,TZMonGit}&
$\circ$&
$\circ$&
$\circ$&
$\circ$&
$\circ$&
$\circ$&
$\circ$&
$\circ$&
$\circ$&
$\circ$&
$\circ$&
$\circ$&
$\circ$&
\stepcounter{computergamescounter}$\bullet$&
$\circ$&
$\circ$&
$\circ$&
$\circ$&
$\circ$&
$\circ$&
$\circ$&
&
$\circ$&
$\bullet$&
$\circ$&
$\circ$&
&
$\bullet$&
$\bullet$&
\\
\midrule
\multirow{6}{*}{\rotatebox[origin=c]{270}{\textbf{2022}}}
&Exploring {Widevine} for Fun and Profit&\addtocounter{totalappscounter}{1}\citep{DBLP:conf/sp/PatatSF22}&
$\circ$&
$\circ$&
\stepcounter{accesscontrolcounter}$\bullet$&
$\circ$&
$\circ$&
$\circ$&
$\circ$&
$\circ$&
$\circ$&
$\circ$&
\stepcounter{securechannelscounter}$\bullet$&
\stepcounter{contentsharingcounter}$\bullet$&
$\circ$&
$\circ$&
$\circ$&
$\circ$&
$\circ$&
$\circ$&
$\circ$&
$\circ$&
$\circ$&
&
$\circ$&
$\bullet$&
$\circ$&
$\circ$&
&
$\circ$&
$\bullet$&
\\
&{MAGE}: Mutual Attestation for a Group of Enclaves without Trusted Third Parties&\addtocounter{totalappscounter}{1}\citep{DBLP:conf/uss/ChenZ22,linux-sgx-mage}&
$\circ$&
$\circ$&
$\circ$&
$\circ$&
$\circ$&
$\circ$&
\stepcounter{attestationtoolcounter}$\bullet$&
$\circ$&
$\circ$&
$\circ$&
$\circ$&
$\circ$&
$\circ$&
$\circ$&
$\circ$&
$\circ$&
$\circ$&
$\circ$&
$\circ$&
$\circ$&
$\circ$&
&
$\bullet$&
$\circ$&
$\circ$&
$\circ$&
&
$\bullet$&
$\bullet$&
\\
&{MMLedger: A ledger for confidential computing shims for tracking memory management system calls}&\addtocounter{totalappscounter}{1}\citep{mmledger}&
$\circ$&
$\circ$&
$\circ$&
$\circ$&
$\circ$&
\stepcounter{memoryprotectioncounter}$\bullet$&
$\circ$&
$\circ$&
$\circ$&
$\circ$&
$\circ$&
$\circ$&
$\circ$&
$\circ$&
$\circ$&
$\circ$&
$\circ$&
$\circ$&
$\circ$&
$\circ$&
$\circ$&
&
$\circ$&
$\bullet$&
$\circ$&
$\circ$&
&
$\bullet$&
$\bullet$&
\\
&{OLIVE:} Oblivious and Differentially Private Federated Learning on {TEE}&\addtocounter{totalappscounter}{1}\citep{DBLP:journals/corr/abs-2202-07165}&
\stepcounter{dataanalyticscounter}$\bullet$&
$\circ$&
$\circ$&
$\circ$&
$\circ$&
$\circ$&
$\circ$&
$\circ$&
$\circ$&
$\circ$&
$\circ$&
$\circ$&
$\circ$&
$\circ$&
$\circ$&
$\circ$&
$\circ$&
$\circ$&
$\circ$&
$\circ$&
$\circ$&
&
$\bullet$&
$\circ$&
$\circ$&
$\circ$&
&
$\circ$&
$\circ$&
\\
&{Supporting Passkeys}&\addtocounter{totalappscounter}{1}\citep{ApplePasskeys}&
$\circ$&
$\circ$&
\stepcounter{accesscontrolcounter}$\bullet$&
$\circ$&
$\circ$&
$\circ$&
$\circ$&
$\circ$&
$\circ$&
$\circ$&
$\circ$&
$\circ$&
$\circ$&
$\circ$&
$\circ$&
$\circ$&
$\circ$&
$\circ$&
$\circ$&
$\circ$&
$\circ$&
&
$\circ$&
$\circ$&
$\circ$&
$\circ$&
&
$\circ$&
$\bullet$&
\\
&Toward a Secure, Rich, and Fair Query Service for Light Clients on Public Blockchains&\addtocounter{totalappscounter}{1}\citep{9512478}&
$\circ$&
$\circ$&
$\circ$&
$\circ$&
$\circ$&
$\circ$&
$\circ$&
$\circ$&
$\circ$&
$\circ$&
$\circ$&
$\circ$&
\stepcounter{smartcontractscounter}$\bullet$&
$\circ$&
$\circ$&
$\circ$&
$\circ$&
$\circ$&
$\circ$&
$\circ$&
$\circ$&
&
$\bullet$&
$\circ$&
$\circ$&
$\circ$&
&
$\circ$&
$\circ$&
\\
 \bottomrule
\end{tabular}%
}
\end{table*}

We gathered a total of \thetotalappscounter{} application use cases.
The categories and the number of references corresponding to each category are the following:
\emph{Data analytics} (\thedataanalyticscounter{}),
\emph{Cloud computing} (\thecloudcomputingcounter{}),
\emph{Access control} (\theaccesscontrolcounter{}),
\emph{Data protection} (\thedataprotectioncounter{}),
\emph{Online payments} (\theonlinepaymentscounter{}),
\emph{Memory protection} (\thememoryprotectioncounter{}),
\emph{Attestation tools} (\theattestationtoolcounter{}),
\emph{Secure storage} (\thesecurestoragecounter{}),
\emph{Network security} (\thenetworksecuritycounter{}),
\emph{Secure channels} (\thesecurechannelscounter{}),
\emph{Content sharing} (\thecontentsharingcounter{}),
\emph{Secure code offloading} (\thesecurecodeoffloadingcounter{}),
\emph{Smart contracts} (\thesmartcontractscounter{}),
\emph{Computer games} (\thecomputergamescounter{}),
\emph{Hardware accelerators} (\thehardwareacceleratorscounter{}),
\emph{Formal methods} (\theformalmethodscounter{}),
\emph{Medical data} (\themedicaldatacounter{}),
\emph{Secure system logging} (\thesecuresystemloggingcounter{}),
\emph{Web search} (\thewebsearchcounter{}),
\emph{Data trading} (\thedatatradingcounter{}),
and \emph{Digital contracts} (\thedigitalcontractscounter{}).

According to \autoref{tab:apps}, the vast majority of TEE applications operate on Intel SGX, ARM TrustZone, or both.
Only a minority of applications operate on other platforms such as AMD SEV, RISC-V, or GPU TEEs.
While most of the references we collected fit within the 21 categories outlined in \autoref{sec:methodology},
five applications did not fit into any of these categories.

On this basis, the majority of TEE applications aim to provide privacy-preserving data analysis (including machine learning applications).
Cloud computing is frequently associated with machine learning applications and is the second-largest TEE usage category in our listing.
Application domains surrounding access control, data protection, online payments, and memory protection are also among the most common use cases for TEEs.
Albeit noticeably less prevalent than the use cases previously stated, attestation tools, secure storage,
network security, secure channels, content sharing, and secure code offloading are all prominent use cases as well with seven references each.
Smart contracts, computer games, hardware accelerators, formal methods, and medical data are also fairly prevalent use cases, with three to five references each.
The remaining categories represent highly specific TEE use cases with few existing applications.
Examples include web search data protection, digital contract signing, and secure system logging.

The number of applications utilising TEEs has steadily increased since 2015.
52 of the \thetotalappscounter{} references we collected are from 2020 or after, and
48 applications have been deployed to actual users, according to our study.
40 of these 48 applications deployed to actual users are licenced under an open-source licence.
Notably, despite this, a large number of proprietary applications with closed-source licences comparable to the
Widevine DRM component \citep{DBLP:conf/sp/PatatSF22} utilise TEEs.
Typically, these proprietary applications are not accompanied by any public documentation or scholarly studies,
hence they are largely absent from our work.

\section{Tools for developing TEE software}\label{sec:frameworks}

\textit{\autoref*{rq:frameworks}: Which SDKs are available for TA development?}

Numerous middleware frameworks are available to assist developers with TEE development, deployment, and maintenance.
To address \autoref*{rq:frameworks}, \autoref{tab:frameworks} combines tools for developing TEE software.
In \autoref{tab:frameworks}, we highlight open-source SDKs that are currently being actively developed.
Four of the open-source frameworks, such as Webinos \citep{Webinos2013}, are deprecated and no longer under active development.
Although there are minor updates, Open-TEE \citep{DBLP:conf/trustcom/McGillionDNA15} is no longer undergoing substantial development.
For our purposes, we consider a project active if there are software updates in 2022, which we assessed on 6 November 2022.

\begin{table}[htb]
\Description[TEE software development tools]
{We show the available TEE software development tools and their language support, hardware support, activity, and whether they are Open-source or not.}
\caption{TEE software development tools and language support ($\bullet$=Yes, $\circ$=No, $\bulletg$=Not mentioned).}\label{tab:frameworks}
\resizebox{1.0\linewidth}{!}{%
\setlength{\tabcolsep}{1pt}
\begin{tabular}{p{0.6\linewidth}ccccccccccccccc}
\toprule
\textbf{Framework} & &
\rotatebox[origin=r]{270}{\textbf{C}}&
\rotatebox[origin=r]{270}{\textbf{C++}}&
\rotatebox[origin=r]{270}{\textbf{Java}}&
\rotatebox[origin=r]{270}{\textbf{Go}}&
\rotatebox[origin=r]{270}{\textbf{Rust}}&
\rotatebox[origin=r]{270}{\textbf{JavaScript}}&&
\rotatebox[origin=r]{270}{\textbf{Intel SGX}}&
\rotatebox[origin=r]{270}{\textbf{AMD SEV}}&
\rotatebox[origin=r]{270}{\textbf{ARM TrustZone}}&
\rotatebox[origin=r]{270}{\textbf{RISC-V}}&&
\rotatebox[origin=r]{270}{\textbf{Active (2022)}}&
\rotatebox[origin=r]{270}{\textbf{Open source}}\\
\midrule
\stepcounter{totalframeworkscounter}%
Asylo &
\citep{Asylo2022} &
$\bulletg$&
$\bullet$&
$\bullet$&
$\bulletg$&
$\bulletg$&
$\bulletg$&
&
$\bullet$&
$\bullet$&
$\circ$&
$\circ$&
&
$\bullet$&
$\bullet$
\stepcounter{totalopensourceframeworkscounter}%
\\
\stepcounter{totalframeworkscounter}%
Confidential Consortium &
\citep{Ccf2022} &
$\bulletg$&
$\bullet$&
$\bulletg$&
$\bulletg$&
$\bulletg$&
$\bullet$&
&
$\bullet$&
$\circ$&
$\circ$&
$\circ$&
&
$\bullet$&
$\bullet$
\stepcounter{totalopensourceframeworkscounter}%
\\
\stepcounter{totalframeworkscounter}%
Edgeless RT &
\citep{EdgelessRT2022} &
$\bullet$&
$\bullet$&
$\bulletg$&
$\bullet$&
$\bullet$&
$\bulletg$&
&
$\bullet$&
$\circ$&
$\circ$&
$\circ$&
&
$\bullet$&
$\bullet$
\stepcounter{totalopensourceframeworkscounter}%
\\
\stepcounter{totalframeworkscounter}%
Intel SGX SDK &
\citep{linux-sgx2022} &
$\bulletg$&
$\bullet$&
$\bulletg$&
$\bulletg$&
$\bulletg$&
$\bulletg$&
&
$\bullet$&
$\circ$&
$\circ$&
$\circ$&
&
$\bullet$&
$\bullet$
\stepcounter{totalopensourceframeworkscounter}%
\\
\stepcounter{totalframeworkscounter}%
Keystone &
\citep{DBLP:conf/eurosys/LeeKSAS20,Keystone2022} &
$\bullet$&
$\bullet$&
$\bulletg$&
$\bulletg$&
$\bulletg$&
$\bulletg$&
&
$\circ$&
$\circ$&
$\circ$&
$\bullet$&
&
$\bullet$&
$\bullet$
\stepcounter{totalopensourceframeworkscounter}%
\\
\stepcounter{totalframeworkscounter}%
Occlum's fork of Intel SGX SDK &
\citep{linux-sgx} &
$\bulletg$&
$\bulletg$&
$\bulletg$&
$\bulletg$&
$\bulletg$&
$\bulletg$&
&
$\bullet$&
$\circ$&
$\circ$&
$\circ$&
&
$\bullet$&
$\bullet$
\stepcounter{totalopensourceframeworkscounter}%
\\
\stepcounter{totalframeworkscounter}%
Open-TEE &
\citep{DBLP:conf/trustcom/McGillionDNA15} &
$\bulletg$&
$\bulletg$&
$\bulletg$&
$\bulletg$&
$\bulletg$&
$\bulletg$&
&
$\bullet$&
$\bullet$&
$\bullet$&
$\circ$&
&
$\bullet$&
$\bullet$
\stepcounter{totalopensourceframeworkscounter}%
\\
\stepcounter{totalframeworkscounter}%
OP-TEE &
\citep{Optee2022,optee_client} &
$\bulletg$&
$\bullet$&
$\bulletg$&
$\bulletg$&
$\bullet$&
$\bulletg$&
&
$\circ$&
$\circ$&
$\bullet$&
$\circ$&
&
$\bullet$&
$\bullet$
\stepcounter{totalopensourceframeworkscounter}%
\\
\stepcounter{totalframeworkscounter}%
Open Enclave SDK &
\citep{OpenEnclave2022} &
$\bullet$&
$\bullet$&
$\bulletg$&
$\bulletg$&
$\bulletg$&
$\bulletg$&
&
$\bullet$&
$\circ$&
$\bullet$&
$\circ$&
&
$\bullet$&
$\bullet$
\stepcounter{totalopensourceframeworkscounter}%
\\
\stepcounter{totalframeworkscounter}%
QSEE SDK &
\citep{QSEEInitialRelease,QSEESDK,DBLP:journals/compsec/KhalidM22} &
$\bulletg$&
$\bulletg$&
$\bulletg$&
$\bulletg$&
$\bulletg$&
$\bulletg$&
&
$\circ$&
$\circ$&
$\bullet$&
$\circ$&
&
$\circ$&
$\circ$
\\
\stepcounter{totalframeworkscounter}%
Samsung Knox SDK &
\citep{SamsungKnox22} &
$\bullet$&
$\bulletg$&
$\bullet$&
$\bulletg$&
$\bulletg$&
$\bulletg$&
&
$\circ$&
$\circ$&
$\bullet$&
$\circ$&
&
$\bullet$&
$\circ$
\\
\stepcounter{totalframeworkscounter}%
Samsung Knox Tizen SDK &
\citep{SamsungKnoxTizenSDK21} &
$\bullet$&
$\bulletg$&
$\bullet$&
$\bulletg$&
$\bulletg$&
$\bulletg$&
&
$\circ$&
$\circ$&
$\bullet$&
$\circ$&
&
$\bullet$&
$\circ$
\\
\stepcounter{totalframeworkscounter}%
Samsung mTower &
\citep{mTower} &
$\bullet$&
$\bullet$&
$\bulletg$&
$\bulletg$&
$\bulletg$&
$\bulletg$&
&
$\circ$&
$\circ$&
$\bullet$&
$\circ$&
&
$\bullet$&
$\bullet$
\stepcounter{totalopensourceframeworkscounter}%
\\
\stepcounter{totalframeworkscounter}%
Samsung TEEGRIS SDK &
\citep{SamsungTeegrisSDK17} &
$\bullet$&
$\bullet$&
$\bulletg$&
$\bulletg$&
$\bulletg$&
$\bulletg$&
&
$\circ$&
$\circ$&
$\bullet$&
$\circ$&
&
$\bullet$&
$\circ$
\\
\stepcounter{totalframeworkscounter}%
Sanctuary &
\citep{DBLP:conf/ndss/BrasserGJSS19,SanctuaryDev} &
$\bulletg$&
$\bulletg$&
$\bulletg$&
$\bulletg$&
$\bulletg$&
$\bulletg$&
&
$\circ$&
$\circ$&
$\bullet$&
$\circ$&
&
$\bullet$&
$\circ$
\\
\stepcounter{totalframeworkscounter}%
Sanctum &
\citep{DBLP:conf/uss/CostanLD16,SanctumOldGit,SanctumNewGit} &
$\bulletg$&
$\bulletg$&
$\bulletg$&
$\bulletg$&
$\bulletg$&
$\bulletg$&
&
$\circ$&
$\circ$&
$\circ$&
$\bullet$&
&
$\circ$&
$\bullet$
\stepcounter{totalopensourceframeworkscounter}%
\\
\stepcounter{totalframeworkscounter}%
SecGear &
\citep{secGear} &
$\bullet$&
$\bullet$&
$\bulletg$&
$\bulletg$&
$\bulletg$&
$\bulletg$&
&
$\bullet$&
$\circ$&
$\circ$&
$\circ$&
&
$\bullet$&
$\bullet$
\stepcounter{totalopensourceframeworkscounter}%
\\
\stepcounter{totalframeworkscounter}%
Teaclave SGX SDK &
\citep{TeaclaveSGX2022} &
$\bulletg$&
$\bulletg$&
$\bulletg$&
$\bulletg$&
$\bullet$&
$\bulletg$&
&
$\bullet$&
$\circ$&
$\circ$&
$\circ$&
&
$\bullet$&
$\bullet$
\stepcounter{totalopensourceframeworkscounter}%
\\
\stepcounter{totalframeworkscounter}%
Teaclave TrustZone SDK &
\citep{TeaclaveTZ2022} &
$\bulletg$&
$\bulletg$&
$\bulletg$&
$\bulletg$&
$\bullet$&
$\bulletg$&
&
$\circ$&
$\circ$&
$\bullet$&
$\circ$&
&
$\bullet$&
$\bullet$
\stepcounter{totalopensourceframeworkscounter}%
\\
\stepcounter{totalframeworkscounter}%
TEEKAP &
\citep{Teekap2022} &
$\bulletg$&
$\bulletg$&
$\bulletg$&
$\bulletg$&
$\bulletg$&
$\bulletg$&
&
$\bullet$&
$\circ$&
$\circ$&
$\circ$&
&
$\circ$&
$\bullet$
\stepcounter{totalopensourceframeworkscounter}%
\\
\stepcounter{totalframeworkscounter}%
Trustonic TEE SDKs &
\citep{Kinibi520a21,KinibiM20,trustonic-tee-user-space} &
$\bulletg$&
$\bulletg$&
$\bulletg$&
$\bulletg$&
$\bulletg$&
$\bulletg$&
&
$\circ$&
$\circ$&
$\bullet$&
$\circ$&
&
$\circ$&
$\circ$
\\
\stepcounter{totalframeworkscounter}%
Trusty TEE &
\citep{TrustyTEE20} &
$\bulletg$&
$\bulletg$&
$\bullet$&
$\bulletg$&
$\bulletg$&
$\bulletg$&
&
$\bullet$&
$\circ$&
$\bullet$&
$\circ$&
&
$\bullet$&
$\bullet$
\stepcounter{totalopensourceframeworkscounter}%
\\
\stepcounter{totalframeworkscounter}%
Webinos &
\citep{DBLP:conf/trust/NamilukoPS13,Webinos2013} &
$\bulletg$&
$\bulletg$&
$\bulletg$&
$\bulletg$&
$\bulletg$&
$\bulletg$&
&
$\circ$&
$\circ$&
$\bullet$&
$\circ$&
&
$\circ$&
$\bullet$
\stepcounter{totalopensourceframeworkscounter}%
\\
 \bottomrule
\end{tabular}%
}
\end{table}

There is a wide selection of frameworks available to software developers for a variety of hardware architectures.
The frameworks mentioned are available as open-source software or as brand-focused commercial solutions from various manufacturers,
such as the Samsung Knox SDK for Samsung Android devices \citep{SamsungKnox22}.
11 of the 23 referred frameworks support Intel SGX, while 13 frameworks support ARM TrustZone, as \autoref{tab:frameworks} shows.
Notably, 21 of the 23 referred frameworks support either Intel SGX or ARM TrustZone, or both.

We researched and compiled a list of supported software languages for active SDK projects.
We obtained this information from the SDKs' documentation and examples.
This is a limitation, as we can only include supported language information from documented open-source SDKs;
thus, these SDKs might have wider non-documented language support.
We found that the main languages supported by active SDKs are C and C++. 12 SDKs support at least one of these two languages.
Four of them also work with Rust, four work with Java, one supports Go (Edgeless RT), and one supports JavaScript (Confidential Consortium).

The frameworks serve a variety of practical purposes in order to facilitate development efforts.
Several frameworks focus on mobile devices and wearables, where the intent is to provide ready-made APIs to support application
development \citep{SamsungKnox22,SamsungKnoxTizenSDK21,SamsungTeegrisSDK17}.
The framework references are also focused on IoT devices or web applications, but due to the wide range of programming language support, the frameworks cover also
many other areas \citep{DBLP:conf/trust/NamilukoPS13,Webinos2013,Ccf2022,KinibiM20}.
Some of the frameworks are focused on or support very niche areas, like Trustonic's Kinibi-520a SDK \citep{Kinibi520a21},
where Symmetric-Multi-Processing enables the development of biometric functions like fingerprint scanning and face recognition.

The choice of development framework by the developer is usually severely constrained by the hardware architecture.
For example, developers of mobile applications can only use options that are compatible with ARM TrustZone.
We find that open-source development frameworks, such as OP-TEE \citep{Optee2022}, Open Enclave SDK \citep{OpenEnclave2022},
Teaclave TrustZone SDK \citep{TeaclaveTZ2022}, and Trusty TEE \citep{TrustyTEE20} support TrustZone at least in some capacity and are still actively maintained.
These frameworks may provide open-source alternatives for mobile application developers, who have traditionally been limited to proprietary
closed-source frameworks, such as the Samsung Knox SDK \citep{SamsungKnox22} or Trustonic's TEE SDKs \citep{Kinibi520a21,KinibiM20,trustonic-tee-user-space}.
Nevertheless, many open-source frameworks only support specific platforms,
so proprietary SDKs may remain the only option for developers on unsupported platforms.
\section{Trusted containers (tcons)}\label{sec:containers}

\textit{\autoref*{rq:containers}: What types of tcons are available?}

\begin{table}[htb]
\Description[Trusted containers]
{We depict the available Trusted containers and their application middleware interfaces, hardware support, activity, and whether they are open source or not.}
\caption{Trusted containers.}\label{tab:containers}
\resizebox{1.0\linewidth}{!}{%
\begin{tabular}{p{0.3\linewidth}cccccccccc}
\toprule
\textbf{Container} & &
\rotatebox[origin=r]{270}{\textbf{libc wrapper}}&
\rotatebox[origin=r]{270}{\textbf{LibOS}}&
\rotatebox[origin=r]{270}{\textbf{WASI}}&&
\rotatebox[origin=r]{270}{\textbf{Intel SGX}}&
\rotatebox[origin=r]{270}{\textbf{AMD SEV}}&&
\rotatebox[origin=r]{270}{\textbf{Active (2022)}}&
\rotatebox[origin=r]{270}{\textbf{Open source}}\\
\midrule
\stepcounter{totalcontainerscounter}%
AccTEE &
\citep{DBLP:conf/middleware/GoltzscheNKK19,AccTEE2020} &
$\circ$&
$\circ$&
$\bullet$&
&
$\bullet$&
$\circ$&
&
$\circ$&
$\bullet$
\stepcounter{totalopensourcecontainerscounter}%
\\
\stepcounter{totalcontainerscounter}%
Anjuna &
\citep{Anjuna2022} &
$\circ$&
$\bullet$&
$\circ$&
&
$\bullet$&
$\bullet$&
&
$\circ$&
$\circ$
\\
\stepcounter{totalcontainerscounter}%
Apache Teaclave &
\citep{TeaclaveIncubating2022} &
$\circ$&
$\circ$&
$\bullet$&
&
$\bullet$&
$\circ$&
&
$\bullet$&
$\bullet$
\stepcounter{totalopensourcecontainerscounter}%
\\
\stepcounter{totalcontainerscounter}%
Chancel &
\citep{DBLP:conf/ndss/AhmadKSSFL21} &
$\bullet$&
$\circ$&
$\circ$&
&
$\bullet$&
$\circ$&
&
$\circ$&
$\circ$
\\
\stepcounter{totalcontainerscounter}%
Decentriq &
\citep{Decentriq2022} &
$\circ$&
$\bullet$&
$\circ$&
&
$\bullet$&
$\circ$&
&
$\circ$&
$\circ$
\\
\stepcounter{totalcontainerscounter}%
Deflection &
\citep{DBLP:conf/dsn/0004WCWLCWSCT21,Deflection2021} &
$\bullet$&
$\circ$&
$\circ$&
&
$\bullet$&
$\circ$&
&
$\circ$&
$\bullet$
\stepcounter{totalopensourcecontainerscounter}%
\\
\stepcounter{totalcontainerscounter}%
EGo SDK &
\citep{Ego2022} &
$\bullet$&
$\circ$&
$\circ$&
&
$\bullet$&
$\circ$&
&
$\bullet$&
$\bullet$
\stepcounter{totalopensourcecontainerscounter}%
\\
\stepcounter{totalcontainerscounter}%
Enarx &
\citep{Enarx2022} &
$\circ$&
$\circ$&
$\bullet$&
&
$\bullet$&
$\bullet$&
&
$\bullet$&
$\bullet$
\stepcounter{totalopensourcecontainerscounter}%
\\
\stepcounter{totalcontainerscounter}%
Fortanix EDP &
\citep{FortanixEDP2022} &
$\circ$&
$\bullet$&
$\circ$&
&
$\bullet$&
$\circ$&
&
$\bullet$&
$\bullet$
\stepcounter{totalopensourcecontainerscounter}%
\\
\stepcounter{totalcontainerscounter}%
GOTEE &
\citep{DBLP:conf/usenix/GhosnLB19,GoTEE2022} &
$\circ$&
$\circ$&
$\circ$&
&
$\bullet$&
$\circ$&
&
$\circ$&
$\bullet$
\stepcounter{totalopensourcecontainerscounter}%
\\
\stepcounter{totalcontainerscounter}%
Gramine &
\citep{DBLP:conf/usenix/TsaiPV17,Gramine2022} &
$\circ$&
$\bullet$&
$\circ$&
&
$\bullet$&
$\circ$&
&
$\bullet$&
$\bullet$
\stepcounter{totalopensourcecontainerscounter}%
\\
\stepcounter{totalcontainerscounter}%
MesaPy &
\citep{DBLP:journals/corr/abs-2005-05996,MesaPy2019} &
$\circ$&
$\circ$&
$\circ$&
&
$\bullet$&
$\circ$&
&
$\circ$&
$\bullet$
\stepcounter{totalopensourcecontainerscounter}%
\\
\stepcounter{totalcontainerscounter}%
Mystikos &
\citep{Mystikos2022} &
$\circ$&
$\bullet$&
$\circ$&
&
$\bullet$&
$\circ$&
&
$\bullet$&
$\bullet$
\stepcounter{totalopensourcecontainerscounter}%
\\
\stepcounter{totalcontainerscounter}%
Occlum &
\citep{DBLP:conf/asplos/ShenTCCWXXY20,Occlum2022} &
$\circ$&
$\bullet$&
$\circ$&
&
$\bullet$&
$\circ$&
&
$\bullet$&
$\bullet$
\stepcounter{totalopensourcecontainerscounter}%
\\
\stepcounter{totalcontainerscounter}%
Ratel &
\citep{DBLP:journals/tissec/CuiSSSY22,Ratel2021} &
$\bullet$&
$\circ$&
$\circ$&
&
$\bullet$&
$\circ$&
&
$\circ$&
$\bullet$
\stepcounter{totalopensourcecontainerscounter}%
\\
\stepcounter{totalcontainerscounter}%
Ryoan &
\citep{DBLP:journals/tocs/HuntZXPW18,Ryoan2021} &
$\bullet$&
$\circ$&
$\circ$&
&
$\bullet$&
$\circ$&
&
$\circ$&
$\bullet$
\stepcounter{totalopensourcecontainerscounter}%
\\
\stepcounter{totalcontainerscounter}%
SCONE &
\citep{DBLP:conf/osdi/ArnautovTGKMPLM16,Scone2022} &
$\bullet$&
$\circ$&
$\circ$&
&
$\bullet$&
$\circ$&
&
$\bullet$&
$\bullet$
\stepcounter{totalopensourcecontainerscounter}%
\\
\stepcounter{totalcontainerscounter}%
SGX-LKL &
\citep{DBLP:journals/corr/abs-1908-11143,SGX-LKL2021} &
$\circ$&
$\bullet$&
$\circ$&
&
$\bullet$&
$\circ$&
&
$\circ$&
$\bullet$
\stepcounter{totalopensourcecontainerscounter}%
\\
\stepcounter{totalcontainerscounter}%
Twine &
\citep{DBLP:conf/icde/MenetreyPFS21,Twine2021} &
$\circ$&
$\circ$&
$\bullet$&
&
$\bullet$&
$\circ$&
&
$\circ$&
$\bullet$
\stepcounter{totalopensourcecontainerscounter}%
\\
\stepcounter{totalcontainerscounter}%
vSGX &
\citep{DBLP:conf/sp/ZhaoLZL22} &
$\circ$&
$\circ$&
$\circ$&
&
$\circ$&
$\bullet$&
&
$\bullet$&
$\bullet$
\stepcounter{totalopensourcecontainerscounter}%
\\
 \bottomrule
\end{tabular}%
}
\end{table}

For an application to function on any TEE technology, the development process must follow framework-specific design solutions.
This makes the procedure difficult and time-intensive for application developers.
In addition, a developer must implement attestation to trust the application.
To address the usability issue with different TEE technologies, a set of tcons
enables either the direct execution of unmodified binary code inside a TEE or
automatic transformation of source code prior to loading it into a TEE executable \citep{DBLP:journals/corr/abs-2109-01923}.
In order to address \autoref*{rq:containers}, \autoref{tab:containers} enumerates tcons.

We collected \total{totalcontainerscounter} distinct containers, identified the supported hardware and application middleware interfaces,
and determined whether or not the project is open source and active.

\total{totalopensourcecontainerscounter} of the \total{totalcontainerscounter} referred tcons are open-source software.
If there are software updates in 2022, we consider the tcon project to be active.
We evaluated this on 13 October 2022.
The open-source tcons saw development activity in the following years:
MesaPy (2018);
AccTEE (2020);
Deflection, GoTEE, Ratel, Ryoan, SGX-LKL, Twine (2021);
vSGX, Enarx, Apache Teaclave, EGo SDK, Fortanix EDP, Gramine, Mystikos, and Occlum (2022).
Accordingly, there are eight active tcon projects.

We discover that 19 of the \total{totalcontainerscounter} tcons support Intel SGX and only three support AMD SEV\null.
In addition, we find no tcons that support TrustZone TEE technology, confining mobile application developers to SDKs.
A recent trend seems to be containers that support multiple hardware architectures.
The objective is to allow developers to adapt the same program to many platforms without having to alter the source code.
Enarx \citep{Enarx2022} is a good example of such a tcon.
Recently published vSGX \citep{DBLP:conf/sp/ZhaoLZL22} supports directly running SGX-enabled applications inside AMD SEV\@.

A system call is an interface between software and the OS through which applications can request services from the OS\null.
Since Intel SGX restricts applications from making system calls,
unmodified applications cannot be executed within an enclave.

Seven tcons utilise library OS (LibOS)\null:
the missing OS interface that either natively
or transparently relays in-enclave system calls to the OS outside the enclave.
The LibOS concept predates TEE technologies by at least a decade,
motivated by applications in the embedded space due to severe resource constraints \citep{DBLP:conf/asplos/PorterBHOH11}.
LibOS is an approach to operating system design and implementation where the
traditional functionality of an OS is provided by a set of libraries. These libraries
are linked directly into the application to create a single address space executable.
By encapsulating the operating system functionality within libraries, it becomes
easier to define and enforce boundaries between different components, while
reducing the TCB\null. This enables developers to implement security policies at
the application level, restricting access to sensitive resources, and preventing
unauthorised access to data or interference with other processes. In addition,
because the OS primitives are included in the application, this removes the need
to invoke system calls and hence, reduces the context switches between user space
and kernel space, thus improving performance.

All of this makes LibOS an ideal candidate for use in a TEE, either as a set
of standalone applications or as a wrapper around existing applications to reduce
the porting effort, e.g., by intercepting system calls from the application and
replacing them with LibOS-specific ones.

Six of the \total{totalcontainerscounter} tcons utilise wrappers around the C standard library (libc) as an application middleware interface.
Executing a system call with libc wrappers, such as EGo SDK,
is equivalent to requesting the untrusted OS to perform the corresponding operations outside of the enclave.

The WebAssembly System Interface (WASI) works in a similar fashion and restricts system calls.
It provides a runtime for WebAssembly (WASM) binary execution within a TEE \citep{DBLP:journals/corr/abs-2109-01923}.
Of the referred \total{totalcontainerscounter} tcons, four utilise WASI\null:
AccTEE, Apache Teaclave, Enarx, and Twine execute WASM binaries within a TEE\null.

From 13 October 2022 to 22 November 2022, we collected and compared the number of Linux system calls against the number of implemented system calls in
Gramine, Occlum, Mystikos, Enarx, and Fortanix EDP\null.
For reference,
the Linux kernel has a total of 451 system calls, including outdated system calls\footnote{\url{https://github.com/torvalds/linux/blob/master/include/uapi/asm-generic/unistd.h}}.

Gramine implements 166 system calls\footnote{\url{https://github.com/gramineproject/gramine/blob/master/libos/include/libos_table.h}}.
Occlum has 159 implemented system calls\footnote{\url{https://github.com/occlum/occlum/blob/master/src/libos/src/syscall/mod.rs}}.
Mystikos implements 102 system calls\footnote{\url{https://github.com/deislabs/mystikos/blob/main/kernel/syscall.c}}.
Enarx implements the \emph{sallyport}\footnote{\url{https://github.com/enarx-archive/sallyport}} proxying service for service calls and executes WASM binary within a TEE \emph{Keep}.
The sallyport protocol implements 31 system calls from a \emph{Keep} to the host\footnote{\url{https://github.com/enarx/enarx/blob/main/crates/sallyport/src/host/syscall.rs}}.
Fortanix EDP -- specifically, its user-call API -- implements 16 system calls\footnote{\url{https://edp.fortanix.com/docs/api/fortanix_sgx_abi/struct.Usercalls.html}},
purposefully kept simple to facilitate security audits.

From a security point of view, these tcons increase the TCB\@.
However, if we examine a realistic TA application like machine learning with Python, it has very few system interactions.
The developer chooses between implementing the whole software stack from scratch with some SDK or using a tcon;
both options have pros and cons.
\section{Testing tcons}\label{sec:benchmark}

\textit{\autoref*{rq:test}: What are the usability implications of porting existing applications to tcons?}

We compare tcons that are actively in development.
A project is deemed active if software updates are released in 2022.
Based on the comparison, we deploy and run a benchmark application within the suitable tcons.
Realistically, a software developer would choose between the eight active tcon projects
based on the functionality they provide.

\subsection{Are tcons easy to use?}

The trusted containers that use WASM (and Wasmtime) promote the phrase ``put your app in a container''.
When we tested this, we discovered restrictions imposed by WASM that contradict this statement.

While testing WASM containers such as Enarx, we discovered the following obstacles:
(1) The selected programming language needs to have native support for WASM development -- for instance, Rust.
(2) Even with a properly chosen programming language, routine standard library operations like threading and networking may require a redesign of the application.
(3) As a result, the majority of Rust's libraries are inoperable by default because they depend on standard libraries.
For example, a programmer cannot use existing HTTP libraries to execute an HTTP GET request.
(4) Instead, low-level code may be a requirement for even a simple task where you would normally just use one line to call a library.
(5) A programmer eventually needs to add \emph{.cargo/config} configurations, macros, and dependencies that are unique to WASM\null.
(6) Development requires mappings between software code and the tcon, for example, pre-opened sockets need to be defined in the \emph{Enarx.toml} configuration file.

\begin{figure}
    \includegraphics[width=\linewidth]{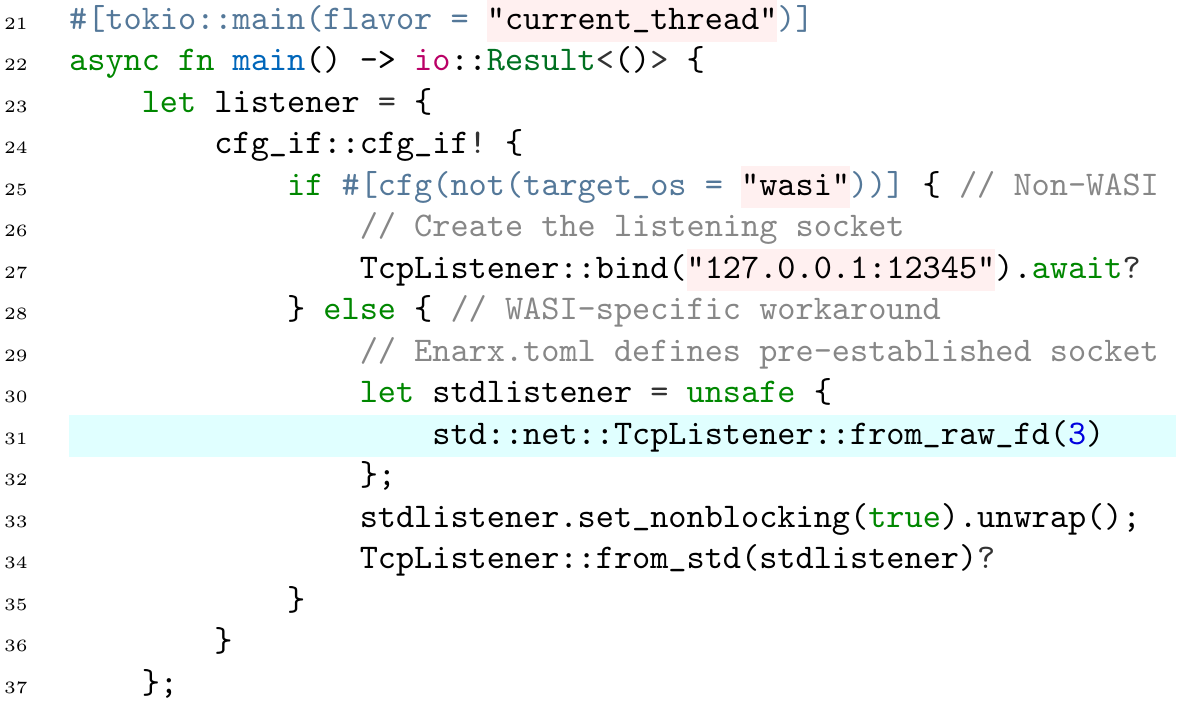}
    \Description[WASM tcon-modifications]
    {We present a code snippet of a TCP proxy application that demonstrates the required WASM and tcon-specific modifications regarding sockets and threading. Line 31 catches the pre-opened socket.}
    \caption{This snippet of a TCP proxy application demonstrates WASM and tcon-specific modifications regarding sockets
    and threading. Line 31 catches the pre-opened socket.}\label{fig:rust-tcp-proxy}
\end{figure}

In certain situations, standard libraries need to be replaced with alternatives that support WASM\null.
For instance, \emph{Tokio}\footnote{\url{https://tokio.rs/}} is an event-driven, non-blocking I/O platform for developing asynchronous Rust applications,
with unstable support for some extra WASM features.
However, not all methods are available.
For example, new sockets cannot be created from within WASM\null.
Instead, the code must catch the sockets that the tcon creates,
as demonstrated by \autoref{fig:rust-tcp-proxy}.

As a test, we rewrote a TCP proxy application using Tokio with unstable WASM support.
In the code, we define WASM build sections with macros and use them to catch the pre-opened sockets.
We utilise the ``current\_thread'' macro for threads instead of using a thread pool.
Until Wasmtime supports a large number of standard library requirements,
it is difficult to simply ``put your app in a container''.

Using LibOS-based tcons, such as Gramine, Mystikos,
and Occlum, we were able to launch diverse applications without modifications.
Therefore, it is easier to utilise LibOS-based tcons than those that require WASM\null.

\subsection{Performance of tcons}

We test the general-purpose containers
Enarx, Gramine, Mystikos, and Occlum
to benchmark and execute a Rust application.
We select these tcons because they are actively developed, can execute Rust applications, and support Intel SGX\null.
As a comparison, we execute the WASM binary without a secure enclave using Wasmtime.
As a second comparison, we test Enarx using AMD SEV hardware instead of Intel SGX hardware.

We forked a paper-based backup scheme application\footnote{\url{https://github.com/cyphar/paperback}} that generates encrypted backups
and splits the secret key into multiple key shards that can be held independently by different users (Shamir's Secret Sharing).
We select this application because it is utilised in the real world, is built in Rust, and Rust supports WASM builds.
Due to the file interface constraints of tcons, we hard-coded the values in order to test the application.

\begin{table}[htb]
\Description[TEE application benchmarking]
{We benchmark average test application runtimes using Enarx, Gramine, Mystikos, and Occlum tcons. In addition, we benchmark runtime without TEE as a comparison.}
\caption{Average application runtimes using Enarx, Gramine, Mystikos, and Occlum. In addition, runtime without TEE as a comparison.}\label{tab:benchmark}
\resizebox{1.0\linewidth}{!}{%
\begin{tabular}{llrr}
  \toprule
  \textbf{TEE hardware} & \textbf{Software} & \textbf{CPU cycles} & \textbf{App.\ Runtime (s)}\ \\
  \midrule
  Without TEE & Wasmtime &    729,059,936 & 0.81 \\
  \midrule\midrule
  AMD SEV     & Enarx    &  1,772,958,278 & 0.42 \\
  \midrule\midrule
              & Enarx    &  8,599,566,984 & 0.39 \\
  Intel SGX   & Gramine  &  1,003,098,599 & 0.55 \\
              & Mystikos &  1,380,584,095 & 0.37 \\
              & Occlum   & 11,702,240,620 & 2.34 \\
  \bottomrule
\end{tabular}
 }
\end{table}

The time to load a container varies greatly,
however, we did not compare this metric because it depends on whether the container performs the attestation process.
In order to enable attestation, we set up an entire Intel Software Guard Extensions Data Center Attestation Primitives
(Intel SGX DCAP)\footnote{\url{https://www.intel.com/content/www/us/en/developer/articles/guide/intel-software-guard-extensions-data-center-attestation-primitives-quick-install-guide.html}}.

We measured the real execution time of the application: the main function prints out the duration of the entire code execution.
The variance was very low, so the average of 100 samples accurately reflects the execution time and the number of CPU cycles.
Notice that CPU cycles take into account the full launch of the tcon and execution of the application, but for the time we measured only the execution time of the application inside the tcon.
This means that the number of runtime seconds does not necessarily correlate with the number of CPU cycles.
We repeated the Intel SGX tests and Wasmtime tests in the same environment.
Unexpectedly, code execution on Occlum takes 2.34 seconds, whereas on Enarx it takes 0.39 seconds while utilising the same WASM binary.
There is no obvious reason why Occlum execution is significantly slower.
In comparison, without the enclave, the execution time is 0.81 seconds, which is noticeably longer than the Enarx execution time.
We used the same build, wasm32-wasi target file, for both execution with Enarx and Wasmtime, yet Enarx is faster, so we suspect that its WASM runtime is lightweight.
With Enarx, the application runtime is similar to that on AMD SEV and Intel SGX hardware, but the number of CPU cycles greatly differs.
The number of CPU cycles is not comparable in this case because these are different pieces of hardware.
As a result, as expected, using a tcon adds overhead to the execution and, when compared to Wasmtime, requires more CPU cycles.
\section{Conclusion}\label{sec:conclusion}
This article organises TEE applications, frameworks, containers, and reviews in order to determine historic use and usability factors.
Our key conclusions are as follows:

\Paragraph{Open-source~TEE~SDKs~help~TA~creation}
Typically, a developer must make laborious framework-specific modifications to the original application in order for it to run within a TEE\null.
We listed \total{totalframeworkscounter} SDKs available to aid developers with TEE deployment,
out of which \total{totalopensourceframeworkscounter} are open-source software.

\Paragraph{Open-source~tcons~are~gaining~popularity}
A trusted container (tcon) solves the usability issue raised in the previous paragraph by enabling either the direct execution of unmodified
binary code within a TEE or the automatic transformation of source code prior to loading a TEE executable.
We provided a list of \total{totalcontainerscounter} tcons that eliminate the need for software developers to use specific SDKs to write TEE-related code,
out of which \total{totalopensourcecontainerscounter} are open-source software.

\Paragraph{Current~tcons~are~not~as~easy~to~use~as~advertised}
Our experiments indicate that tcons are not as simple to utilise as advertised.
Particularly WASM-based tcons impose strict limitations, necessitating a rewrite of the software's source code and the creation of separate configuration files.
In addition, the application must be written in a language that supports WASM natively.

\Paragraph{We~benchmarked~tcons}
We benchmarked Enarx, Gramine, Mystikos, and Occlum tcons with the Intel SGX backend.
As a comparison, we also ran a WASM binary without a TEE using Wasmtime and Enarx with an AMD SEV backend.
Using the identical WASM binary, code execution varied from 2.34 seconds with Occlum to 0.39 seconds with Enarx.
According to the measurements, tcons add overhead to the execution and need 1.4 to 16 times more CPU cycles than Wasmtime.

\Paragraph{Intel~SGX~and~ARM~TrustZone~are~the~most~researched}
Most of the publications demonstrate application use cases, and Intel SGX is the most popular hardware for applications.
In fact, 93 out of \thetotalappscounter{} TEE applications utilise either Intel SGX, ARM TrustZone, or both.
Only a small number of applications can run on other platforms such as AMD SEV, RISC-V, or GPU TEEs.

\Paragraph{Current~tcons~support~primarily~Intel~SGX~or~AMD~SEV\@}
The choice of SDK and tcon by the developer is severely constrained by the hardware architecture.
For instance, mobile application developers are restricted to options that are compatible with ARM TrustZone,
which means there are no tcons available and a limited number of SDKs to choose from, the majority of which are closed-source frameworks.
Some recent tcons, such as Enarx \citep{Enarx2022} and vSGX \citep{DBLP:conf/sp/ZhaoLZL22}, enable the execution of the same application within TEEs based on
multiple hardware architectures without requiring source code modifications (in theory).
Typically, though, tcons only support Intel SGX, AMD SEV, or both.

\Paragraph{Data~analytics~is~the~most~common~application~category~for~open-source~TAs}
Additionally, we examined the primary elements of the execution and the data people attempt to secure with their TAs.
We gathered a total of \thetotalappscounter{} application use cases in \autoref{tab:apps}.
\emph{Data analytics} (\thedataanalyticscounter{} references),
\emph{Cloud computing} (\thecloudcomputingcounter{} references),
and \emph{Access control} (\theaccesscontrolcounter{} references)
are the most common of the 21 primary drivers to use TEE\null.

\Paragraph{RISC-V,~Sanctum,~and~Keystone}
According to academic references, RISC-V TEE technologies are \textit{interesting}, but few publications are available about them.
The scientific community is ideally suited to pursue the objective of open-source hardware, which is undeniably a concrete development step.
The objective is to create a secure and trustworthy hardware-backed enclave for RISC-V\null.
Sanctum \citep{DBLP:conf/uss/CostanLD16} and Keystone \citep{DBLP:conf/eurosys/LeeKSAS20} are seminal steps in this direction,
yet we are unaware of any deployments.
This lack of mainstream hardware inhibits the growth of the surrounding software ecosystem,
somewhat analogous to TrustZone-based TEE technologies, such as On-board Credentials (ObC) \citep{DBLP:conf/ccs/EkbergAKR08} in the 2000s:
it is clear that ObC predates unified TEE software architectures, such as the GlobalPlatform API \citep{GlobalPlatformAPI},
yet such standardisation and unification efforts arguably emerged too late to prevent fragmentation of the software ecosystem.
In summary, as a community, we should steer TA software development in a consistent and narrow fashion,
and to achieve this, we need mainstream hardware available with TEE-relevant hardware-assisted security features that are open source and accessible to developers.
\section{Future work}\label{sec:future}

\Paragraph{RISC-V~and~applying~lessons~learned}
Several factors have negatively influenced the adoption of TEEs in the past. Moreover, we identified a key research topic
based on the analysis of real-world threats and effective mitigation techniques that are most relevant for TEE implementations.
With RISC-V as an emerging technology, the standardisation of TEE mechanisms for RISC-V is an excellent opportunity to not only apply
valuable lessons learned but to drive the development toward a secure and useable TEE\null.

\begin{acks}
This project has received funding from the European Union's Horizon 2020 research and innovation
programme under grant agreement No 952622 (SPIRS),
and grant agreement No 804476 (SCARE).
Supported in part by the Cybersecurity Research Award granted by
the Technology Innovation Institute (TII) in UAE and Technology Innovation Institute's Secure Systems Research Center (SSRC) in UAE\@.
\end{acks}

\bibliographystyle{ACM-Reference-Format}

\begin{thebibliography}{223}

%
%
%
%
%
%
%
%
%
%
%
%
%
%
%
%

\ifx \showCODEN    \undefined \def \showCODEN     #1{\unskip}     \fi
\ifx \showDOI      \undefined \def \showDOI       #1{#1}\fi
\ifx \showISBNx    \undefined \def \showISBNx     #1{\unskip}     \fi
\ifx \showISBNxiii \undefined \def \showISBNxiii  #1{\unskip}     \fi
\ifx \showISSN     \undefined \def \showISSN      #1{\unskip}     \fi
\ifx \showLCCN     \undefined \def \showLCCN      #1{\unskip}     \fi
\ifx \shownote     \undefined \def \shownote      #1{#1}          \fi
\ifx \showarticletitle \undefined \def \showarticletitle #1{#1}   \fi
\ifx \showURL      \undefined \def \showURL       {\relax}        \fi
%
%
\providecommand\bibfield[2]{#2}
\providecommand\bibinfo[2]{#2}
\providecommand\natexlab[1]{#1}
\providecommand\showeprint[2][]{arXiv:#2}

\bibitem[\protect\citeauthoryear{Ahmad, Kim, Seo, Shin, Fonseca, and Lee}{Ahmad
  et~al\mbox{.}}{2021}]%
        {DBLP:conf/ndss/AhmadKSSFL21}
\bibfield{author}{\bibinfo{person}{Adil Ahmad}, \bibinfo{person}{Juhee Kim},
  \bibinfo{person}{Jaebaek Seo}, \bibinfo{person}{Insik Shin},
  \bibinfo{person}{Pedro Fonseca}, {and} \bibinfo{person}{Byoungyoung Lee}.}
  \bibinfo{year}{2021}\natexlab{}.
\newblock \showarticletitle{{CHANCEL}: Efficient Multi-client Isolation Under
  Adversarial Programs}. In \bibinfo{booktitle}{\emph{{NDSS}}}.
  \bibinfo{publisher}{The Internet Society}.
\newblock
\urldef\tempurl%
\url{https://www.ndss-symposium.org/ndss-paper/chancel-efficient-multi-client-isolation-under-adversarial-programs/}
\showURL{%
\tempurl}


\bibitem[\protect\citeauthoryear{Ahn, Lee, and Kim}{Ahn et~al\mbox{.}}{2020}]%
        {DBLP:journals/wpc/AhnLK20}
\bibfield{author}{\bibinfo{person}{Jaehwan Ahn}, \bibinfo{person}{Il{-}Gu Lee},
  {and} \bibinfo{person}{Myungchul Kim}.} \bibinfo{year}{2020}\natexlab{}.
\newblock \showarticletitle{Design and Implementation of Hardware-Based Remote
  Attestation for a Secure {Internet} of {Things}}.
\newblock \bibinfo{journal}{\emph{Wirel. Pers. Commun.}} \bibinfo{volume}{114},
  \bibinfo{number}{1} (\bibinfo{year}{2020}), \bibinfo{pages}{295--327}.
\newblock
\urldef\tempurl%
\url{https://doi.org/10.1007/s11277-020-07364-5}
\showURL{%
\tempurl}


\bibitem[\protect\citeauthoryear{Akram, Akella, Peisert, and
  Lowe{-}Power}{Akram et~al\mbox{.}}{2022}]%
        {DBLP:conf/seed/AkramAPL22}
\bibfield{author}{\bibinfo{person}{Ayaz Akram}, \bibinfo{person}{Venkatesh
  Akella}, \bibinfo{person}{Sean Peisert}, {and} \bibinfo{person}{Jason
  Lowe{-}Power}.} \bibinfo{year}{2022}\natexlab{}.
\newblock \showarticletitle{{SoK}: Limitations of Confidential Computing via
  {TEEs} for High-Performance Compute Systems}. In
  \bibinfo{booktitle}{\emph{{SEED}}}. \bibinfo{publisher}{{IEEE}},
  \bibinfo{pages}{121--132}.
\newblock
\urldef\tempurl%
\url{https://doi.org/10.1109/SEED55351.2022.00018}
\showURL{%
\tempurl}


\bibitem[\protect\citeauthoryear{Amjad, Kamara, and Moataz}{Amjad
  et~al\mbox{.}}{2019}]%
        {DBLP:conf/eurosec/AmjadKM19}
\bibfield{author}{\bibinfo{person}{Ghous Amjad}, \bibinfo{person}{Seny Kamara},
  {and} \bibinfo{person}{Tarik Moataz}.} \bibinfo{year}{2019}\natexlab{}.
\newblock \showarticletitle{Forward and Backward Private Searchable Encryption
  with {SGX}}. In \bibinfo{booktitle}{\emph{{EuroSec}}}.
  \bibinfo{publisher}{{ACM}}, \bibinfo{pages}{4:1--4:6}.
\newblock
\urldef\tempurl%
\url{https://doi.org/10.1145/3301417.3312496}
\showURL{%
\tempurl}


\bibitem[\protect\citeauthoryear{{AndroidOpen Source Project}}{{AndroidOpen
  Source Project}}{2016}]%
        {TrustyTEE20}
\bibfield{author}{\bibinfo{person}{{AndroidOpen Source Project}}.}
  \bibinfo{year}{2016}\natexlab{}.
\newblock \bibinfo{title}{{Trusty TEE}}.
\newblock
  \bibinfo{howpublished}{\url{https://source.android.com/security/trusty}}.
\newblock
\newblock
\shownote{{Latest rel.\ 2020}.}


\bibitem[\protect\citeauthoryear{Anjuna}{Anjuna}{2018}]%
        {Anjuna2022}
\bibfield{author}{\bibinfo{person}{Anjuna}.} \bibinfo{year}{2018}\natexlab{}.
\newblock \bibinfo{title}{{Anjuna Confidential Cloud Software}}.
\newblock
\newblock
\urldef\tempurl%
\url{https://www.anjuna.io/}
\showURL{%
\tempurl}
\newblock
\shownote{Latest rel.\ 2022.}


\bibitem[\protect\citeauthoryear{ApplePasskeys}{ApplePasskeys}{2022}]%
        {ApplePasskeys}
\bibfield{author}{\bibinfo{person}{ApplePasskeys}.}
  \bibinfo{year}{2022}\natexlab{}.
\newblock \bibinfo{title}{{Supporting Passkeys}}.
\newblock
  \bibinfo{howpublished}{\url{https://developer.apple.com/documentation/authenticationservices/public-private_key_authentication/supporting_passkeys}}.
\newblock


\bibitem[\protect\citeauthoryear{Arfaoui, Gharout, and Traor{\'{e}}}{Arfaoui
  et~al\mbox{.}}{2014}]%
        {DBLP:conf/mobilecloud/ArfaouiGT14}
\bibfield{author}{\bibinfo{person}{Ghada Arfaoui}, \bibinfo{person}{Said
  Gharout}, {and} \bibinfo{person}{Jacques Traor{\'{e}}}.}
  \bibinfo{year}{2014}\natexlab{}.
\newblock \showarticletitle{Trusted Execution Environments: A Look under the
  Hood}. In \bibinfo{booktitle}{\emph{MobileCloud}}. \bibinfo{publisher}{{IEEE}
  Computer Society}, \bibinfo{pages}{259--266}.
\newblock
\urldef\tempurl%
\url{https://doi.org/10.1109/MobileCloud.2014.47}
\showURL{%
\tempurl}


\bibitem[\protect\citeauthoryear{Arnautov, Brito, Felber, Fetzer, Gregor,
  Krahn, Ozga, Martin, Schiavoni, Silva, Tenorio, and Thummel}{Arnautov
  et~al\mbox{.}}{2018}]%
        {DBLP:conf/srds/ArnautovBFFGKOM18}
\bibfield{author}{\bibinfo{person}{Sergei Arnautov}, \bibinfo{person}{Andrey
  Brito}, \bibinfo{person}{Pascal Felber}, \bibinfo{person}{Christof Fetzer},
  \bibinfo{person}{Franz Gregor}, \bibinfo{person}{Robert Krahn},
  \bibinfo{person}{Wojciech Ozga}, \bibinfo{person}{Andr{\'{e}} Martin},
  \bibinfo{person}{Valerio Schiavoni}, \bibinfo{person}{F{\'{a}}bio Silva},
  \bibinfo{person}{Marcus Tenorio}, {and} \bibinfo{person}{Nikolaus Thummel}.}
  \bibinfo{year}{2018}\natexlab{}.
\newblock \showarticletitle{{PubSub-SGX}: Exploiting Trusted Execution
  Environments for Privacy-Preserving Publish/Subscribe Systems}. In
  \bibinfo{booktitle}{\emph{{SRDS}}}. \bibinfo{publisher}{{IEEE} Computer
  Society}, \bibinfo{pages}{123--132}.
\newblock
\urldef\tempurl%
\url{https://doi.org/10.1109/SRDS.2018.00023}
\showURL{%
\tempurl}


\bibitem[\protect\citeauthoryear{Arnautov, Trach, Gregor, Knauth, Martin,
  Priebe, Lind, Muthukumaran, O'Keeffe, Stillwell, Goltzsche, Eyers, Kapitza,
  Pietzuch, and Fetzer}{Arnautov et~al\mbox{.}}{2016}]%
        {DBLP:conf/osdi/ArnautovTGKMPLM16}
\bibfield{author}{\bibinfo{person}{Sergei Arnautov}, \bibinfo{person}{Bohdan
  Trach}, \bibinfo{person}{Franz Gregor}, \bibinfo{person}{Thomas Knauth},
  \bibinfo{person}{Andr{\'{e}} Martin}, \bibinfo{person}{Christian Priebe},
  \bibinfo{person}{Joshua Lind}, \bibinfo{person}{Divya Muthukumaran},
  \bibinfo{person}{Dan O'Keeffe}, \bibinfo{person}{Mark Stillwell},
  \bibinfo{person}{David Goltzsche}, \bibinfo{person}{David~M. Eyers},
  \bibinfo{person}{R{\"{u}}diger Kapitza}, \bibinfo{person}{Peter~R. Pietzuch},
  {and} \bibinfo{person}{Christof Fetzer}.} \bibinfo{year}{2016}\natexlab{}.
\newblock \showarticletitle{{SCONE}: Secure {Linux} Containers with {Intel}
  {SGX}}. In \bibinfo{booktitle}{\emph{{OSDI}}}. \bibinfo{publisher}{{USENIX}
  Association}, \bibinfo{pages}{689--703}.
\newblock
\urldef\tempurl%
\url{https://www.usenix.org/conference/osdi16/technical-sessions/presentation/arnautov}
\showURL{%
\tempurl}


\bibitem[\protect\citeauthoryear{Asokan}{Asokan}{2019}]%
        {DBLP:conf/ccs/Asokan19}
\bibfield{author}{\bibinfo{person}{N. Asokan}.}
  \bibinfo{year}{2019}\natexlab{}.
\newblock \showarticletitle{Hardware-assisted Trusted Execution Environments:
  Look Back, Look Ahead}. In \bibinfo{booktitle}{\emph{{ACM} {CCS}}}.
  \bibinfo{publisher}{{ACM}}, \bibinfo{pages}{1687}.
\newblock
\urldef\tempurl%
\url{https://doi.org/10.1145/3319535.3364969}
\showURL{%
\tempurl}


\bibitem[\protect\citeauthoryear{Asvadishirehjini, Kantarcioglu, and
  Malin}{Asvadishirehjini et~al\mbox{.}}{2020}]%
        {DBLP:journals/corr/abs-2010-08855}
\bibfield{author}{\bibinfo{person}{Aref Asvadishirehjini},
  \bibinfo{person}{Murat Kantarcioglu}, {and} \bibinfo{person}{Bradley~A.
  Malin}.} \bibinfo{year}{2020}\natexlab{}.
\newblock \showarticletitle{{GOAT:} {GPU} Outsourcing of Deep Learning Training
  With Asynchronous Probabilistic Integrity Verification Inside Trusted
  Execution Environment}.
\newblock \bibinfo{journal}{\emph{CoRR}}  \bibinfo{volume}{abs/2010.08855}
  (\bibinfo{year}{2020}).
\newblock
\urldef\tempurl%
\url{https://arxiv.org/abs/2010.08855}
\showURL{%
\tempurl}


\bibitem[\protect\citeauthoryear{{Atlas Runtime}}{{Atlas Runtime}}{2022}]%
        {AtlasRuntime}
\bibfield{author}{\bibinfo{person}{{Atlas Runtime}}.}
  \bibinfo{year}{2022}\natexlab{}.
\newblock \bibinfo{title}{{Atlas: Automated Scale-out of Trust-Oblivious
  Systems to Trusted Execution Environments}}.
\newblock
\newblock
\urldef\tempurl%
\url{https://github.com/atlas-runtime/applications/}
\showURL{%
\tempurl}


\bibitem[\protect\citeauthoryear{Aublin, Kelbert, O'Keeffe, Muthukumaran,
  Priebe, Lind, Krahn, Fetzer, Eyers, and Pietzuch}{Aublin
  et~al\mbox{.}}{2018}]%
        {DBLP:conf/eurosys/AublinKOMPLKFEP18}
\bibfield{author}{\bibinfo{person}{Pierre{-}Louis Aublin},
  \bibinfo{person}{Florian Kelbert}, \bibinfo{person}{Dan O'Keeffe},
  \bibinfo{person}{Divya Muthukumaran}, \bibinfo{person}{Christian Priebe},
  \bibinfo{person}{Joshua Lind}, \bibinfo{person}{Robert Krahn},
  \bibinfo{person}{Christof Fetzer}, \bibinfo{person}{David~M. Eyers}, {and}
  \bibinfo{person}{Peter~R. Pietzuch}.} \bibinfo{year}{2018}\natexlab{}.
\newblock \showarticletitle{{LibSEAL}: revealing service integrity violations
  using trusted execution}. In \bibinfo{booktitle}{\emph{EuroSys}}.
  \bibinfo{publisher}{{ACM}}, \bibinfo{pages}{24:1--24:15}.
\newblock
\urldef\tempurl%
\url{https://doi.org/10.1145/3190508.3190547}
\showURL{%
\tempurl}


\bibitem[\protect\citeauthoryear{Azab, Ning, Shah, Chen, Bhutkar, Ganesh, Ma,
  and Shen}{Azab et~al\mbox{.}}{2014}]%
        {DBLP:conf/ccs/AzabNSCBGMS14}
\bibfield{author}{\bibinfo{person}{Ahmed~M. Azab}, \bibinfo{person}{Peng Ning},
  \bibinfo{person}{Jitesh Shah}, \bibinfo{person}{Quan Chen},
  \bibinfo{person}{Rohan Bhutkar}, \bibinfo{person}{Guruprasad Ganesh},
  \bibinfo{person}{Jia Ma}, {and} \bibinfo{person}{Wenbo Shen}.}
  \bibinfo{year}{2014}\natexlab{}.
\newblock \showarticletitle{Hypervision Across Worlds: Real-time Kernel
  Protection from the {ARM} {TrustZone} Secure World}. In
  \bibinfo{booktitle}{\emph{{ACM} {CCS}}}. \bibinfo{publisher}{{ACM}},
  \bibinfo{pages}{90--102}.
\newblock
\urldef\tempurl%
\url{https://doi.org/10.1145/2660267.2660350}
\showURL{%
\tempurl}


\bibitem[\protect\citeauthoryear{Bahmani, Brasser, Dessouky, Jauernig, Klimmek,
  Sadeghi, and Stapf}{Bahmani et~al\mbox{.}}{2021}]%
        {DBLP:conf/uss/BahmaniBDJKSS21}
\bibfield{author}{\bibinfo{person}{Raad Bahmani}, \bibinfo{person}{Ferdinand
  Brasser}, \bibinfo{person}{Ghada Dessouky}, \bibinfo{person}{Patrick
  Jauernig}, \bibinfo{person}{Matthias Klimmek}, \bibinfo{person}{Ahmad{-}Reza
  Sadeghi}, {and} \bibinfo{person}{Emmanuel Stapf}.}
  \bibinfo{year}{2021}\natexlab{}.
\newblock \showarticletitle{{CURE}: A Security Architecture with CUstomizable
  and Resilient Enclaves}. In \bibinfo{booktitle}{\emph{{USENIX} Sec.}}
  \bibinfo{publisher}{{USENIX} Association}, \bibinfo{pages}{1073--1090}.
\newblock
\urldef\tempurl%
\url{https://www.usenix.org/conference/usenixsecurity21/presentation/bahmani}
\showURL{%
\tempurl}


\bibitem[\protect\citeauthoryear{Bauman and Lin}{Bauman and Lin}{2016}]%
        {DBLP:conf/middleware/BaumanL16}
\bibfield{author}{\bibinfo{person}{Erick Bauman} {and}
  \bibinfo{person}{Zhiqiang Lin}.} \bibinfo{year}{2016}\natexlab{}.
\newblock \showarticletitle{A Case for Protecting Computer Games With {SGX}}.
  In \bibinfo{booktitle}{\emph{{SysTEX}}}. \bibinfo{publisher}{{ACM}},
  \bibinfo{pages}{4:1--4:6}.
\newblock
\urldef\tempurl%
\url{https://doi.org/10.1145/3007788.3007792}
\showURL{%
\tempurl}


\bibitem[\protect\citeauthoryear{BitObscuro}{BitObscuro}{2020}]%
        {ObscuroGit}
\bibfield{author}{\bibinfo{person}{BitObscuro}.}
  \bibinfo{year}{2020}\natexlab{}.
\newblock \bibinfo{title}{{Obscuro}}.
\newblock
\newblock
\urldef\tempurl%
\url{https://github.com/BitObscuro/Obscuro}
\showURL{%
\tempurl}


\bibitem[\protect\citeauthoryear{Bogaard}{Bogaard}{2019}]%
        {FuzzingOPTee19}
\bibfield{author}{\bibinfo{person}{Martijn Bogaard}.}
  \bibinfo{year}{2019}\natexlab{}.
\newblock \bibinfo{title}{{Fuzzing OP-TEE with AFL}}.
\newblock
  \bibinfo{howpublished}{\url{https://static.linaro.org/connect/san19/presentations/san19-225.pdf}}.
\newblock


\bibitem[\protect\citeauthoryear{Brasser, Gens, Jauernig, Sadeghi, and
  Stapf}{Brasser et~al\mbox{.}}{2019}]%
        {DBLP:conf/ndss/BrasserGJSS19}
\bibfield{author}{\bibinfo{person}{Ferdinand Brasser}, \bibinfo{person}{David
  Gens}, \bibinfo{person}{Patrick Jauernig}, \bibinfo{person}{Ahmad{-}Reza
  Sadeghi}, {and} \bibinfo{person}{Emmanuel Stapf}.}
  \bibinfo{year}{2019}\natexlab{}.
\newblock \showarticletitle{{SANCTUARY}: {ARMing} {TrustZone} with User-space
  Enclaves}. In \bibinfo{booktitle}{\emph{{NDSS}}}. \bibinfo{publisher}{The
  Internet Society}.
\newblock
\urldef\tempurl%
\url{https://www.ndss-symposium.org/ndss-paper/sanctuary-arming-trustzone-with-user-space-enclaves/}
\showURL{%
\tempurl}


\bibitem[\protect\citeauthoryear{Brenner and Kapitza}{Brenner and
  Kapitza}{2019}]%
        {DBLP:conf/systor/BrennerK19}
\bibfield{author}{\bibinfo{person}{Stefan Brenner} {and}
  \bibinfo{person}{R{\"{u}}diger Kapitza}.} \bibinfo{year}{2019}\natexlab{}.
\newblock \showarticletitle{Trust more, serverless}. In
  \bibinfo{booktitle}{\emph{{SYSTOR}}}. \bibinfo{publisher}{{ACM}},
  \bibinfo{pages}{33--43}.
\newblock
\urldef\tempurl%
\url{https://doi.org/10.1145/3319647.3325825}
\showURL{%
\tempurl}


\bibitem[\protect\citeauthoryear{{Bytecode Alliance}}{{Bytecode
  Alliance}}{2019}]%
        {wasm-micro-runtime}
\bibfield{author}{\bibinfo{person}{{Bytecode Alliance}}.}
  \bibinfo{year}{2019}\natexlab{}.
\newblock \bibinfo{title}{{WebAssembly Micro Runtime (WAMR)}}.
\newblock
\newblock
\urldef\tempurl%
\url{https://github.com/bytecodealliance/wasm-micro-runtime}
\showURL{%
\tempurl}
\newblock
\shownote{Latest rel.\ 2022.}


\bibitem[\protect\citeauthoryear{Cai, Xu, Zhou, and Wang}{Cai
  et~al\mbox{.}}{2022}]%
        {9512478}
\bibfield{author}{\bibinfo{person}{Chengjun Cai}, \bibinfo{person}{Lei Xu},
  \bibinfo{person}{Anxin Zhou}, {and} \bibinfo{person}{Cong Wang}.}
  \bibinfo{year}{2022}\natexlab{}.
\newblock \showarticletitle{Toward a Secure, Rich, and Fair Query Service for
  Light Clients on Public Blockchains}.
\newblock \bibinfo{journal}{\emph{{IEEE} Trans. Dependable Secur. Comput.}}
  \bibinfo{volume}{19}, \bibinfo{number}{6} (\bibinfo{year}{2022}),
  \bibinfo{pages}{3640--3655}.
\newblock
\urldef\tempurl%
\url{https://doi.org/10.1109/TDSC.2021.3103382}
\showURL{%
\tempurl}


\bibitem[\protect\citeauthoryear{Cerdeira, Santos, Fonseca, and Pinto}{Cerdeira
  et~al\mbox{.}}{2020}]%
        {DBLP:conf/sp/Cerdeira0FP20}
\bibfield{author}{\bibinfo{person}{David Cerdeira}, \bibinfo{person}{Nuno
  Santos}, \bibinfo{person}{Pedro Fonseca}, {and} \bibinfo{person}{Sandro
  Pinto}.} \bibinfo{year}{2020}\natexlab{}.
\newblock \showarticletitle{{SoK}: Understanding the Prevailing Security
  Vulnerabilities in {TrustZone}-assisted {TEE} Systems}. In
  \bibinfo{booktitle}{\emph{{IEEE} S{\&}P}}. \bibinfo{publisher}{{IEEE}},
  \bibinfo{pages}{1416--1432}.
\newblock
\urldef\tempurl%
\url{https://doi.org/10.1109/SP40000.2020.00061}
\showURL{%
\tempurl}


\bibitem[\protect\citeauthoryear{Chandra}{Chandra}{2017}]%
        {secure-analytics-sgx}
\bibfield{author}{\bibinfo{person}{Swarup Chandra}.}
  \bibinfo{year}{2017}\natexlab{}.
\newblock \bibinfo{title}{{Securing Data Analytics on SGX with Randomization}}.
\newblock
\newblock
\urldef\tempurl%
\url{https://github.com/swarupchandra/secure-analytics-sgx}
\showURL{%
\tempurl}


\bibitem[\protect\citeauthoryear{Chandra, Karande, Lin, Khan, Kantarcioglu, and
  Thuraisingham}{Chandra et~al\mbox{.}}{2017}]%
        {DBLP:conf/esorics/ChandraKLKKT17}
\bibfield{author}{\bibinfo{person}{Swarup Chandra}, \bibinfo{person}{Vishal
  Karande}, \bibinfo{person}{Zhiqiang Lin}, \bibinfo{person}{Latifur Khan},
  \bibinfo{person}{Murat Kantarcioglu}, {and} \bibinfo{person}{Bhavani~M.
  Thuraisingham}.} \bibinfo{year}{2017}\natexlab{}.
\newblock \showarticletitle{Securing Data Analytics on {SGX} with
  Randomization}. In \bibinfo{booktitle}{\emph{{ESORICS}}}
  \emph{(\bibinfo{series}{LNCS}, Vol.~\bibinfo{volume}{10492})}.
  \bibinfo{publisher}{Springer}, \bibinfo{pages}{352--369}.
\newblock
\urldef\tempurl%
\url{https://doi.org/10.1007/978-3-319-66402-6_21}
\showURL{%
\tempurl}


\bibitem[\protect\citeauthoryear{Chang, Jiang, Chen, Xiang, Cheng, and
  Alelaiwi}{Chang et~al\mbox{.}}{2017}]%
        {DBLP:journals/cluster/ChangJCXCA17}
\bibfield{author}{\bibinfo{person}{Rui Chang}, \bibinfo{person}{Liehui Jiang},
  \bibinfo{person}{Wenzhi Chen}, \bibinfo{person}{Yang Xiang},
  \bibinfo{person}{Yuxia Cheng}, {and} \bibinfo{person}{Abdulhameed Alelaiwi}.}
  \bibinfo{year}{2017}\natexlab{}.
\newblock \showarticletitle{{MIPE}: a practical memory integrity protection
  method in a trusted execution environment}.
\newblock \bibinfo{journal}{\emph{Clust. Comput.}} \bibinfo{volume}{20},
  \bibinfo{number}{2} (\bibinfo{year}{2017}), \bibinfo{pages}{1075--1087}.
\newblock
\urldef\tempurl%
\url{https://doi.org/10.1007/s10586-017-0833-4}
\showURL{%
\tempurl}


\bibitem[\protect\citeauthoryear{Chen}{Chen}{2019}]%
        {linux-sgx-mage}
\bibfield{author}{\bibinfo{person}{Guoxing Chen}.}
  \bibinfo{year}{2019}\natexlab{}.
\newblock \bibinfo{title}{{MAGE: Mutual Attestation for a Group of Enclaves
  without Trusted Third Parties}}.
\newblock
\newblock
\urldef\tempurl%
\url{https://github.com/donnod/linux-sgx-mage}
\showURL{%
\tempurl}
\newblock
\shownote{Latest rel.\ 2021.}


\bibitem[\protect\citeauthoryear{Chen and Zhang}{Chen and Zhang}{2022}]%
        {DBLP:conf/uss/ChenZ22}
\bibfield{author}{\bibinfo{person}{Guoxing Chen} {and} \bibinfo{person}{Yinqian
  Zhang}.} \bibinfo{year}{2022}\natexlab{}.
\newblock \showarticletitle{{MAGE}: Mutual Attestation for a Group of Enclaves
  without Trusted Third Parties}. In \bibinfo{booktitle}{\emph{{USENIX} Sec.}}
  \bibinfo{publisher}{{USENIX} Association}, \bibinfo{pages}{4095--4110}.
\newblock
\urldef\tempurl%
\url{https://www.usenix.org/conference/usenixsecurity22/presentation/chen-guoxing}
\showURL{%
\tempurl}


\bibitem[\protect\citeauthoryear{Chen, Zhang, and Lai}{Chen
  et~al\mbox{.}}{2019}]%
        {DBLP:conf/ccs/ChenZL19}
\bibfield{author}{\bibinfo{person}{Guoxing Chen}, \bibinfo{person}{Yinqian
  Zhang}, {and} \bibinfo{person}{Ten{-}Hwang Lai}.}
  \bibinfo{year}{2019}\natexlab{}.
\newblock \showarticletitle{{OPERA}: Open Remote Attestation for {Intel's}
  Secure Enclaves}. In \bibinfo{booktitle}{\emph{{ACM} {CCS}}}.
  \bibinfo{publisher}{{ACM}}, \bibinfo{pages}{2317--2331}.
\newblock
\urldef\tempurl%
\url{https://doi.org/10.1145/3319535.3354220}
\showURL{%
\tempurl}


\bibitem[\protect\citeauthoryear{Chen, Yuan, and Xia}{Chen
  et~al\mbox{.}}{2021}]%
        {9566165}
\bibfield{author}{\bibinfo{person}{Lili Chen}, \bibinfo{person}{Rui Yuan},
  {and} \bibinfo{person}{Yubin Xia}.} \bibinfo{year}{2021}\natexlab{}.
\newblock \showarticletitle{{Tora}: A Trusted Blockchain Oracle Based on a
  Decentralized {TEE} Network}. In \bibinfo{booktitle}{\emph{{JCC}}}.
  \bibinfo{pages}{28--33}.
\newblock
\urldef\tempurl%
\url{https://doi.org/10.1109/JCC53141.2021.00016}
\showDOI{\tempurl}


\bibitem[\protect\citeauthoryear{Chen, Luo, Li, Xiang, Liu, and Li}{Chen
  et~al\mbox{.}}{2020}]%
        {DBLP:journals/isci/ChenLLXLL20}
\bibfield{author}{\bibinfo{person}{Yu Chen}, \bibinfo{person}{Fang Luo},
  \bibinfo{person}{Tong Li}, \bibinfo{person}{Tao Xiang},
  \bibinfo{person}{Zheli Liu}, {and} \bibinfo{person}{Jin Li}.}
  \bibinfo{year}{2020}\natexlab{}.
\newblock \showarticletitle{A training-integrity privacy-preserving federated
  learning scheme with trusted execution environment}.
\newblock \bibinfo{journal}{\emph{Inf. Sci.}}  \bibinfo{volume}{522}
  (\bibinfo{year}{2020}), \bibinfo{pages}{69--79}.
\newblock
\urldef\tempurl%
\url{https://doi.org/10.1016/j.ins.2020.02.037}
\showURL{%
\tempurl}


\bibitem[\protect\citeauthoryear{Costan}{Costan}{2015}]%
        {SanctumOldGit}
\bibfield{author}{\bibinfo{person}{V. Costan}.}
  \bibinfo{year}{2015}\natexlab{}.
\newblock \bibinfo{title}{{Sanctum}}.
\newblock
\newblock
\urldef\tempurl%
\url{https://github.com/pwnall/sanctum}
\showURL{%
\tempurl}
\newblock
\shownote{{Latest rel.\ 2019}.}


\bibitem[\protect\citeauthoryear{Costan, Lebedev, and Devadas}{Costan
  et~al\mbox{.}}{2016}]%
        {DBLP:conf/uss/CostanLD16}
\bibfield{author}{\bibinfo{person}{Victor Costan}, \bibinfo{person}{Ilia~A.
  Lebedev}, {and} \bibinfo{person}{Srinivas Devadas}.}
  \bibinfo{year}{2016}\natexlab{}.
\newblock \showarticletitle{{Sanctum}: Minimal Hardware Extensions for Strong
  Software Isolation}. In \bibinfo{booktitle}{\emph{{USENIX} Sec.}}
  \bibinfo{publisher}{{USENIX} Association}, \bibinfo{pages}{857--874}.
\newblock
\urldef\tempurl%
\url{https://www.usenix.org/conference/usenixsecurity16/technical-sessions/presentation/costan}
\showURL{%
\tempurl}


\bibitem[\protect\citeauthoryear{Cui, Shinde, Sen, Saxena, and Yuan}{Cui
  et~al\mbox{.}}{2022}]%
        {DBLP:journals/tissec/CuiSSSY22}
\bibfield{author}{\bibinfo{person}{Jinhua Cui}, \bibinfo{person}{Shweta
  Shinde}, \bibinfo{person}{Satyaki Sen}, \bibinfo{person}{Prateek Saxena},
  {and} \bibinfo{person}{Pinghai Yuan}.} \bibinfo{year}{2022}\natexlab{}.
\newblock \showarticletitle{Dynamic Binary Translation for {SGX} Enclaves}.
\newblock \bibinfo{journal}{\emph{{ACM} Trans. Priv. Secur.}}
  \bibinfo{volume}{25}, \bibinfo{number}{4} (\bibinfo{year}{2022}),
  \bibinfo{pages}{32:1--32:40}.
\newblock
\urldef\tempurl%
\url{https://doi.org/10.1145/3532862}
\showURL{%
\tempurl}


\bibitem[\protect\citeauthoryear{CyFI-Lab-Public}{CyFI-Lab-Public}{2016}]%
        {RetroScope}
\bibfield{author}{\bibinfo{person}{CyFI-Lab-Public}.}
  \bibinfo{year}{2016}\natexlab{}.
\newblock \bibinfo{title}{{RetroScope: Android memory forensics framework}}.
\newblock
\newblock
\urldef\tempurl%
\url{https://github.com/CyFI-Lab-Public/RetroScope}
\showURL{%
\tempurl}


\bibitem[\protect\citeauthoryear{da~Rocha, Valadares, Perkusich,
  Gorg{\^{o}}nio, Pagno, and Will}{da~Rocha et~al\mbox{.}}{2020}]%
        {DBLP:conf/closer/RochaVPGPW20}
\bibfield{author}{\bibinfo{person}{Marciano da Rocha}, \bibinfo{person}{Dalton
  C{\'{e}}zane~Gomes Valadares}, \bibinfo{person}{Angelo Perkusich},
  \bibinfo{person}{Kyller~Costa Gorg{\^{o}}nio}, \bibinfo{person}{Rodrigo~Tomaz
  Pagno}, {and} \bibinfo{person}{Newton~Carlos Will}.}
  \bibinfo{year}{2020}\natexlab{}.
\newblock \showarticletitle{Secure Cloud Storage with Client-side Encryption
  using a Trusted Execution Environment}. In
  \bibinfo{booktitle}{\emph{{CLOSER}}}. \bibinfo{publisher}{{SCITEPRESS}},
  \bibinfo{pages}{31--43}.
\newblock
\urldef\tempurl%
\url{https://doi.org/10.5220/0009130600310043}
\showURL{%
\tempurl}


\bibitem[\protect\citeauthoryear{Dai, Wang, Wang, Lin, Zou, and Jin}{Dai
  et~al\mbox{.}}{2021}]%
        {DBLP:journals/jpdc/DaiWWLZJ21}
\bibfield{author}{\bibinfo{person}{Weiqi Dai}, \bibinfo{person}{Qinyuan Wang},
  \bibinfo{person}{Zeli Wang}, \bibinfo{person}{Xiaobin Lin},
  \bibinfo{person}{Deqing Zou}, {and} \bibinfo{person}{Hai Jin}.}
  \bibinfo{year}{2021}\natexlab{}.
\newblock \showarticletitle{{TrustZone}-based secure lightweight wallet for
  hyperledger fabric}.
\newblock \bibinfo{journal}{\emph{J. Parallel Distributed Comput.}}
  \bibinfo{volume}{149} (\bibinfo{year}{2021}), \bibinfo{pages}{66--75}.
\newblock
\urldef\tempurl%
\url{https://doi.org/10.1016/j.jpdc.2020.11.001}
\showURL{%
\tempurl}


\bibitem[\protect\citeauthoryear{Dangwal, Cowan, Alaghi, Lee, Reagen, and
  Trippel}{Dangwal et~al\mbox{.}}{2020}]%
        {DBLP:journals/corr/abs-2105-00378}
\bibfield{author}{\bibinfo{person}{Deeksha Dangwal}, \bibinfo{person}{Meghan
  Cowan}, \bibinfo{person}{Armin Alaghi}, \bibinfo{person}{Vincent~T. Lee},
  \bibinfo{person}{Brandon Reagen}, {and} \bibinfo{person}{Caroline Trippel}.}
  \bibinfo{year}{2020}\natexlab{}.
\newblock \showarticletitle{{SoK}: Opportunities for Software-Hardware-Security
  Codesign for Next Generation Secure Computing}. In
  \bibinfo{booktitle}{\emph{{HASP}}}. \bibinfo{publisher}{{ACM}},
  \bibinfo{pages}{8:1--8:9}.
\newblock
\urldef\tempurl%
\url{https://doi.org/10.1145/3458903.3458911}
\showURL{%
\tempurl}


\bibitem[\protect\citeauthoryear{Decentriq}{Decentriq}{2021}]%
        {Decentriq2022}
\bibfield{author}{\bibinfo{person}{Decentriq}.}
  \bibinfo{year}{2021}\natexlab{}.
\newblock \bibinfo{title}{Decentriq}.
\newblock
\newblock
\urldef\tempurl%
\url{https://www.decentriq.com/}
\showURL{%
\tempurl}
\newblock
\shownote{Latest rel.\ 2022.}


\bibitem[\protect\citeauthoryear{{deislabs}}{{deislabs}}{2021}]%
        {Mystikos2022}
\bibfield{author}{\bibinfo{person}{{deislabs}}.}
  \bibinfo{year}{2021}\natexlab{}.
\newblock \bibinfo{title}{{Mystikos}}.
\newblock
\newblock
\urldef\tempurl%
\url{https://github.com/deislabs/mystikos}
\showURL{%
\tempurl}
\newblock
\shownote{Latest rel.\ 2022.}


\bibitem[\protect\citeauthoryear{Dhar, Puddu, Kostiainen, and Capkun}{Dhar
  et~al\mbox{.}}{2020}]%
        {DBLP:conf/codaspy/DharPKC20}
\bibfield{author}{\bibinfo{person}{Aritra Dhar}, \bibinfo{person}{Ivan Puddu},
  \bibinfo{person}{Kari Kostiainen}, {and} \bibinfo{person}{Srdjan Capkun}.}
  \bibinfo{year}{2020}\natexlab{}.
\newblock \showarticletitle{{ProximiTEE}: Hardened {SGX} Attestation by
  Proximity Verification}. In \bibinfo{booktitle}{\emph{{CODASPY}}}.
  \bibinfo{publisher}{{ACM}}, \bibinfo{pages}{5--16}.
\newblock
\urldef\tempurl%
\url{https://doi.org/10.1145/3374664.3375726}
\showURL{%
\tempurl}


\bibitem[\protect\citeauthoryear{{Distributed Systems group at IBR, TU
  Braunschweig}}{{Distributed Systems group at IBR, TU Braunschweig}}{2020}]%
        {AccTEE2020}
\bibfield{author}{\bibinfo{person}{{Distributed Systems group at IBR, TU
  Braunschweig}}.} \bibinfo{year}{2020}\natexlab{}.
\newblock \bibinfo{title}{{AccTEE: A WebAssembly-based Two-way Sandbox for
  Trusted Resource Accounting}}.
\newblock
\newblock
\urldef\tempurl%
\url{https://github.com/ibr-ds/AccTEE}
\showURL{%
\tempurl}


\bibitem[\protect\citeauthoryear{Djoko}{Djoko}{2020}]%
        {NexusGit}
\bibfield{author}{\bibinfo{person}{Briand Djoko}.}
  \bibinfo{year}{2020}\natexlab{}.
\newblock \bibinfo{title}{{Secure cloud access/usage control using client-side
  SGX}}.
\newblock
\newblock
\urldef\tempurl%
\url{https://github.com/sporgj/nexus-code}
\showURL{%
\tempurl}


\bibitem[\protect\citeauthoryear{Djoko, Lange, and Lee}{Djoko
  et~al\mbox{.}}{2019}]%
        {DBLP:conf/dsn/DjokoLL19}
\bibfield{author}{\bibinfo{person}{Judicael~Briand Djoko},
  \bibinfo{person}{Jack Lange}, {and} \bibinfo{person}{Adam~J. Lee}.}
  \bibinfo{year}{2019}\natexlab{}.
\newblock \showarticletitle{{NeXUS}: Practical and Secure Access Control on
  Untrusted Storage Platforms using Client-Side {SGX}}. In
  \bibinfo{booktitle}{\emph{{DSN}}}. \bibinfo{publisher}{{IEEE}},
  \bibinfo{pages}{401--413}.
\newblock
\urldef\tempurl%
\url{https://doi.org/10.1109/DSN.2019.00049}
\showURL{%
\tempurl}


\bibitem[\protect\citeauthoryear{Dokmai}{Dokmai}{2022}]%
        {SmacGit}
\bibfield{author}{\bibinfo{person}{Ko Dokmai}.}
  \bibinfo{year}{2022}\natexlab{}.
\newblock \bibinfo{title}{{SMac: Secure Genotype Imputation in Intel SGX}}.
\newblock
\newblock
\urldef\tempurl%
\url{https://github.com/ndokmai/sgx-genotype-imputation}
\showURL{%
\tempurl}


\bibitem[\protect\citeauthoryear{Dokmai, Kockan, Zhu, Wang, Sahinalp, and
  Cho}{Dokmai et~al\mbox{.}}{2021}]%
        {DOKMAI2021983}
\bibfield{author}{\bibinfo{person}{Natnatee Dokmai}, \bibinfo{person}{Can
  Kockan}, \bibinfo{person}{Kaiyuan Zhu}, \bibinfo{person}{XiaoFeng Wang},
  \bibinfo{person}{S.~Cenk Sahinalp}, {and} \bibinfo{person}{Hyunghoon Cho}.}
  \bibinfo{year}{2021}\natexlab{}.
\newblock \showarticletitle{Privacy-preserving genotype imputation in a trusted
  execution environment}.
\newblock \bibinfo{journal}{\emph{Cell Systems}} \bibinfo{volume}{12},
  \bibinfo{number}{10} (\bibinfo{year}{2021}), \bibinfo{pages}{983--993.e7}.
\newblock
\showISSN{2405-4712}
\urldef\tempurl%
\url{https://doi.org/10.1016/j.cels.2021.08.001}
\showDOI{\tempurl}


\bibitem[\protect\citeauthoryear{Dong}{Dong}{2021}]%
        {QSEESDK}
\bibfield{author}{\bibinfo{person}{David Dong}.}
  \bibinfo{year}{2021}\natexlab{}.
\newblock \bibinfo{title}{{Build TA images on different TEE}}.
\newblock
\newblock
\urldef\tempurl%
\url{https://dqdongg.com/android/fingerprint/2021/02/03/Fingerprint-build-ta.html}
\showURL{%
\tempurl}


\bibitem[\protect\citeauthoryear{Duan, Wang, Yuan, Zhou, Wang, and Ren}{Duan
  et~al\mbox{.}}{2019}]%
        {DBLP:conf/ccs/DuanWYZW019}
\bibfield{author}{\bibinfo{person}{Huayi Duan}, \bibinfo{person}{Cong Wang},
  \bibinfo{person}{Xingliang Yuan}, \bibinfo{person}{Yajin Zhou},
  \bibinfo{person}{Qian Wang}, {and} \bibinfo{person}{Kui Ren}.}
  \bibinfo{year}{2019}\natexlab{}.
\newblock \showarticletitle{{LightBox}: Full-stack Protected Stateful Middlebox
  at Lightning Speed}. In \bibinfo{booktitle}{\emph{{ACM} {CCS}}}.
  \bibinfo{publisher}{{ACM}}, \bibinfo{pages}{2351--2367}.
\newblock
\urldef\tempurl%
\url{https://doi.org/10.1145/3319535.3339814}
\showURL{%
\tempurl}


\bibitem[\protect\citeauthoryear{{Edgeless Systems}}{{Edgeless
  Systems}}{2020}]%
        {EdgelessRT2022}
\bibfield{author}{\bibinfo{person}{{Edgeless Systems}}.}
  \bibinfo{year}{2020}\natexlab{}.
\newblock \bibinfo{title}{{Edgeless RT}}.
\newblock
\newblock
\urldef\tempurl%
\url{https://github.com/edgelesssys/edgelessrt}
\showURL{%
\tempurl}
\newblock
\shownote{{Latest rel.\ 2022}.}


\bibitem[\protect\citeauthoryear{{Edgeless Systems}}{{Edgeless
  Systems}}{2021}]%
        {Ego2022}
\bibfield{author}{\bibinfo{person}{{Edgeless Systems}}.}
  \bibinfo{year}{2021}\natexlab{}.
\newblock \bibinfo{title}{{Welcome to EGo}}.
\newblock
\newblock
\urldef\tempurl%
\url{https://docs.edgeless.systems/ego}
\showURL{%
\tempurl}
\newblock
\shownote{Latest rel.\ 2022.}


\bibitem[\protect\citeauthoryear{Ekberg, Asokan, Kostiainen, and
  Rantala}{Ekberg et~al\mbox{.}}{2008}]%
        {DBLP:conf/ccs/EkbergAKR08}
\bibfield{author}{\bibinfo{person}{Jan{-}Erik Ekberg}, \bibinfo{person}{N.
  Asokan}, \bibinfo{person}{Kari Kostiainen}, {and} \bibinfo{person}{Aarne
  Rantala}.} \bibinfo{year}{2008}\natexlab{}.
\newblock \showarticletitle{Scheduling execution of credentials in constrained
  secure environments}. In \bibinfo{booktitle}{\emph{{STC}}}.
  \bibinfo{publisher}{{ACM}}, \bibinfo{pages}{61--70}.
\newblock
\urldef\tempurl%
\url{https://doi.org/10.1145/1456455.1456465}
\showURL{%
\tempurl}


\bibitem[\protect\citeauthoryear{Ekberg, Kostiainen, and Asokan}{Ekberg
  et~al\mbox{.}}{2014}]%
        {DBLP:journals/ieeesp/EkbergKA14}
\bibfield{author}{\bibinfo{person}{Jan{-}Erik Ekberg}, \bibinfo{person}{Kari
  Kostiainen}, {and} \bibinfo{person}{N. Asokan}.}
  \bibinfo{year}{2014}\natexlab{}.
\newblock \showarticletitle{The Untapped Potential of Trusted Execution
  Environments on Mobile Devices}.
\newblock \bibinfo{journal}{\emph{{IEEE} Secur. Priv.}} \bibinfo{volume}{12},
  \bibinfo{number}{4} (\bibinfo{year}{2014}), \bibinfo{pages}{29--37}.
\newblock
\urldef\tempurl%
\url{https://doi.org/10.1109/MSP.2014.38}
\showURL{%
\tempurl}


\bibitem[\protect\citeauthoryear{Enarx}{Enarx}{2021a}]%
        {Enarx2022}
\bibfield{author}{\bibinfo{person}{Enarx}.} \bibinfo{year}{2021}\natexlab{a}.
\newblock \bibinfo{title}{{Enarx}}.
\newblock
\newblock
\urldef\tempurl%
\url{https://github.com/enarx/enarx}
\showURL{%
\tempurl}
\newblock
\shownote{Latest rel.\ 2022.}


\bibitem[\protect\citeauthoryear{Enarx}{Enarx}{2021b}]%
        {enarx-shim}
\bibfield{author}{\bibinfo{person}{Enarx}.} \bibinfo{year}{2021}\natexlab{b}.
\newblock \bibinfo{title}{{Enarx Shim SGX}}.
\newblock
\newblock
\urldef\tempurl%
\url{https://github.com/enarx/enarx-shim-sgx}
\showURL{%
\tempurl}


\bibitem[\protect\citeauthoryear{Enarx}{Enarx}{2022}]%
        {mmledger}
\bibfield{author}{\bibinfo{person}{Enarx}.} \bibinfo{year}{2022}\natexlab{}.
\newblock \bibinfo{title}{{MMLedger: A ledger for confidential computing shims
  for tracking memory management system calls}}.
\newblock
\newblock
\urldef\tempurl%
\url{https://github.com/enarx/mmledger}
\showURL{%
\tempurl}


\bibitem[\protect\citeauthoryear{Fei, Yan, Ding, and Xie}{Fei
  et~al\mbox{.}}{2021}]%
        {DBLP:journals/csur/FeiYDX21}
\bibfield{author}{\bibinfo{person}{Shufan Fei}, \bibinfo{person}{Zheng Yan},
  \bibinfo{person}{Wenxiu Ding}, {and} \bibinfo{person}{Haomeng Xie}.}
  \bibinfo{year}{2021}\natexlab{}.
\newblock \showarticletitle{Security Vulnerabilities of {SGX} and
  Countermeasures: A Survey}.
\newblock \bibinfo{journal}{\emph{{ACM} Comput. Surv.}} \bibinfo{volume}{54},
  \bibinfo{number}{6} (\bibinfo{year}{2021}), \bibinfo{pages}{126:1--126:36}.
\newblock
\urldef\tempurl%
\url{https://doi.org/10.1145/3456631}
\showURL{%
\tempurl}


\bibitem[\protect\citeauthoryear{Feng, Lu, Du, Yang, Jiang, Xia, Zang, and
  Chen}{Feng et~al\mbox{.}}{2021}]%
        {DBLP:conf/osdi/FengLDYJXZ021}
\bibfield{author}{\bibinfo{person}{Erhu Feng}, \bibinfo{person}{Xu Lu},
  \bibinfo{person}{Dong Du}, \bibinfo{person}{Bicheng Yang},
  \bibinfo{person}{Xueqiang Jiang}, \bibinfo{person}{Yubin Xia},
  \bibinfo{person}{Binyu Zang}, {and} \bibinfo{person}{Haibo Chen}.}
  \bibinfo{year}{2021}\natexlab{}.
\newblock \showarticletitle{Scalable Memory Protection in the {PENGLAI}
  Enclave}. In \bibinfo{booktitle}{\emph{{OSDI}}}. \bibinfo{publisher}{{USENIX}
  Association}, \bibinfo{pages}{275--294}.
\newblock
\urldef\tempurl%
\url{https://www.usenix.org/conference/osdi21/presentation/feng}
\showURL{%
\tempurl}


\bibitem[\protect\citeauthoryear{Ferraiuolo, Baumann, Hawblitzel, and
  Parno}{Ferraiuolo et~al\mbox{.}}{2017}]%
        {DBLP:conf/sosp/FerraiuoloBHP17}
\bibfield{author}{\bibinfo{person}{Andrew Ferraiuolo}, \bibinfo{person}{Andrew
  Baumann}, \bibinfo{person}{Chris Hawblitzel}, {and} \bibinfo{person}{Bryan
  Parno}.} \bibinfo{year}{2017}\natexlab{}.
\newblock \showarticletitle{{Komodo}: Using verification to disentangle
  secure-enclave hardware from software}. In
  \bibinfo{booktitle}{\emph{{SOSP}}}. \bibinfo{publisher}{{ACM}},
  \bibinfo{pages}{287--305}.
\newblock
\urldef\tempurl%
\url{https://doi.org/10.1145/3132747.3132782}
\showURL{%
\tempurl}


\bibitem[\protect\citeauthoryear{Fisch, Vinayagamurthy, Boneh, and
  Gorbunov}{Fisch et~al\mbox{.}}{2017}]%
        {DBLP:conf/ccs/FischVBG17}
\bibfield{author}{\bibinfo{person}{Ben Fisch}, \bibinfo{person}{Dhinakaran
  Vinayagamurthy}, \bibinfo{person}{Dan Boneh}, {and} \bibinfo{person}{Sergey
  Gorbunov}.} \bibinfo{year}{2017}\natexlab{}.
\newblock \showarticletitle{{IRON}: Functional Encryption using {Intel} {SGX}}.
  In \bibinfo{booktitle}{\emph{{ACM} {CCS}}}. \bibinfo{publisher}{{ACM}},
  \bibinfo{pages}{765--782}.
\newblock
\urldef\tempurl%
\url{https://doi.org/10.1145/3133956.3134106}
\showURL{%
\tempurl}


\bibitem[\protect\citeauthoryear{Fischer, Lesjak, Pirker, and Steger}{Fischer
  et~al\mbox{.}}{2019}]%
        {fischer2019rpc}
\bibfield{author}{\bibinfo{person}{Thomas Fischer}, \bibinfo{person}{Christian
  Lesjak}, \bibinfo{person}{Dominic Pirker}, {and} \bibinfo{person}{Christian
  Steger}.} \bibinfo{year}{2019}\natexlab{}.
\newblock \showarticletitle{{RPC} Based Framework for Partitioning {IoT}
  Security Software for Trusted Execution Environments}. In
  \bibinfo{booktitle}{\emph{{IEMCON}}}. \bibinfo{pages}{430--435}.
\newblock
\urldef\tempurl%
\url{https://doi.org/10.1109/IEMCON.2019.8936247}
\showDOI{\tempurl}


\bibitem[\protect\citeauthoryear{Fortanix}{Fortanix}{2016}]%
        {FortanixEDP2022}
\bibfield{author}{\bibinfo{person}{Fortanix}.} \bibinfo{year}{2016}\natexlab{}.
\newblock \bibinfo{title}{{Fortanix Rust Enclave Development Platform}}.
\newblock
\newblock
\urldef\tempurl%
\url{https://github.com/fortanix/rust-sgx}
\showURL{%
\tempurl}
\newblock
\shownote{Latest rel.\ 2022.}


\bibitem[\protect\citeauthoryear{Fuhry, Hirschoff, Koesnadi, and
  Kerschbaum}{Fuhry et~al\mbox{.}}{2020}]%
        {DBLP:conf/dsn/FuhryHKK20}
\bibfield{author}{\bibinfo{person}{Benny Fuhry}, \bibinfo{person}{Lina
  Hirschoff}, \bibinfo{person}{Samuel Koesnadi}, {and} \bibinfo{person}{Florian
  Kerschbaum}.} \bibinfo{year}{2020}\natexlab{}.
\newblock \showarticletitle{{SeGShare}: Secure Group File Sharing in the Cloud
  using Enclaves}. In \bibinfo{booktitle}{\emph{{DSN}}}.
  \bibinfo{publisher}{{IEEE}}, \bibinfo{pages}{476--488}.
\newblock
\urldef\tempurl%
\url{https://doi.org/10.1109/DSN48063.2020.00061}
\showURL{%
\tempurl}


\bibitem[\protect\citeauthoryear{Gao}{Gao}{2021}]%
        {Teekap2022}
\bibfield{author}{\bibinfo{person}{Mingyuan Gao}.}
  \bibinfo{year}{2021}\natexlab{}.
\newblock \bibinfo{title}{TEEKAP}.
\newblock
\newblock
\urldef\tempurl%
\url{https://github.com/MingyuanGao/TEEKAP}
\showURL{%
\tempurl}
\newblock
\shownote{{Latest rel.\ 2022}.}


\bibitem[\protect\citeauthoryear{Gao, Dang, and Chang}{Gao
  et~al\mbox{.}}{2021}]%
        {DBLP:conf/acsac/GaoDC21}
\bibfield{author}{\bibinfo{person}{Mingyuan Gao}, \bibinfo{person}{Hung Dang},
  {and} \bibinfo{person}{Ee{-}Chien Chang}.} \bibinfo{year}{2021}\natexlab{}.
\newblock \showarticletitle{{TEEKAP}: Self-Expiring Data Capsule using Trusted
  Execution Environment}. In \bibinfo{booktitle}{\emph{{ACSAC}}}.
  \bibinfo{publisher}{{ACM}}, \bibinfo{pages}{235--247}.
\newblock
\urldef\tempurl%
\url{https://doi.org/10.1145/3485832.3485919}
\showURL{%
\tempurl}


\bibitem[\protect\citeauthoryear{Georgios}{Georgios}{2021}]%
        {AtlasMScThesis}
\bibfield{author}{\bibinfo{person}{Anagnopoulos Georgios}.}
  \bibinfo{year}{2021}\natexlab{}.
\newblock \emph{\bibinfo{title}{Atlas: Automated Scale-out of Trust-Oblivious
  Systems to Trusted Execution Environments}}.
\newblock \bibinfo{thesistype}{Master's\ thesis}. \bibinfo{school}{University
  of Crete}.
\newblock
\urldef\tempurl%
\url{https://elocus.lib.uoc.gr/dlib/e/6/1/metadata-dlib-1637579552-223704-1365.tkl}
\showURL{%
\tempurl}


\bibitem[\protect\citeauthoryear{Ghosn, Larus, and Bugnion}{Ghosn
  et~al\mbox{.}}{2019}]%
        {DBLP:conf/usenix/GhosnLB19}
\bibfield{author}{\bibinfo{person}{Adrien Ghosn}, \bibinfo{person}{James~R.
  Larus}, {and} \bibinfo{person}{Edouard Bugnion}.}
  \bibinfo{year}{2019}\natexlab{}.
\newblock \showarticletitle{Secured Routines: Language-based Construction of
  Trusted Execution Environments}. In \bibinfo{booktitle}{\emph{{USENIX}
  {ATC}}}. \bibinfo{publisher}{{USENIX} Association},
  \bibinfo{pages}{571--586}.
\newblock
\urldef\tempurl%
\url{https://www.usenix.org/conference/atc19/presentation/ghosn}
\showURL{%
\tempurl}


\bibitem[\protect\citeauthoryear{{GlobalPlatform Device
  Technology}}{{GlobalPlatform Device Technology}}{2010}]%
        {GlobalPlatformAPI}
\bibfield{author}{\bibinfo{person}{{GlobalPlatform Device Technology}}.}
  \bibinfo{year}{2010}\natexlab{}.
\newblock \bibinfo{booktitle}{\emph{{TEE Client API Specification Version
  1.0}}}.
\newblock \bibinfo{type}{{T}echnical {R}eport}.
  \bibinfo{institution}{GlobalPlatform}.
\newblock
\urldef\tempurl%
\url{https://globalplatform.org/wp-content/uploads/2010/07/TEE_Client_API_Specification-V1.0.pdf}
\showURL{%
\tempurl}


\bibitem[\protect\citeauthoryear{Goltzsche, Nieke, Knauth, and
  Kapitza}{Goltzsche et~al\mbox{.}}{2019}]%
        {DBLP:conf/middleware/GoltzscheNKK19}
\bibfield{author}{\bibinfo{person}{David Goltzsche}, \bibinfo{person}{Manuel
  Nieke}, \bibinfo{person}{Thomas Knauth}, {and} \bibinfo{person}{R{\"{u}}diger
  Kapitza}.} \bibinfo{year}{2019}\natexlab{}.
\newblock \showarticletitle{{AccTEE}: A {WebAssembly}-based Two-way Sandbox for
  Trusted Resource Accounting}. In \bibinfo{booktitle}{\emph{Middleware}}.
  \bibinfo{publisher}{{ACM}}, \bibinfo{pages}{123--135}.
\newblock
\urldef\tempurl%
\url{https://doi.org/10.1145/3361525.3361541}
\showURL{%
\tempurl}


\bibitem[\protect\citeauthoryear{Goltzsche, Wulf, Muthukumaran, Rieck,
  Pietzuch, and Kapitza}{Goltzsche et~al\mbox{.}}{2017}]%
        {DBLP:conf/eurosec/GoltzscheWMRPK17}
\bibfield{author}{\bibinfo{person}{David Goltzsche}, \bibinfo{person}{Colin
  Wulf}, \bibinfo{person}{Divya Muthukumaran}, \bibinfo{person}{Konrad Rieck},
  \bibinfo{person}{Peter~R. Pietzuch}, {and} \bibinfo{person}{R{\"{u}}diger
  Kapitza}.} \bibinfo{year}{2017}\natexlab{}.
\newblock \showarticletitle{{TrustJS}: Trusted Client-side Execution of
  {JavaScript}}. In \bibinfo{booktitle}{\emph{{EuroSec}}}.
  \bibinfo{publisher}{{ACM}}, \bibinfo{pages}{7:1--7:6}.
\newblock
\urldef\tempurl%
\url{https://doi.org/10.1145/3065913.3065917}
\showURL{%
\tempurl}


\bibitem[\protect\citeauthoryear{Google}{Google}{2018}]%
        {Asylo2022}
\bibfield{author}{\bibinfo{person}{Google}.} \bibinfo{year}{2018}\natexlab{}.
\newblock \bibinfo{title}{Asylo}.
\newblock
\newblock
\urldef\tempurl%
\url{https://github.com/google/asylo}
\showURL{%
\tempurl}
\newblock
\shownote{{Latest rel.\ 2022}.}


\bibitem[\protect\citeauthoryear{{Google Git}}{{Google Git}}{2013}]%
        {QSEEInitialRelease}
\bibfield{author}{\bibinfo{person}{{Google Git}}.}
  \bibinfo{year}{2013}\natexlab{}.
\newblock \bibinfo{title}{{qseecom: Add qseecom Driver}}.
\newblock
\newblock
\urldef\tempurl%
\url{https://android.googlesource.com/kernel/msm/+/d316c3dc0464e9703234bc1631700d832b2695bc}
\showURL{%
\tempurl}


\bibitem[\protect\citeauthoryear{Hanel}{Hanel}{2021}]%
        {Kinibi520a21}
\bibfield{author}{\bibinfo{person}{Lukas Hanel}.}
  \bibinfo{year}{2021}\natexlab{}.
\newblock \bibinfo{title}{{Kinibi-520a: The latest Trustonic Trusted Execution
  Environment (TEE)}}.
\newblock
  \bibinfo{howpublished}{\url{https://www.trustonic.com/technical-articles/kinibi-520a-the-latest-trusted-execution-environment-tee/}}.
\newblock


\bibitem[\protect\citeauthoryear{Hu, Chen, Joung, Carlak, Feng, Mao, and
  Liu}{Hu et~al\mbox{.}}{2020}]%
        {DBLP:conf/codaspy/HuCJCFML20}
\bibfield{author}{\bibinfo{person}{Shengtuo Hu}, \bibinfo{person}{Qi~Alfred
  Chen}, \bibinfo{person}{Jiwon Joung}, \bibinfo{person}{Can Carlak},
  \bibinfo{person}{Yiheng Feng}, \bibinfo{person}{Z.~Morley Mao}, {and}
  \bibinfo{person}{Henry~X. Liu}.} \bibinfo{year}{2020}\natexlab{}.
\newblock \showarticletitle{{CVShield}: Guarding Sensor Data in Connected
  Vehicle with Trusted Execution Environment}. In
  \bibinfo{booktitle}{\emph{{AutoSec}}}. \bibinfo{publisher}{{ACM}},
  \bibinfo{pages}{1--4}.
\newblock
\urldef\tempurl%
\url{https://doi.org/10.1145/3375706.3380552}
\showURL{%
\tempurl}


\bibitem[\protect\citeauthoryear{Hua, Yu, Gu, Xia, Chen, and Zang}{Hua
  et~al\mbox{.}}{2021}]%
        {DBLP:journals/chinaf/HuaYGXCZ21}
\bibfield{author}{\bibinfo{person}{Zhichao Hua}, \bibinfo{person}{Yang Yu},
  \bibinfo{person}{Jinyu Gu}, \bibinfo{person}{Yubin Xia},
  \bibinfo{person}{Haibo Chen}, {and} \bibinfo{person}{Binyu Zang}.}
  \bibinfo{year}{2021}\natexlab{}.
\newblock \showarticletitle{{TZ-Container}: protecting container from untrusted
  {OS} with {ARM} {TrustZone}}.
\newblock \bibinfo{journal}{\emph{Sci. China Inf. Sci.}} \bibinfo{volume}{64},
  \bibinfo{number}{9} (\bibinfo{year}{2021}).
\newblock
\urldef\tempurl%
\url{https://doi.org/10.1007/s11432-019-2707-6}
\showURL{%
\tempurl}


\bibitem[\protect\citeauthoryear{Hunt, Jia, Miller, Szekely, Hu, Rossbach, and
  Witchel}{Hunt et~al\mbox{.}}{2020}]%
        {DBLP:conf/nsdi/HuntJMSHRW20}
\bibfield{author}{\bibinfo{person}{Tyler Hunt}, \bibinfo{person}{Zhipeng Jia},
  \bibinfo{person}{Vance Miller}, \bibinfo{person}{Ariel Szekely},
  \bibinfo{person}{Yige Hu}, \bibinfo{person}{Christopher~J. Rossbach}, {and}
  \bibinfo{person}{Emmett Witchel}.} \bibinfo{year}{2020}\natexlab{}.
\newblock \showarticletitle{{Telekine}: Secure Computing with Cloud {GPUs}}. In
  \bibinfo{booktitle}{\emph{{NSDI}}}. \bibinfo{publisher}{{USENIX}
  Association}, \bibinfo{pages}{817--833}.
\newblock
\urldef\tempurl%
\url{https://www.usenix.org/conference/nsdi20/presentation/hunt}
\showURL{%
\tempurl}


\bibitem[\protect\citeauthoryear{Hunt, Zhu, Xu, Peter, and Witchel}{Hunt
  et~al\mbox{.}}{2018}]%
        {DBLP:journals/tocs/HuntZXPW18}
\bibfield{author}{\bibinfo{person}{Tyler Hunt}, \bibinfo{person}{Zhiting Zhu},
  \bibinfo{person}{Yuanzhong Xu}, \bibinfo{person}{Simon Peter}, {and}
  \bibinfo{person}{Emmett Witchel}.} \bibinfo{year}{2018}\natexlab{}.
\newblock \showarticletitle{{Ryoan}: A Distributed Sandbox for Untrusted
  Computation on Secret Data}.
\newblock \bibinfo{journal}{\emph{{ACM} Trans. Comput. Syst.}}
  \bibinfo{volume}{35}, \bibinfo{number}{4} (\bibinfo{year}{2018}),
  \bibinfo{pages}{13:1--13:32}.
\newblock
\urldef\tempurl%
\url{https://doi.org/10.1145/3231594}
\showURL{%
\tempurl}


\bibitem[\protect\citeauthoryear{{Intel Corporation}}{{Intel
  Corporation}}{2016}]%
        {linux-sgx2022}
\bibfield{author}{\bibinfo{person}{{Intel Corporation}}.}
  \bibinfo{year}{2016}\natexlab{}.
\newblock \bibinfo{title}{{Intel(R) Software Guard Extensions for Linux* OS}}.
\newblock
\newblock
\urldef\tempurl%
\url{https://github.com/intel/linux-sgx}
\showURL{%
\tempurl}
\newblock
\shownote{{Latest rel.\ 2022}.}


\bibitem[\protect\citeauthoryear{{IPADS}}{{IPADS}}{2021}]%
        {PenglaiGit}
\bibfield{author}{\bibinfo{person}{{IPADS}}.} \bibinfo{year}{2021}\natexlab{}.
\newblock \bibinfo{title}{{Penglai: Scalable Trusted Execution Environment for
  RISC-V}}.
\newblock
\newblock
\urldef\tempurl%
\url{https://github.com/Penglai-Enclave/Penglai-Enclave-sPMP}
\showURL{%
\tempurl}


\bibitem[\protect\citeauthoryear{Jang, Tang, Kim, Sethumadhavan, and Huh}{Jang
  et~al\mbox{.}}{2019}]%
        {DBLP:conf/asplos/JangTKSH19}
\bibfield{author}{\bibinfo{person}{Insu Jang}, \bibinfo{person}{Adrian Tang},
  \bibinfo{person}{Taehoon Kim}, \bibinfo{person}{Simha Sethumadhavan}, {and}
  \bibinfo{person}{Jaehyuk Huh}.} \bibinfo{year}{2019}\natexlab{}.
\newblock \showarticletitle{Heterogeneous Isolated Execution for Commodity
  {GPUs}}. In \bibinfo{booktitle}{\emph{{ASPLOS}}}. \bibinfo{publisher}{{ACM}},
  \bibinfo{pages}{455--468}.
\newblock
\urldef\tempurl%
\url{https://doi.org/10.1145/3297858.3304021}
\showURL{%
\tempurl}


\bibitem[\protect\citeauthoryear{Jang, Kong, Kim, Kim, and Kang}{Jang
  et~al\mbox{.}}{2015}]%
        {DBLP:conf/ndss/JangKKKK15}
\bibfield{author}{\bibinfo{person}{Jin~Soo Jang}, \bibinfo{person}{Sunjune
  Kong}, \bibinfo{person}{Minsu Kim}, \bibinfo{person}{Daegyeong Kim}, {and}
  \bibinfo{person}{Brent~ByungHoon Kang}.} \bibinfo{year}{2015}\natexlab{}.
\newblock \showarticletitle{{SeCReT}: Secure Channel between Rich Execution
  Environment and Trusted Execution Environment}. In
  \bibinfo{booktitle}{\emph{{NDSS}}}. \bibinfo{publisher}{The Internet
  Society}.
\newblock
\urldef\tempurl%
\url{https://www.ndss-symposium.org/ndss2015/secret-secure-channel-between-rich-execution-environment-and-trusted-execution-environment}
\showURL{%
\tempurl}


\bibitem[\protect\citeauthoryear{Jeon and Kim}{Jeon and Kim}{2021}]%
        {DBLP:journals/compsec/JeonK21a}
\bibfield{author}{\bibinfo{person}{Sanghoon Jeon} {and}
  \bibinfo{person}{Huy~Kang Kim}.} \bibinfo{year}{2021}\natexlab{}.
\newblock \showarticletitle{{TZMon}: Improving mobile game security with {ARM}
  {TrustZone}}.
\newblock \bibinfo{journal}{\emph{Comput. Secur.}}  \bibinfo{volume}{109}
  (\bibinfo{year}{2021}).
\newblock
\urldef\tempurl%
\url{https://doi.org/10.1016/j.cose.2021.102391}
\showURL{%
\tempurl}


\bibitem[\protect\citeauthoryear{Jeon}{Jeon}{2018}]%
        {TZMonGit}
\bibfield{author}{\bibinfo{person}{Sanghoon~(Kevin) Jeon}.}
  \bibinfo{year}{2018}\natexlab{}.
\newblock \bibinfo{title}{{TZMon: Improving mobile game security with ARM
  trustzone}}.
\newblock
\newblock
\urldef\tempurl%
\url{https://github.com/kppw99/TZMon}
\showURL{%
\tempurl}


\bibitem[\protect\citeauthoryear{{Joel Snyder}}{{Joel Snyder}}{2021}]%
        {Biometric-Authentication}
\bibfield{author}{\bibinfo{person}{{Joel Snyder}}.}
  \bibinfo{year}{2021}\natexlab{}.
\newblock \bibinfo{title}{{Using biometrics for authentication in Android}}.
\newblock
\newblock
\urldef\tempurl%
\url{https://insights.samsung.com/2021/04/21/using-biometrics-for-authentication-in-android-2}
\showURL{%
\tempurl}


\bibitem[\protect\citeauthoryear{Jseam}{Jseam}{2021}]%
        {ToraRepo}
\bibfield{author}{\bibinfo{person}{Jseam}.} \bibinfo{year}{2021}\natexlab{}.
\newblock \bibinfo{title}{{Tora-Zilliqa}}.
\newblock
\newblock
\urldef\tempurl%
\url{https://issueantenna.com/repo/JSeam2/Tora-Zilliqa}
\showURL{%
\tempurl}


\bibitem[\protect\citeauthoryear{{kaist-ina}}{{kaist-ina}}{2019}]%
        {SgxTorGit}
\bibfield{author}{\bibinfo{person}{{kaist-ina}}.}
  \bibinfo{year}{2019}\natexlab{}.
\newblock \bibinfo{title}{{SGX-Tor}}.
\newblock
\newblock
\urldef\tempurl%
\url{https://github.com/kaist-ina/SGX-Tor}
\showURL{%
\tempurl}


\bibitem[\protect\citeauthoryear{Kang, Xue, Jia, Wang, Kim, Youn, Kang, Lim,
  Jacob, and Huang}{Kang et~al\mbox{.}}{2021}]%
        {DBLP:conf/micro/KangXJWKYKLJ021}
\bibfield{author}{\bibinfo{person}{Luyi Kang}, \bibinfo{person}{Yuqi Xue},
  \bibinfo{person}{Weiwei Jia}, \bibinfo{person}{Xiaohao Wang},
  \bibinfo{person}{Jongryool Kim}, \bibinfo{person}{Changhwan Youn},
  \bibinfo{person}{Myeong~Joon Kang}, \bibinfo{person}{Hyung~Jin Lim},
  \bibinfo{person}{Bruce~L. Jacob}, {and} \bibinfo{person}{Jian Huang}.}
  \bibinfo{year}{2021}\natexlab{}.
\newblock \showarticletitle{{IceClave}: A Trusted Execution Environment for
  In-Storage Computing}. In \bibinfo{booktitle}{\emph{{MICRO}}}.
  \bibinfo{publisher}{{ACM}}, \bibinfo{pages}{199--211}.
\newblock
\urldef\tempurl%
\url{https://doi.org/10.1145/3466752.3480109}
\showURL{%
\tempurl}


\bibitem[\protect\citeauthoryear{Karande, Bauman, Lin, and Khan}{Karande
  et~al\mbox{.}}{2017}]%
        {DBLP:conf/ccs/KarandeBLK17}
\bibfield{author}{\bibinfo{person}{Vishal Karande}, \bibinfo{person}{Erick
  Bauman}, \bibinfo{person}{Zhiqiang Lin}, {and} \bibinfo{person}{Latifur
  Khan}.} \bibinfo{year}{2017}\natexlab{}.
\newblock \showarticletitle{{SGX-Log}: Securing System Logs With {SGX}}. In
  \bibinfo{booktitle}{\emph{AsiaCCS}}. \bibinfo{publisher}{{ACM}},
  \bibinfo{pages}{19--30}.
\newblock
\urldef\tempurl%
\url{https://doi.org/10.1145/3052973.3053034}
\showURL{%
\tempurl}


\bibitem[\protect\citeauthoryear{Kato, Cao, and Yoshikawa}{Kato
  et~al\mbox{.}}{2022}]%
        {DBLP:journals/corr/abs-2202-07165}
\bibfield{author}{\bibinfo{person}{Fumiyuki Kato}, \bibinfo{person}{Yang Cao},
  {and} \bibinfo{person}{Masatoshi Yoshikawa}.}
  \bibinfo{year}{2022}\natexlab{}.
\newblock \showarticletitle{{OLIVE:} Oblivious and Differentially Private
  Federated Learning on Trusted Execution Environment}.
\newblock \bibinfo{journal}{\emph{CoRR}}  \bibinfo{volume}{abs/2202.07165}
  (\bibinfo{year}{2022}).
\newblock
\urldef\tempurl%
\url{https://arxiv.org/abs/2202.07165}
\showURL{%
\tempurl}


\bibitem[\protect\citeauthoryear{{Keystone Enclave}}{{Keystone
  Enclave}}{2018}]%
        {Keystone2022}
\bibfield{author}{\bibinfo{person}{{Keystone Enclave}}.}
  \bibinfo{year}{2018}\natexlab{}.
\newblock \bibinfo{title}{{Keystone: An Open-Source Secure Enclave Framework
  for RISC-V Processors}}.
\newblock
\newblock
\urldef\tempurl%
\url{https://github.com/keystone-enclave/keystone}
\showURL{%
\tempurl}
\newblock
\shownote{{Latest rel.\ 2022}.}


\bibitem[\protect\citeauthoryear{Khalid and Masood}{Khalid and Masood}{2022}]%
        {DBLP:journals/compsec/KhalidM22}
\bibfield{author}{\bibinfo{person}{Fatima Khalid} {and} \bibinfo{person}{Ammar
  Masood}.} \bibinfo{year}{2022}\natexlab{}.
\newblock \showarticletitle{Vulnerability analysis of {Qualcomm} Secure
  Execution Environment ({QSEE})}.
\newblock \bibinfo{journal}{\emph{Comput. Secur.}}  \bibinfo{volume}{116}
  (\bibinfo{year}{2022}).
\newblock
\urldef\tempurl%
\url{https://doi.org/10.1016/j.cose.2022.102628}
\showURL{%
\tempurl}


\bibitem[\protect\citeauthoryear{Kim, Han, Ha, Kim, and Han}{Kim
  et~al\mbox{.}}{2017}]%
        {DBLP:conf/nsdi/KimHHKH17}
\bibfield{author}{\bibinfo{person}{Seong~Min Kim}, \bibinfo{person}{Juhyeng
  Han}, \bibinfo{person}{Jaehyeong Ha}, \bibinfo{person}{Taesoo Kim}, {and}
  \bibinfo{person}{Dongsu Han}.} \bibinfo{year}{2017}\natexlab{}.
\newblock \showarticletitle{Enhancing Security and Privacy of {Tor's} Ecosystem
  by Using Trusted Execution Environments}. In
  \bibinfo{booktitle}{\emph{{NSDI}}}. \bibinfo{publisher}{{USENIX}
  Association}, \bibinfo{pages}{145--161}.
\newblock
\urldef\tempurl%
\url{https://www.usenix.org/conference/nsdi17/technical-sessions/presentation/kim-seongmin}
\showURL{%
\tempurl}


\bibitem[\protect\citeauthoryear{Kim}{Kim}{2020}]%
        {ShieldStore}
\bibfield{author}{\bibinfo{person}{Taehoon Kim}.}
  \bibinfo{year}{2020}\natexlab{}.
\newblock \bibinfo{title}{{ShieldStore}}.
\newblock
\newblock
\urldef\tempurl%
\url{https://github.com/cocoppang/ShieldStore}
\showURL{%
\tempurl}


\bibitem[\protect\citeauthoryear{Kim, Park, Woo, Jeon, and Huh}{Kim
  et~al\mbox{.}}{2019}]%
        {DBLP:conf/eurosys/KimPWJH19}
\bibfield{author}{\bibinfo{person}{Taehoon Kim}, \bibinfo{person}{Joongun
  Park}, \bibinfo{person}{Jaewook Woo}, \bibinfo{person}{Seungheun Jeon}, {and}
  \bibinfo{person}{Jaehyuk Huh}.} \bibinfo{year}{2019}\natexlab{}.
\newblock \showarticletitle{{ShieldStore}: Shielded In-memory Key-value Storage
  with {SGX}}. In \bibinfo{booktitle}{\emph{EuroSys}}.
  \bibinfo{publisher}{{ACM}}, \bibinfo{pages}{14:1--14:15}.
\newblock
\urldef\tempurl%
\url{https://doi.org/10.1145/3302424.3303951}
\showURL{%
\tempurl}


\bibitem[\protect\citeauthoryear{Koutroumpouchos, Ntantogian, and
  Xenakis}{Koutroumpouchos et~al\mbox{.}}{2021}]%
        {DBLP:journals/sensors/Koutroumpouchos21}
\bibfield{author}{\bibinfo{person}{Nikolaos Koutroumpouchos},
  \bibinfo{person}{Christoforos Ntantogian}, {and} \bibinfo{person}{Christos
  Xenakis}.} \bibinfo{year}{2021}\natexlab{}.
\newblock \showarticletitle{Building Trust for Smart Connected Devices: The
  Challenges and Pitfalls of {TrustZone}}.
\newblock \bibinfo{journal}{\emph{Sensors}} \bibinfo{volume}{21},
  \bibinfo{number}{2} (\bibinfo{year}{2021}), \bibinfo{pages}{520}.
\newblock
\urldef\tempurl%
\url{https://doi.org/10.3390/s21020520}
\showURL{%
\tempurl}


\bibitem[\protect\citeauthoryear{Krawiecka, Kurnikov, Paverd, Mannan, and
  Asokan}{Krawiecka et~al\mbox{.}}{2018}]%
        {DBLP:conf/www/KrawieckaKPMA18}
\bibfield{author}{\bibinfo{person}{Klaudia Krawiecka}, \bibinfo{person}{Arseny
  Kurnikov}, \bibinfo{person}{Andrew Paverd}, \bibinfo{person}{Mohammad
  Mannan}, {and} \bibinfo{person}{N. Asokan}.} \bibinfo{year}{2018}\natexlab{}.
\newblock \showarticletitle{{SafeKeeper}: Protecting Web Passwords using
  Trusted Execution Environments}. In \bibinfo{booktitle}{\emph{{WWW}}}.
  \bibinfo{publisher}{{ACM}}, \bibinfo{pages}{349--358}.
\newblock
\urldef\tempurl%
\url{https://doi.org/10.1145/3178876.3186101}
\showURL{%
\tempurl}


\bibitem[\protect\citeauthoryear{{Large-Scale Data {\&} Systems (LSDS)
  Group}}{{Large-Scale Data {\&} Systems (LSDS) Group}}{2021}]%
        {LibSealGit}
\bibfield{author}{\bibinfo{person}{{Large-Scale Data {\&} Systems (LSDS)
  Group}}.} \bibinfo{year}{2021}\natexlab{}.
\newblock \bibinfo{title}{{LibSEAL}}.
\newblock
\newblock
\urldef\tempurl%
\url{https://github.com/lsds/LibSEAL}
\showURL{%
\tempurl}


\bibitem[\protect\citeauthoryear{{Large-Scale Data {\&} Systems (LSDS)
  Group}}{{Large-Scale Data {\&} Systems (LSDS) Group}}{2022}]%
        {SGX-LKL2021}
\bibfield{author}{\bibinfo{person}{{Large-Scale Data {\&} Systems (LSDS)
  Group}}.} \bibinfo{year}{2022}\natexlab{}.
\newblock \bibinfo{title}{{SGX-LKL-OE (Open Enclave Edition)}}.
\newblock
\newblock
\urldef\tempurl%
\url{https://github.com/lsds/sgx-lkl}
\showURL{%
\tempurl}


\bibitem[\protect\citeauthoryear{Lebedev}{Lebedev}{2019}]%
        {SanctumNewGit}
\bibfield{author}{\bibinfo{person}{Ilia Lebedev}.}
  \bibinfo{year}{2019}\natexlab{}.
\newblock \bibinfo{title}{{The MIT Sanctum processor system}}.
\newblock
\newblock
\urldef\tempurl%
\url{https://github.com/ilebedev/sanctum}
\showURL{%
\tempurl}
\newblock
\shownote{{Latest rel.\ 2020}.}


\bibitem[\protect\citeauthoryear{Lee, Kohlbrenner, Shinde, Asanovic, and
  Song}{Lee et~al\mbox{.}}{2020b}]%
        {DBLP:conf/eurosys/LeeKSAS20}
\bibfield{author}{\bibinfo{person}{Dayeol Lee}, \bibinfo{person}{David
  Kohlbrenner}, \bibinfo{person}{Shweta Shinde}, \bibinfo{person}{Krste
  Asanovic}, {and} \bibinfo{person}{Dawn Song}.}
  \bibinfo{year}{2020}\natexlab{b}.
\newblock \showarticletitle{{Keystone}: an open framework for architecting
  trusted execution environments}. In \bibinfo{booktitle}{\emph{EuroSys}}.
  \bibinfo{publisher}{{ACM}}, \bibinfo{pages}{38:1--38:16}.
\newblock
\urldef\tempurl%
\url{https://doi.org/10.1145/3342195.3387532}
\showURL{%
\tempurl}


\bibitem[\protect\citeauthoryear{Lee, Jo, Choi, Kim, Park, and Lee}{Lee
  et~al\mbox{.}}{2020a}]%
        {DBLP:journals/access/LeeJCKPL20}
\bibfield{author}{\bibinfo{person}{Seungho Lee}, \bibinfo{person}{Hyo~Jin Jo},
  \bibinfo{person}{Wonsuk Choi}, \bibinfo{person}{Hyoseung Kim},
  \bibinfo{person}{Jong~Hwan Park}, {and} \bibinfo{person}{Dong~Hoon Lee}.}
  \bibinfo{year}{2020}\natexlab{a}.
\newblock \showarticletitle{Fine-Grained Access Control-Enabled Logging Method
  on {ARM} {TrustZone}}.
\newblock \bibinfo{journal}{\emph{{IEEE} Access}}  \bibinfo{volume}{8}
  (\bibinfo{year}{2020}), \bibinfo{pages}{81348--81364}.
\newblock
\urldef\tempurl%
\url{https://doi.org/10.1109/ACCESS.2020.2991431}
\showURL{%
\tempurl}


\bibitem[\protect\citeauthoryear{Li, Weng, Li, Wu, Weng, Li, and Deng}{Li
  et~al\mbox{.}}{2021a}]%
        {DBLP:journals/iacr/LiWLWWLD21}
\bibfield{author}{\bibinfo{person}{Ming Li}, \bibinfo{person}{Jian Weng},
  \bibinfo{person}{Yi Li}, \bibinfo{person}{Yongdong Wu},
  \bibinfo{person}{Jiasi Weng}, \bibinfo{person}{Dingcheng Li}, {and}
  \bibinfo{person}{Robert~H. Deng}.} \bibinfo{year}{2021}\natexlab{a}.
\newblock \showarticletitle{{IvyCross}: A Trustworthy and Privacy-preserving
  Framework for Blockchain Interoperability}.
\newblock \bibinfo{journal}{\emph{{IACR} Cryptol. ePrint Arch.}}
  (\bibinfo{year}{2021}), \bibinfo{pages}{1244}.
\newblock
\urldef\tempurl%
\url{https://eprint.iacr.org/2021/1244}
\showURL{%
\tempurl}


\bibitem[\protect\citeauthoryear{Li, Zhu, Zhang, Tan, Xia, Angel, and Chen}{Li
  et~al\mbox{.}}{2021b}]%
        {DBLP:conf/osdi/LiZZ0XA021}
\bibfield{author}{\bibinfo{person}{Mingyu Li}, \bibinfo{person}{Jinhao Zhu},
  \bibinfo{person}{Tianxu Zhang}, \bibinfo{person}{Cheng Tan},
  \bibinfo{person}{Yubin Xia}, \bibinfo{person}{Sebastian Angel}, {and}
  \bibinfo{person}{Haibo Chen}.} \bibinfo{year}{2021}\natexlab{b}.
\newblock \showarticletitle{Bringing Decentralized Search to Decentralized
  Services}. In \bibinfo{booktitle}{\emph{{OSDI}}}.
  \bibinfo{publisher}{{USENIX} Association}, \bibinfo{pages}{331--347}.
\newblock
\urldef\tempurl%
\url{https://www.usenix.org/conference/osdi21/presentation/li}
\showURL{%
\tempurl}


\bibitem[\protect\citeauthoryear{Li, Luo, Miyazaki, and Guo}{Li
  et~al\mbox{.}}{2020}]%
        {DBLP:conf/icc/LiLM020}
\bibfield{author}{\bibinfo{person}{Peng Li}, \bibinfo{person}{Xiaofei Luo},
  \bibinfo{person}{Toshiaki Miyazaki}, {and} \bibinfo{person}{Song Guo}.}
  \bibinfo{year}{2020}\natexlab{}.
\newblock \showarticletitle{Privacy-preserving Payment Channel Networks using
  Trusted Execution Environment}. In \bibinfo{booktitle}{\emph{{ICC}}}.
  \bibinfo{publisher}{{IEEE}}, \bibinfo{pages}{1--6}.
\newblock
\urldef\tempurl%
\url{https://doi.org/10.1109/ICC40277.2020.9149447}
\showURL{%
\tempurl}


\bibitem[\protect\citeauthoryear{Li, Li, Chen, and Xia}{Li
  et~al\mbox{.}}{2015}]%
        {DBLP:conf/mobisys/LiLCX15}
\bibfield{author}{\bibinfo{person}{Wenhao Li}, \bibinfo{person}{Haibo Li},
  \bibinfo{person}{Haibo Chen}, {and} \bibinfo{person}{Yubin Xia}.}
  \bibinfo{year}{2015}\natexlab{}.
\newblock \showarticletitle{{AdAttester}: Secure Online Mobile Advertisement
  Attestation Using {TrustZone}}. In \bibinfo{booktitle}{\emph{MobiSys}}.
  \bibinfo{publisher}{{ACM}}, \bibinfo{pages}{75--88}.
\newblock
\urldef\tempurl%
\url{https://doi.org/10.1145/2742647.2742676}
\showURL{%
\tempurl}


\bibitem[\protect\citeauthoryear{{Linaro Security Working Group}}{{Linaro
  Security Working Group}}{2019}]%
        {kmgk}
\bibfield{author}{\bibinfo{person}{{Linaro Security Working Group}}.}
  \bibinfo{year}{2019}\natexlab{}.
\newblock \bibinfo{title}{{OP-TEE based keymaster and gatekeeper HIDL HAL}}.
\newblock
\newblock
\urldef\tempurl%
\url{https://github.com/linaro-swg/kmgk}
\showURL{%
\tempurl}
\newblock
\shownote{Latest rel.\ 2021.}


\bibitem[\protect\citeauthoryear{Lind, Eyal, Pietzuch, and Sirer}{Lind
  et~al\mbox{.}}{2016}]%
        {DBLP:journals/corr/LindEPS16}
\bibfield{author}{\bibinfo{person}{Joshua Lind}, \bibinfo{person}{Ittay Eyal},
  \bibinfo{person}{Peter~R. Pietzuch}, {and} \bibinfo{person}{Emin~G{\"{u}}n
  Sirer}.} \bibinfo{year}{2016}\natexlab{}.
\newblock \showarticletitle{Teechan: Payment Channels Using Trusted Execution
  Environments}.
\newblock \bibinfo{journal}{\emph{CoRR}}  \bibinfo{volume}{abs/1612.07766}
  (\bibinfo{year}{2016}).
\newblock
\urldef\tempurl%
\url{http://arxiv.org/abs/1612.07766}
\showURL{%
\tempurl}


\bibitem[\protect\citeauthoryear{Lind, Naor, Eyal, Kelbert, Sirer, and
  Pietzuch}{Lind et~al\mbox{.}}{2019}]%
        {DBLP:conf/sosp/LindNEKSP19}
\bibfield{author}{\bibinfo{person}{Joshua Lind}, \bibinfo{person}{Oded Naor},
  \bibinfo{person}{Ittay Eyal}, \bibinfo{person}{Florian Kelbert},
  \bibinfo{person}{Emin~G{\"{u}}n Sirer}, {and} \bibinfo{person}{Peter~R.
  Pietzuch}.} \bibinfo{year}{2019}\natexlab{}.
\newblock \showarticletitle{{Teechain}: a secure payment network with
  asynchronous blockchain access}. In \bibinfo{booktitle}{\emph{{SOSP}}}.
  \bibinfo{publisher}{{ACM}}, \bibinfo{pages}{63--79}.
\newblock
\urldef\tempurl%
\url{https://doi.org/10.1145/3341301.3359627}
\showURL{%
\tempurl}


\bibitem[\protect\citeauthoryear{Liu}{Liu}{2021}]%
        {Deflection2021}
\bibfield{author}{\bibinfo{person}{Weijie Liu}.}
  \bibinfo{year}{2021}\natexlab{}.
\newblock \bibinfo{title}{{Deflection (CAT-SGX)}}.
\newblock
\newblock
\urldef\tempurl%
\url{https://github.com/StanPlatinum/Deflection}
\showURL{%
\tempurl}


\bibitem[\protect\citeauthoryear{Liu, Chen, Wang, Li, Zhang, Wang, and
  Tang}{Liu et~al\mbox{.}}{2021a}]%
        {DBLP:journals/corr/abs-2109-01923}
\bibfield{author}{\bibinfo{person}{Weijie Liu}, \bibinfo{person}{Hongbo Chen},
  \bibinfo{person}{XiaoFeng Wang}, \bibinfo{person}{Zhi Li},
  \bibinfo{person}{Danfeng Zhang}, \bibinfo{person}{Wenhao Wang}, {and}
  \bibinfo{person}{Haixu Tang}.} \bibinfo{year}{2021}\natexlab{a}.
\newblock \showarticletitle{Understanding {TEE} Containers, Easy to Use? Hard
  to Trust}.
\newblock \bibinfo{journal}{\emph{CoRR}}  \bibinfo{volume}{abs/2109.01923}
  (\bibinfo{year}{2021}).
\newblock
\urldef\tempurl%
\url{https://arxiv.org/abs/2109.01923}
\showURL{%
\tempurl}


\bibitem[\protect\citeauthoryear{Liu, Wang, Chen, Wang, Lu, Chen, Wang, Shen,
  Chen, and Tang}{Liu et~al\mbox{.}}{2021b}]%
        {DBLP:conf/dsn/0004WCWLCWSCT21}
\bibfield{author}{\bibinfo{person}{Weijie Liu}, \bibinfo{person}{Wenhao Wang},
  \bibinfo{person}{Hongbo Chen}, \bibinfo{person}{Xiaofeng Wang},
  \bibinfo{person}{Yaosong Lu}, \bibinfo{person}{Kai Chen},
  \bibinfo{person}{Xinyu Wang}, \bibinfo{person}{Qintao Shen},
  \bibinfo{person}{Yi Chen}, {and} \bibinfo{person}{Haixu Tang}.}
  \bibinfo{year}{2021}\natexlab{b}.
\newblock \showarticletitle{Practical and Efficient in-Enclave Verification of
  Privacy Compliance}. In \bibinfo{booktitle}{\emph{{DSN}}}.
  \bibinfo{publisher}{{IEEE}}, \bibinfo{pages}{413--425}.
\newblock
\urldef\tempurl%
\url{https://doi.org/10.1109/DSN48987.2021.00052}
\showURL{%
\tempurl}


\bibitem[\protect\citeauthoryear{{LSDS Group}}{{LSDS Group}}{2019}]%
        {TeechainGit}
\bibfield{author}{\bibinfo{person}{{LSDS Group}}.}
  \bibinfo{year}{2019}\natexlab{}.
\newblock \bibinfo{title}{{Teechain: A Secure Payment Network with Asynchronous
  Blockchain Access}}.
\newblock
\newblock
\urldef\tempurl%
\url{https://github.com/lsds/Teechain}
\showURL{%
\tempurl}


\bibitem[\protect\citeauthoryear{Matetic, Schneider, Miller, Juels, and
  Capkun}{Matetic et~al\mbox{.}}{2018}]%
        {DBLP:conf/uss/MateticSMJC18}
\bibfield{author}{\bibinfo{person}{Sinisa Matetic}, \bibinfo{person}{Moritz
  Schneider}, \bibinfo{person}{Andrew Miller}, \bibinfo{person}{Ari Juels},
  {and} \bibinfo{person}{Srdjan Capkun}.} \bibinfo{year}{2018}\natexlab{}.
\newblock \showarticletitle{{DelegaTEE}: Brokered Delegation Using Trusted
  Execution Environments}. In \bibinfo{booktitle}{\emph{{USENIX} Sec.}}
  \bibinfo{publisher}{{USENIX} Association}, \bibinfo{pages}{1387--1403}.
\newblock
\urldef\tempurl%
\url{https://www.usenix.org/conference/usenixsecurity18/presentation/matetic}
\showURL{%
\tempurl}


\bibitem[\protect\citeauthoryear{Matetic, W{\"{u}}st, Schneider, Kostiainen,
  Karame, and Capkun}{Matetic et~al\mbox{.}}{2019}]%
        {DBLP:conf/uss/MateticWSKKC19}
\bibfield{author}{\bibinfo{person}{Sinisa Matetic}, \bibinfo{person}{Karl
  W{\"{u}}st}, \bibinfo{person}{Moritz Schneider}, \bibinfo{person}{Kari
  Kostiainen}, \bibinfo{person}{Ghassan Karame}, {and} \bibinfo{person}{Srdjan
  Capkun}.} \bibinfo{year}{2019}\natexlab{}.
\newblock \showarticletitle{{BITE}: {Bitcoin} Lightweight Client Privacy using
  Trusted Execution}. In \bibinfo{booktitle}{\emph{{USENIX} Sec.}}
  \bibinfo{publisher}{{USENIX} Association}, \bibinfo{pages}{783--800}.
\newblock
\urldef\tempurl%
\url{https://www.usenix.org/conference/usenixsecurity19/presentation/matetic}
\showURL{%
\tempurl}


\bibitem[\protect\citeauthoryear{McGillion, Dettenborn, Nyman, and
  Asokan}{McGillion et~al\mbox{.}}{2015}]%
        {DBLP:conf/trustcom/McGillionDNA15}
\bibfield{author}{\bibinfo{person}{Brian McGillion}, \bibinfo{person}{Tanel
  Dettenborn}, \bibinfo{person}{Thomas Nyman}, {and} \bibinfo{person}{N.
  Asokan}.} \bibinfo{year}{2015}\natexlab{}.
\newblock \showarticletitle{{Open-TEE} - An Open Virtual Trusted Execution
  Environment}. In \bibinfo{booktitle}{\emph{{TrustCom}}}.
  \bibinfo{publisher}{{IEEE}}, \bibinfo{pages}{400--407}.
\newblock
\urldef\tempurl%
\url{https://doi.org/10.1109/Trustcom.2015.400}
\showURL{%
\tempurl}


\bibitem[\protect\citeauthoryear{M{\'{e}}n{\'{e}}trey}{M{\'{e}}n{\'{e}}trey}{2022}]%
        {Twine2021}
\bibfield{author}{\bibinfo{person}{J{\"{a}}mes M{\'{e}}n{\'{e}}trey}.}
  \bibinfo{year}{2022}\natexlab{}.
\newblock \bibinfo{title}{{Twine: An Embedded Trusted Runtime for
  WebAssembly}}.
\newblock
\newblock
\urldef\tempurl%
\url{https://github.com/JamesMenetrey/unine-twine}
\showURL{%
\tempurl}


\bibitem[\protect\citeauthoryear{M{\'{e}}n{\'{e}}trey, G{\"{o}}ttel, Khurshid,
  Pasin, Felber, Schiavoni, and Raza}{M{\'{e}}n{\'{e}}trey
  et~al\mbox{.}}{2022}]%
        {DBLP:conf/dais/MenetreyGKPFSR22}
\bibfield{author}{\bibinfo{person}{J{\"{a}}mes M{\'{e}}n{\'{e}}trey},
  \bibinfo{person}{Christian G{\"{o}}ttel}, \bibinfo{person}{Anum Khurshid},
  \bibinfo{person}{Marcelo Pasin}, \bibinfo{person}{Pascal Felber},
  \bibinfo{person}{Valerio Schiavoni}, {and} \bibinfo{person}{Shahid Raza}.}
  \bibinfo{year}{2022}\natexlab{}.
\newblock \showarticletitle{Attestation Mechanisms for Trusted Execution
  Environments Demystified}. In \bibinfo{booktitle}{\emph{{DAIS}}}
  \emph{(\bibinfo{series}{LNCS}, Vol.~\bibinfo{volume}{13272})}.
  \bibinfo{publisher}{Springer}, \bibinfo{pages}{95--113}.
\newblock
\urldef\tempurl%
\url{https://doi.org/10.1007/978-3-031-16092-9_7}
\showURL{%
\tempurl}


\bibitem[\protect\citeauthoryear{M{\'{e}}n{\'{e}}trey, Pasin, Felber, and
  Schiavoni}{M{\'{e}}n{\'{e}}trey et~al\mbox{.}}{2021}]%
        {DBLP:conf/icde/MenetreyPFS21}
\bibfield{author}{\bibinfo{person}{J{\"{a}}mes M{\'{e}}n{\'{e}}trey},
  \bibinfo{person}{Marcelo Pasin}, \bibinfo{person}{Pascal Felber}, {and}
  \bibinfo{person}{Valerio Schiavoni}.} \bibinfo{year}{2021}\natexlab{}.
\newblock \showarticletitle{{Twine}: An Embedded Trusted Runtime for
  {WebAssembly}}. In \bibinfo{booktitle}{\emph{{ICDE}}}.
  \bibinfo{publisher}{{IEEE}}, \bibinfo{pages}{205--216}.
\newblock
\urldef\tempurl%
\url{https://doi.org/10.1109/ICDE51399.2021.00025}
\showURL{%
\tempurl}


\bibitem[\protect\citeauthoryear{{MesaLock Linux}}{{MesaLock Linux}}{2019}]%
        {MesaPy2019}
\bibfield{author}{\bibinfo{person}{{MesaLock Linux}}.}
  \bibinfo{year}{2019}\natexlab{}.
\newblock \bibinfo{title}{{MesaPy: A Memory-Safe Python Implementation based on
  PyPy}}.
\newblock
\newblock
\urldef\tempurl%
\url{https://github.com/mesalock-linux/mesapy}
\showURL{%
\tempurl}


\bibitem[\protect\citeauthoryear{{Microsoft}}{{Microsoft}}{2017}]%
        {Komodo}
\bibfield{author}{\bibinfo{person}{{Microsoft}}.}
  \bibinfo{year}{2017}\natexlab{}.
\newblock \bibinfo{title}{{Project Komodo}}.
\newblock
\newblock
\urldef\tempurl%
\url{https://github.com/Microsoft/Komodo}
\showURL{%
\tempurl}


\bibitem[\protect\citeauthoryear{Microsoft}{Microsoft}{2019}]%
        {Ccf2022}
\bibfield{author}{\bibinfo{person}{Microsoft}.}
  \bibinfo{year}{2019}\natexlab{}.
\newblock \bibinfo{title}{{The Confidential Consortium Framework}}.
\newblock
\newblock
\urldef\tempurl%
\url{https://github.com/microsoft/CCF}
\showURL{%
\tempurl}
\newblock
\shownote{{Latest rel.\ 2022}.}


\bibitem[\protect\citeauthoryear{Miranda}{Miranda}{2021}]%
        {S2DedupGit}
\bibfield{author}{\bibinfo{person}{Mariana Miranda}.}
  \bibinfo{year}{2021}\natexlab{}.
\newblock \bibinfo{title}{{S2Dedup}}.
\newblock
\newblock
\urldef\tempurl%
\url{https://github.com/mmm97/S2Dedup}
\showURL{%
\tempurl}


\bibitem[\protect\citeauthoryear{Miranda, Esteves, Portela, and Paulo}{Miranda
  et~al\mbox{.}}{2021}]%
        {DBLP:conf/systor/MirandaEPP21}
\bibfield{author}{\bibinfo{person}{Mariana Miranda},
  \bibinfo{person}{T{\^{a}}nia Esteves}, \bibinfo{person}{Bernardo Portela},
  {and} \bibinfo{person}{Jo{\~{a}}o Paulo}.} \bibinfo{year}{2021}\natexlab{}.
\newblock \showarticletitle{{S2Dedup}: {SGX}-enabled secure deduplication}. In
  \bibinfo{booktitle}{\emph{{SYSTOR}}}. \bibinfo{publisher}{{ACM}},
  \bibinfo{pages}{14:1--14:12}.
\newblock
\urldef\tempurl%
\url{https://doi.org/10.1145/3456727.3463773}
\showURL{%
\tempurl}


\bibitem[\protect\citeauthoryear{Mirzamohammadi, Liu, Huang, Sani, Agarwal, and
  Kim}{Mirzamohammadi et~al\mbox{.}}{2020}]%
        {DBLP:conf/mobisys/MirzamohammadiL20}
\bibfield{author}{\bibinfo{person}{Saeed Mirzamohammadi},
  \bibinfo{person}{Yuxin~(Myles) Liu}, \bibinfo{person}{Tianmei~Ann Huang},
  \bibinfo{person}{Ardalan~Amiri Sani}, \bibinfo{person}{Sharad Agarwal}, {and}
  \bibinfo{person}{Sung Eun~(Summer) Kim}.} \bibinfo{year}{2020}\natexlab{}.
\newblock \showarticletitle{Tabellion: secure legal contracts on mobile
  devices}. In \bibinfo{booktitle}{\emph{MobiSys}}. \bibinfo{publisher}{{ACM}},
  \bibinfo{pages}{220--233}.
\newblock
\urldef\tempurl%
\url{https://doi.org/10.1145/3386901.3389027}
\showURL{%
\tempurl}


\bibitem[\protect\citeauthoryear{Mo, Haddadi, Katevas, Marin, Perino, and
  Kourtellis}{Mo et~al\mbox{.}}{2021}]%
        {DBLP:conf/mobisys/MoHKMPK21}
\bibfield{author}{\bibinfo{person}{Fan Mo}, \bibinfo{person}{Hamed Haddadi},
  \bibinfo{person}{Kleomenis Katevas}, \bibinfo{person}{Eduard Marin},
  \bibinfo{person}{Diego Perino}, {and} \bibinfo{person}{Nicolas Kourtellis}.}
  \bibinfo{year}{2021}\natexlab{}.
\newblock \showarticletitle{{PPFL}: privacy-preserving federated learning with
  trusted execution environments}. In \bibinfo{booktitle}{\emph{MobiSys}}.
  \bibinfo{publisher}{{ACM}}, \bibinfo{pages}{94--108}.
\newblock
\urldef\tempurl%
\url{https://doi.org/10.1145/3458864.3466628}
\showURL{%
\tempurl}


\bibitem[\protect\citeauthoryear{Mo, Shamsabadi, Katevas, Demetriou,
  Leontiadis, Cavallaro, and Haddadi}{Mo et~al\mbox{.}}{2020}]%
        {DBLP:conf/mobisys/MoSKDLCH20}
\bibfield{author}{\bibinfo{person}{Fan Mo}, \bibinfo{person}{Ali~Shahin
  Shamsabadi}, \bibinfo{person}{Kleomenis Katevas}, \bibinfo{person}{Soteris
  Demetriou}, \bibinfo{person}{Ilias Leontiadis}, \bibinfo{person}{Andrea
  Cavallaro}, {and} \bibinfo{person}{Hamed Haddadi}.}
  \bibinfo{year}{2020}\natexlab{}.
\newblock \showarticletitle{{DarkneTZ}: towards model privacy at the edge using
  trusted execution environments}. In \bibinfo{booktitle}{\emph{MobiSys}}.
  \bibinfo{publisher}{{ACM}}, \bibinfo{pages}{161--174}.
\newblock
\urldef\tempurl%
\url{https://doi.org/10.1145/3386901.3388946}
\showURL{%
\tempurl}


\bibitem[\protect\citeauthoryear{Mo}{Mo}{2020}]%
        {darknetz}
\bibfield{author}{\bibinfo{person}{Fan~Vincent Mo}.}
  \bibinfo{year}{2020}\natexlab{}.
\newblock \bibinfo{title}{{DarkneTZ: Towards Model Privacy at the Edge using
  Trusted Execution Environments}}.
\newblock
\newblock
\urldef\tempurl%
\url{https://github.com/mofanv/darknetz}
\showURL{%
\tempurl}


\bibitem[\protect\citeauthoryear{Mo}{Mo}{2021}]%
        {PPFLgit}
\bibfield{author}{\bibinfo{person}{Fan~Vincent Mo}.}
  \bibinfo{year}{2021}\natexlab{}.
\newblock \bibinfo{title}{{Privacy-preserving Federated Learning with Trusted
  Execution Environments}}.
\newblock
\newblock
\urldef\tempurl%
\url{https://github.com/mofanv/PPFL}
\showURL{%
\tempurl}


\bibitem[\protect\citeauthoryear{{MobileCoin Foundation}}{{MobileCoin
  Foundation}}{2020}]%
        {mobilecoin}
\bibfield{author}{\bibinfo{person}{{MobileCoin Foundation}}.}
  \bibinfo{year}{2020}\natexlab{}.
\newblock \bibinfo{title}{{MobileCoin: Private payments for mobile devices}}.
\newblock
\newblock
\urldef\tempurl%
\url{https://github.com/mobilecoinfoundation/mobilecoin}
\showURL{%
\tempurl}
\newblock
\shownote{Latest rel.\ 2022.}


\bibitem[\protect\citeauthoryear{Mondal, More, Rooparaghunath, and
  Gupta}{Mondal et~al\mbox{.}}{2021}]%
        {DBLP:conf/eurosp/MondalMRG21}
\bibfield{author}{\bibinfo{person}{Arup Mondal}, \bibinfo{person}{Yash More},
  \bibinfo{person}{Ruthu~Hulikal Rooparaghunath}, {and}
  \bibinfo{person}{Debayan Gupta}.} \bibinfo{year}{2021}\natexlab{}.
\newblock \showarticletitle{Poster: {FLATEE}: Federated Learning Across Trusted
  Execution Environments}. In \bibinfo{booktitle}{\emph{EuroS{\&}P}}.
  \bibinfo{publisher}{{IEEE}}, \bibinfo{pages}{707--709}.
\newblock
\urldef\tempurl%
\url{https://doi.org/10.1109/EuroSP51992.2021.00054}
\showURL{%
\tempurl}


\bibitem[\protect\citeauthoryear{M{\"{u}}ller}{M{\"{u}}ller}{2019}]%
        {TZ4FabricGit}
\bibfield{author}{\bibinfo{person}{Christina M{\"{u}}ller}.}
  \bibinfo{year}{2019}\natexlab{}.
\newblock \bibinfo{title}{{Hyperledger Fabric chaincode execution with
  OP-TEE}}.
\newblock
\newblock
\urldef\tempurl%
\url{https://github.com/piachristel/open-source-fabric-optee-chaincode}
\showURL{%
\tempurl}


\bibitem[\protect\citeauthoryear{M{\"{u}}ller, Brandenburger, Cachin, Felber,
  G{\"{o}}ttel, and Schiavoni}{M{\"{u}}ller et~al\mbox{.}}{2020}]%
        {DBLP:conf/srds/MullerBCFGS20}
\bibfield{author}{\bibinfo{person}{Christina M{\"{u}}ller},
  \bibinfo{person}{Marcus Brandenburger}, \bibinfo{person}{Christian Cachin},
  \bibinfo{person}{Pascal Felber}, \bibinfo{person}{Christian G{\"{o}}ttel},
  {and} \bibinfo{person}{Valerio Schiavoni}.} \bibinfo{year}{2020}\natexlab{}.
\newblock \showarticletitle{{TZ4Fabric}: Executing Smart Contracts with {ARM}
  {TrustZone}}. In \bibinfo{booktitle}{\emph{{SRDS}}}.
  \bibinfo{publisher}{{IEEE}}, \bibinfo{pages}{31--40}.
\newblock
\urldef\tempurl%
\url{https://doi.org/10.1109/SRDS51746.2020.00011}
\showURL{%
\tempurl}


\bibitem[\protect\citeauthoryear{Namiluko, Paverd, and de~Souza}{Namiluko
  et~al\mbox{.}}{2013}]%
        {DBLP:conf/trust/NamilukoPS13}
\bibfield{author}{\bibinfo{person}{Cornelius Namiluko},
  \bibinfo{person}{Andrew~J. Paverd}, {and} \bibinfo{person}{Tulio de Souza}.}
  \bibinfo{year}{2013}\natexlab{}.
\newblock \showarticletitle{Towards Enhancing Web Application Security Using
  Trusted Execution}. In \bibinfo{booktitle}{\emph{{WASH}}}
  \emph{(\bibinfo{series}{{CEUR} Workshop Proceedings},
  Vol.~\bibinfo{volume}{1011})}. \bibinfo{publisher}{CEUR-WS.org}.
\newblock
\urldef\tempurl%
\url{http://ceur-ws.org/Vol-1011/4.pdf}
\showURL{%
\tempurl}


\bibitem[\protect\citeauthoryear{Ndolo, Henningsen, and Florian}{Ndolo
  et~al\mbox{.}}{2021}]%
        {DBLP:journals/corr/abs-2111-12364}
\bibfield{author}{\bibinfo{person}{Charmaine Ndolo},
  \bibinfo{person}{Sebastian~A. Henningsen}, {and} \bibinfo{person}{Martin
  Florian}.} \bibinfo{year}{2021}\natexlab{}.
\newblock \showarticletitle{Crawling the MobileCoin Quorum System}.
\newblock \bibinfo{journal}{\emph{CoRR}}  \bibinfo{volume}{abs/2111.12364}
  (\bibinfo{year}{2021}).
\newblock
\urldef\tempurl%
\url{https://arxiv.org/abs/2111.12364}
\showURL{%
\tempurl}


\bibitem[\protect\citeauthoryear{Novella}{Novella}{2022}]%
        {TzPapersGit}
\bibfield{author}{\bibinfo{person}{Eduardo Novella}.}
  \bibinfo{year}{2022}\natexlab{}.
\newblock \bibinfo{title}{{A curated list of public TEE resources for learning
  how to reverse-engineer and achieve trusted code execution on ARM devices}}.
\newblock
\newblock
\urldef\tempurl%
\url{https://github.com/enovella/TEE-reversing}
\showURL{%
\tempurl}


\bibitem[\protect\citeauthoryear{{Occlum team}}{{Occlum team}}{2020}]%
        {linux-sgx}
\bibfield{author}{\bibinfo{person}{{Occlum team}}.}
  \bibinfo{year}{2020}\natexlab{}.
\newblock \bibinfo{title}{{Intel(R) Software Guard Extensions for Linux}}.
\newblock
\newblock
\urldef\tempurl%
\url{https://github.com/occlum/linux-sgx}
\showURL{%
\tempurl}
\newblock
\shownote{Latest rel.\ 2022.}


\bibitem[\protect\citeauthoryear{{Occlum team}}{{Occlum team}}{2022}]%
        {Occlum2022}
\bibfield{author}{\bibinfo{person}{{Occlum team}}.}
  \bibinfo{year}{2022}\natexlab{}.
\newblock \bibinfo{title}{{Occlum}}.
\newblock
\newblock
\urldef\tempurl%
\url{https://github.com/occlum/occlum}
\showURL{%
\tempurl}


\bibitem[\protect\citeauthoryear{Oh, Nam, Jeon, Cho, and Paek}{Oh
  et~al\mbox{.}}{2021}]%
        {DBLP:journals/access/OhNJCP21}
\bibfield{author}{\bibinfo{person}{Hyunyoung Oh}, \bibinfo{person}{Kevin Nam},
  \bibinfo{person}{Seongil Jeon}, \bibinfo{person}{Yeongpil Cho}, {and}
  \bibinfo{person}{Yunheung Paek}.} \bibinfo{year}{2021}\natexlab{}.
\newblock \showarticletitle{{MeetGo}: A Trusted Execution Environment for
  Remote Applications on {FPGA}}.
\newblock \bibinfo{journal}{\emph{{IEEE} Access}}  \bibinfo{volume}{9}
  (\bibinfo{year}{2021}), \bibinfo{pages}{51313--51324}.
\newblock
\urldef\tempurl%
\url{https://doi.org/10.1109/ACCESS.2021.3069223}
\showURL{%
\tempurl}


\bibitem[\protect\citeauthoryear{Ohrimenko, Schuster, Fournet, Mehta, Nowozin,
  Vaswani, and Costa}{Ohrimenko et~al\mbox{.}}{2016}]%
        {DBLP:conf/uss/OhrimenkoSFMNVC16}
\bibfield{author}{\bibinfo{person}{Olga Ohrimenko}, \bibinfo{person}{Felix
  Schuster}, \bibinfo{person}{C{\'{e}}dric Fournet}, \bibinfo{person}{Aastha
  Mehta}, \bibinfo{person}{Sebastian Nowozin}, \bibinfo{person}{Kapil Vaswani},
  {and} \bibinfo{person}{Manuel Costa}.} \bibinfo{year}{2016}\natexlab{}.
\newblock \showarticletitle{Oblivious Multi-Party Machine Learning on Trusted
  Processors}. In \bibinfo{booktitle}{\emph{{USENIX} Sec.}}
  \bibinfo{publisher}{{USENIX} Association}, \bibinfo{pages}{619--636}.
\newblock
\urldef\tempurl%
\url{https://www.usenix.org/conference/usenixsecurity16/technical-sessions/presentation/ohrimenko}
\showURL{%
\tempurl}


\bibitem[\protect\citeauthoryear{OP-TEE}{OP-TEE}{2015}]%
        {optee_client}
\bibfield{author}{\bibinfo{person}{OP-TEE}.} \bibinfo{year}{2015}\natexlab{}.
\newblock \bibinfo{title}{{OP-TEE Client API}}.
\newblock
\newblock
\urldef\tempurl%
\url{https://github.com/OP-TEE/optee_client}
\showURL{%
\tempurl}
\newblock
\shownote{Latest rel.\ 2022.}


\bibitem[\protect\citeauthoryear{{Open Enclave}}{{Open Enclave}}{2018}]%
        {OpenEnclave2022}
\bibfield{author}{\bibinfo{person}{{Open Enclave}}.}
  \bibinfo{year}{2018}\natexlab{}.
\newblock \bibinfo{title}{{Open Enclave SDK}}.
\newblock
\newblock
\urldef\tempurl%
\url{https://github.com/openenclave/openenclave}
\showURL{%
\tempurl}
\newblock
\shownote{{Latest rel.\ 2022}.}


\bibitem[\protect\citeauthoryear{openEuler}{openEuler}{2020}]%
        {secGear}
\bibfield{author}{\bibinfo{person}{openEuler}.}
  \bibinfo{year}{2020}\natexlab{}.
\newblock \bibinfo{title}{{secGear}}.
\newblock
\newblock
\urldef\tempurl%
\url{https://github.com/openeuler-mirror/secGear}
\showURL{%
\tempurl}
\newblock
\shownote{Latest rel.\ 2022.}


\bibitem[\protect\citeauthoryear{{Operating Systems and
  Architecture}}{{Operating Systems and Architecture}}{2022}]%
        {Ryoan2021}
\bibfield{author}{\bibinfo{person}{{Operating Systems and Architecture}}.}
  \bibinfo{year}{2022}\natexlab{}.
\newblock \bibinfo{title}{{Ryoan: A distributed sandbox for untrusted
  computation on secret data}}.
\newblock
\newblock
\urldef\tempurl%
\url{https://github.com/ut-osa/ryoan}
\showURL{%
\tempurl}


\bibitem[\protect\citeauthoryear{Paccagnella, Datta, Hassan, Bates, Fletcher,
  Miller, and Tian}{Paccagnella et~al\mbox{.}}{2020}]%
        {DBLP:conf/ndss/PaccagnellaDH0F20}
\bibfield{author}{\bibinfo{person}{Riccardo Paccagnella},
  \bibinfo{person}{Pubali Datta}, \bibinfo{person}{Wajih~Ul Hassan},
  \bibinfo{person}{Adam Bates}, \bibinfo{person}{Christopher~W. Fletcher},
  \bibinfo{person}{Andrew Miller}, {and} \bibinfo{person}{Dave Tian}.}
  \bibinfo{year}{2020}\natexlab{}.
\newblock \showarticletitle{{Custos}: Practical Tamper-Evident Auditing of
  Operating Systems Using Trusted Execution}. In
  \bibinfo{booktitle}{\emph{{NDSS}}}. \bibinfo{publisher}{The Internet
  Society}.
\newblock
\urldef\tempurl%
\url{https://www.ndss-symposium.org/ndss-paper/custos-practical-tamper-evident-auditing-of-operating-systems-using-trusted-execution/}
\showURL{%
\tempurl}


\bibitem[\protect\citeauthoryear{Park, Zhai, Lu, and Lin}{Park
  et~al\mbox{.}}{2019}]%
        {DBLP:conf/usenix/ParkZLL19}
\bibfield{author}{\bibinfo{person}{Heejin Park}, \bibinfo{person}{Shuang Zhai},
  \bibinfo{person}{Long Lu}, {and} \bibinfo{person}{Felix~Xiaozhu Lin}.}
  \bibinfo{year}{2019}\natexlab{}.
\newblock \showarticletitle{{StreamBox-TZ}: Secure Stream Analytics at the Edge
  with {TrustZone}}. In \bibinfo{booktitle}{\emph{{USENIX} {ATC}}}.
  \bibinfo{publisher}{{USENIX} Association}, \bibinfo{pages}{537--554}.
\newblock
\urldef\tempurl%
\url{https://www.usenix.org/conference/atc19/presentation/park-heejin}
\showURL{%
\tempurl}


\bibitem[\protect\citeauthoryear{Park, Ahmad, and Lee}{Park
  et~al\mbox{.}}{2020}]%
        {DBLP:conf/ccs/ParkAL20}
\bibfield{author}{\bibinfo{person}{Seonghyun Park}, \bibinfo{person}{Adil
  Ahmad}, {and} \bibinfo{person}{Byoungyoung Lee}.}
  \bibinfo{year}{2020}\natexlab{}.
\newblock \showarticletitle{{BlackMirror}: Preventing Wallhacks in {3D} Online
  {FPS} Games}. In \bibinfo{booktitle}{\emph{{ACM} {CCS}}}.
  \bibinfo{publisher}{{ACM}}, \bibinfo{pages}{987--1000}.
\newblock
\urldef\tempurl%
\url{https://doi.org/10.1145/3372297.3417890}
\showURL{%
\tempurl}


\bibitem[\protect\citeauthoryear{Patat, Sabt, and Fouque}{Patat
  et~al\mbox{.}}{2022}]%
        {DBLP:conf/sp/PatatSF22}
\bibfield{author}{\bibinfo{person}{Gwendal Patat}, \bibinfo{person}{Mohamed
  Sabt}, {and} \bibinfo{person}{Pierre{-}Alain Fouque}.}
  \bibinfo{year}{2022}\natexlab{}.
\newblock \showarticletitle{Exploring {Widevine} for Fun and Profit}. In
  \bibinfo{booktitle}{\emph{{SP} Workshops}}. \bibinfo{publisher}{{IEEE}},
  \bibinfo{pages}{277--288}.
\newblock
\urldef\tempurl%
\url{https://doi.org/10.1109/SPW54247.2022.9833867}
\showURL{%
\tempurl}


\bibitem[\protect\citeauthoryear{Pires}{Pires}{2019}]%
        {ScbrGit}
\bibfield{author}{\bibinfo{person}{Rafael Pires}.}
  \bibinfo{year}{2019}\natexlab{}.
\newblock \bibinfo{title}{{Secure content-based routing (SCBR)}}.
\newblock
\newblock
\urldef\tempurl%
\url{https://github.com/rafaelppires/scbr}
\showURL{%
\tempurl}


\bibitem[\protect\citeauthoryear{Pires, Goltzsche, Mokhtar, Bouchenak, Boutet,
  Felber, Kapitza, Pasin, and Schiavoni}{Pires et~al\mbox{.}}{2018}]%
        {DBLP:conf/icdcs/GoltzschePMBBFK18}
\bibfield{author}{\bibinfo{person}{Rafael Pires}, \bibinfo{person}{David
  Goltzsche}, \bibinfo{person}{Sonia~Ben Mokhtar}, \bibinfo{person}{Sara
  Bouchenak}, \bibinfo{person}{Antoine Boutet}, \bibinfo{person}{Pascal
  Felber}, \bibinfo{person}{R{\"{u}}diger Kapitza}, \bibinfo{person}{Marcelo
  Pasin}, {and} \bibinfo{person}{Valerio Schiavoni}.}
  \bibinfo{year}{2018}\natexlab{}.
\newblock \showarticletitle{{CYCLOSA}: Decentralizing Private Web Search
  through {SGX}-Based Browser Extensions}. In
  \bibinfo{booktitle}{\emph{{ICDCS}}}. \bibinfo{publisher}{{IEEE} Computer
  Society}, \bibinfo{pages}{467--477}.
\newblock
\urldef\tempurl%
\url{https://doi.org/10.1109/ICDCS.2018.00053}
\showURL{%
\tempurl}


\bibitem[\protect\citeauthoryear{Pires, Pasin, Felber, and Fetzer}{Pires
  et~al\mbox{.}}{2016}]%
        {DBLP:conf/middleware/PiresPFF16}
\bibfield{author}{\bibinfo{person}{Rafael Pires}, \bibinfo{person}{Marcelo
  Pasin}, \bibinfo{person}{Pascal Felber}, {and} \bibinfo{person}{Christof
  Fetzer}.} \bibinfo{year}{2016}\natexlab{}.
\newblock \showarticletitle{Secure Content-Based Routing Using {Intel} Software
  Guard Extensions}. In \bibinfo{booktitle}{\emph{Middleware}}.
  \bibinfo{publisher}{{ACM}}, \bibinfo{pages}{1--10}.
\newblock
\urldef\tempurl%
\url{https://doi.org/10.1145/2988336.2988346}
\showURL{%
\tempurl}


\bibitem[\protect\citeauthoryear{Poddar, Lan, Popa, and Ratnasamy}{Poddar
  et~al\mbox{.}}{2018}]%
        {DBLP:conf/nsdi/PoddarLPR18}
\bibfield{author}{\bibinfo{person}{Rishabh Poddar}, \bibinfo{person}{Chang
  Lan}, \bibinfo{person}{Raluca~Ada Popa}, {and} \bibinfo{person}{Sylvia
  Ratnasamy}.} \bibinfo{year}{2018}\natexlab{}.
\newblock \showarticletitle{{SafeBricks}: Shielding Network Functions in the
  Cloud}. In \bibinfo{booktitle}{\emph{{NSDI}}}. \bibinfo{publisher}{{USENIX}
  Association}, \bibinfo{pages}{201--216}.
\newblock
\urldef\tempurl%
\url{https://www.usenix.org/conference/nsdi18/presentation/poddar}
\showURL{%
\tempurl}


\bibitem[\protect\citeauthoryear{Porter, Boyd{-}Wickizer, Howell, Olinsky, and
  Hunt}{Porter et~al\mbox{.}}{2011}]%
        {DBLP:conf/asplos/PorterBHOH11}
\bibfield{author}{\bibinfo{person}{Donald~E. Porter}, \bibinfo{person}{Silas
  Boyd{-}Wickizer}, \bibinfo{person}{Jon Howell}, \bibinfo{person}{Reuben
  Olinsky}, {and} \bibinfo{person}{Galen~C. Hunt}.}
  \bibinfo{year}{2011}\natexlab{}.
\newblock \showarticletitle{Rethinking the library {OS} from the top down}. In
  \bibinfo{booktitle}{\emph{{ASPLOS}}}. \bibinfo{publisher}{{ACM}},
  \bibinfo{pages}{291--304}.
\newblock
\urldef\tempurl%
\url{https://doi.org/10.1145/1950365.1950399}
\showURL{%
\tempurl}


\bibitem[\protect\citeauthoryear{Prado}{Prado}{2022}]%
        {embeddedbits}
\bibfield{author}{\bibinfo{person}{Sergio Prado}.}
  \bibinfo{year}{2022}\natexlab{}.
\newblock \bibinfo{title}{{Introduction to Trusted Execution Environment and
  ARM's TrustZone}}.
\newblock
  \bibinfo{howpublished}{\url{https://embeddedbits.org/introduction-to-trusted-execution-environment-tee-arm-trustzone/}}.
\newblock
\newblock
\shownote{Accessed: 2022-06-02.}


\bibitem[\protect\citeauthoryear{Priebe, Muthukumaran, Lind, Zhu, Cui,
  Sartakov, and Pietzuch}{Priebe et~al\mbox{.}}{2019}]%
        {DBLP:journals/corr/abs-1908-11143}
\bibfield{author}{\bibinfo{person}{Christian Priebe}, \bibinfo{person}{Divya
  Muthukumaran}, \bibinfo{person}{Joshua Lind}, \bibinfo{person}{Huanzhou Zhu},
  \bibinfo{person}{Shujie Cui}, \bibinfo{person}{Vasily~A. Sartakov}, {and}
  \bibinfo{person}{Peter~R. Pietzuch}.} \bibinfo{year}{2019}\natexlab{}.
\newblock \showarticletitle{{SGX-LKL:} Securing the Host {OS} Interface for
  Trusted Execution}.
\newblock \bibinfo{journal}{\emph{CoRR}}  \bibinfo{volume}{abs/1908.11143}
  (\bibinfo{year}{2019}).
\newblock
\urldef\tempurl%
\url{http://arxiv.org/abs/1908.11143}
\showURL{%
\tempurl}


\bibitem[\protect\citeauthoryear{Quoc, Gregor, Arnautov, Kunkel, Bhatotia, and
  Fetzer}{Quoc et~al\mbox{.}}{2020}]%
        {DBLP:conf/middleware/QuocGAKBF20}
\bibfield{author}{\bibinfo{person}{Do~Le Quoc}, \bibinfo{person}{Franz Gregor},
  \bibinfo{person}{Sergei Arnautov}, \bibinfo{person}{Roland Kunkel},
  \bibinfo{person}{Pramod Bhatotia}, {and} \bibinfo{person}{Christof Fetzer}.}
  \bibinfo{year}{2020}\natexlab{}.
\newblock \showarticletitle{{secureTF}: {A} Secure {TensorFlow} Framework}. In
  \bibinfo{booktitle}{\emph{Middleware}}. \bibinfo{publisher}{{ACM}},
  \bibinfo{pages}{44--59}.
\newblock
\urldef\tempurl%
\url{https://doi.org/10.1145/3423211.3425687}
\showURL{%
\tempurl}


\bibitem[\protect\citeauthoryear{{ratel-enclave}}{{ratel-enclave}}{2022}]%
        {Ratel2021}
\bibfield{author}{\bibinfo{person}{{ratel-enclave}}.}
  \bibinfo{year}{2022}\natexlab{}.
\newblock \bibinfo{title}{{Ratel - a new framework for instruction-level
  interposition on enclaved applications}}.
\newblock
\newblock
\urldef\tempurl%
\url{https://github.com/ratel-enclave/ratel}
\showURL{%
\tempurl}


\bibitem[\protect\citeauthoryear{Riscure}{Riscure}{2019}]%
        {optee_fuzzer}
\bibfield{author}{\bibinfo{person}{Riscure}.} \bibinfo{year}{2019}\natexlab{}.
\newblock \bibinfo{title}{{OP-TEE Fuzzer}}.
\newblock
\newblock
\urldef\tempurl%
\url{https://github.com/Riscure/optee_fuzzer}
\showURL{%
\tempurl}
\newblock
\shownote{Latest rel.\ 2021.}


\bibitem[\protect\citeauthoryear{SafeKeeper}{SafeKeeper}{2018}]%
        {SafeKeeperGit}
\bibfield{author}{\bibinfo{person}{SafeKeeper}.}
  \bibinfo{year}{2018}\natexlab{}.
\newblock \bibinfo{title}{{SafeKeeper - Protecting Web passwords using Trusted
  Execution Environments}}.
\newblock
\newblock
\urldef\tempurl%
\url{https://github.com/SafeKeeper/safekeeper-server}
\showURL{%
\tempurl}


\bibitem[\protect\citeauthoryear{Saltaformaggio, Bhatia, Zhang, Xu, and
  III}{Saltaformaggio et~al\mbox{.}}{2016}]%
        {DBLP:conf/uss/SaltaformaggioB16}
\bibfield{author}{\bibinfo{person}{Brendan Saltaformaggio},
  \bibinfo{person}{Rohit Bhatia}, \bibinfo{person}{Xiangyu Zhang},
  \bibinfo{person}{Dongyan Xu}, {and} \bibinfo{person}{Golden G.~Richard III}.}
  \bibinfo{year}{2016}\natexlab{}.
\newblock \showarticletitle{Screen after Previous Screens: Spatial-Temporal
  Recreation of {Android} App Displays from Memory Images}. In
  \bibinfo{booktitle}{\emph{{USENIX} Sec.}} \bibinfo{publisher}{{USENIX}
  Association}, \bibinfo{pages}{1137--1151}.
\newblock
\urldef\tempurl%
\url{https://www.usenix.org/conference/usenixsecurity16/technical-sessions/presentation/saltaformaggio}
\showURL{%
\tempurl}


\bibitem[\protect\citeauthoryear{{sam1013}}{{sam1013}}{2019}]%
        {TimberVGit}
\bibfield{author}{\bibinfo{person}{{sam1013}}.}
  \bibinfo{year}{2019}\natexlab{}.
\newblock \bibinfo{title}{{TIMBER-V}}.
\newblock
\newblock
\urldef\tempurl%
\url{https://github.com/sam1013/timberv-riscv-tools/tree/timberv}
\showURL{%
\tempurl}


\bibitem[\protect\citeauthoryear{Samsung}{Samsung}{2017}]%
        {SamsungTeegrisSDK17}
\bibfield{author}{\bibinfo{person}{Samsung}.} \bibinfo{year}{2017}\natexlab{}.
\newblock \bibinfo{title}{{SAMSUNG TEEGRIS SDK}}.
\newblock
\newblock
\urldef\tempurl%
\url{https://developer.samsung.com/teegris/overview.html}
\showURL{%
\tempurl}


\bibitem[\protect\citeauthoryear{Samsung}{Samsung}{2018a}]%
        {SamsungKnox22}
\bibfield{author}{\bibinfo{person}{Samsung}.} \bibinfo{year}{2018}\natexlab{a}.
\newblock \bibinfo{title}{{Knox SDK}}.
\newblock
  \bibinfo{howpublished}{\url{https://developer.samsungknox.com/knox-sdk}}.
\newblock
\newblock
\shownote{{Latest rel.\ 2022}.}


\bibitem[\protect\citeauthoryear{Samsung}{Samsung}{2018b}]%
        {TizenFX}
\bibfield{author}{\bibinfo{person}{Samsung}.} \bibinfo{year}{2018}\natexlab{b}.
\newblock \bibinfo{title}{{TizenFX}}.
\newblock
\newblock
\urldef\tempurl%
\url{https://github.com/Samsung/TizenFX}
\showURL{%
\tempurl}
\newblock
\shownote{Latest rel.\ 2022.}


\bibitem[\protect\citeauthoryear{Samsung}{Samsung}{2019a}]%
        {mTower}
\bibfield{author}{\bibinfo{person}{Samsung}.} \bibinfo{year}{2019}\natexlab{a}.
\newblock \bibinfo{title}{{mTower}}.
\newblock
\newblock
\urldef\tempurl%
\url{https://github.com/Samsung/mTower}
\showURL{%
\tempurl}
\newblock
\shownote{Latest rel.\ 2022.}


\bibitem[\protect\citeauthoryear{Samsung}{Samsung}{2019b}]%
        {SamsungKnoxTizenSDK21}
\bibfield{author}{\bibinfo{person}{Samsung}.} \bibinfo{year}{2019}\natexlab{b}.
\newblock \bibinfo{title}{{Welcome to the Knox Tizen SDK for Wearables}}.
\newblock
  \bibinfo{howpublished}{\url{https://docs.samsungknox.com/dev/knox-tizen-sdk/index.htm}}.
\newblock
\newblock
\shownote{{Latest rel.\ 2021}.}


\bibitem[\protect\citeauthoryear{Sanctuary}{Sanctuary}{2021}]%
        {SanctuaryDev}
\bibfield{author}{\bibinfo{person}{Sanctuary}.}
  \bibinfo{year}{2021}\natexlab{}.
\newblock \bibinfo{title}{{Next-Generation Security. Sanctuary}}.
\newblock
  \bibinfo{howpublished}{\url{https://sanctuary.dev/en/solutions/security-services/}}.
\newblock


\bibitem[\protect\citeauthoryear{Sardar, Quoc, and Fetzer}{Sardar
  et~al\mbox{.}}{2020}]%
        {DBLP:conf/dsd/SardarQF20}
\bibfield{author}{\bibinfo{person}{Muhammad~Usama Sardar},
  \bibinfo{person}{Do~Le Quoc}, {and} \bibinfo{person}{Christof Fetzer}.}
  \bibinfo{year}{2020}\natexlab{}.
\newblock \showarticletitle{Towards Formalization of Enhanced Privacy {ID}
  ({EPID})-based Remote Attestation in {Intel} {SGX}}. In
  \bibinfo{booktitle}{\emph{{DSD}}}. \bibinfo{publisher}{{IEEE}},
  \bibinfo{pages}{604--607}.
\newblock
\urldef\tempurl%
\url{https://doi.org/10.1109/DSD51259.2020.00099}
\showURL{%
\tempurl}


\bibitem[\protect\citeauthoryear{Schiavoni}{Schiavoni}{2022}]%
        {SgxPapersGit}
\bibfield{author}{\bibinfo{person}{Valerio Schiavoni}.}
  \bibinfo{year}{2022}\natexlab{}.
\newblock \bibinfo{title}{{sgx-papers}}.
\newblock
\newblock
\urldef\tempurl%
\url{https://github.com/vschiavoni/sgx-papers}
\showURL{%
\tempurl}


\bibitem[\protect\citeauthoryear{Schneider, Masti, Shinde, Capkun, and
  Perez}{Schneider et~al\mbox{.}}{2022}]%
        {DBLP:journals/corr/abs-2205-12742}
\bibfield{author}{\bibinfo{person}{Moritz Schneider},
  \bibinfo{person}{Ramya~Jayaram Masti}, \bibinfo{person}{Shweta Shinde},
  \bibinfo{person}{Srdjan Capkun}, {and} \bibinfo{person}{Ronald Perez}.}
  \bibinfo{year}{2022}\natexlab{}.
\newblock \showarticletitle{{SoK}: Hardware-supported Trusted Execution
  Environments}.
\newblock \bibinfo{journal}{\emph{CoRR}}  \bibinfo{volume}{abs/2205.12742}
  (\bibinfo{year}{2022}).
\newblock
\urldef\tempurl%
\url{https://doi.org/10.48550/arXiv.2205.12742}
\showDOI{\tempurl}


\bibitem[\protect\citeauthoryear{Schuster, Costa, Fournet, Gkantsidis, Peinado,
  Mainar{-}Ruiz, and Russinovich}{Schuster et~al\mbox{.}}{2015}]%
        {DBLP:conf/sp/SchusterCFGPMR15}
\bibfield{author}{\bibinfo{person}{Felix Schuster}, \bibinfo{person}{Manuel
  Costa}, \bibinfo{person}{C{\'{e}}dric Fournet}, \bibinfo{person}{Christos
  Gkantsidis}, \bibinfo{person}{Marcus Peinado}, \bibinfo{person}{Gloria
  Mainar{-}Ruiz}, {and} \bibinfo{person}{Mark Russinovich}.}
  \bibinfo{year}{2015}\natexlab{}.
\newblock \showarticletitle{{VC3}: Trustworthy Data Analytics in the Cloud
  Using {SGX}}. In \bibinfo{booktitle}{\emph{{IEEE} S{\&}P}}.
  \bibinfo{publisher}{{IEEE} Computer Society}, \bibinfo{pages}{38--54}.
\newblock
\urldef\tempurl%
\url{https://doi.org/10.1109/SP.2015.10}
\showURL{%
\tempurl}


\bibitem[\protect\citeauthoryear{Schwarz and Rossow}{Schwarz and
  Rossow}{2020}]%
        {DBLP:conf/uss/SchwarzR20}
\bibfield{author}{\bibinfo{person}{Fabian Schwarz} {and}
  \bibinfo{person}{Christian Rossow}.} \bibinfo{year}{2020}\natexlab{}.
\newblock \showarticletitle{{SENG}, the {SGX}-Enforcing Network Gateway:
  Authorizing Communication from Shielded Clients}. In
  \bibinfo{booktitle}{\emph{{USENIX} Sec.}} \bibinfo{publisher}{{USENIX}
  Association}, \bibinfo{pages}{753--770}.
\newblock
\urldef\tempurl%
\url{https://www.usenix.org/conference/usenixsecurity20/presentation/schwarz}
\showURL{%
\tempurl}


\bibitem[\protect\citeauthoryear{{Scontain}}{{Scontain}}{2022}]%
        {Scone2022}
\bibfield{author}{\bibinfo{person}{{Scontain}}.}
  \bibinfo{year}{2022}\natexlab{}.
\newblock \bibinfo{title}{{SCONE Confidential Computing}}.
\newblock
\newblock
\urldef\tempurl%
\url{https://sconedocs.github.io/}
\showURL{%
\tempurl}


\bibitem[\protect\citeauthoryear{{SCRT Labs}}{{SCRT Labs}}{2020}]%
        {SafeTraceGit}
\bibfield{author}{\bibinfo{person}{{SCRT Labs}}.}
  \bibinfo{year}{2020}\natexlab{}.
\newblock \bibinfo{title}{{SafeTrace: COVID-19 Self-reporting with Privacy}}.
\newblock
\newblock
\urldef\tempurl%
\url{https://github.com/scrtlabs/SafeTrace}
\showURL{%
\tempurl}


\bibitem[\protect\citeauthoryear{Segarra, Delgado{-}Gonzalo, Lemay, Aublin,
  Pietzuch, and Schiavoni}{Segarra et~al\mbox{.}}{2019}]%
        {DBLP:conf/dais/SegarraDLAPS19}
\bibfield{author}{\bibinfo{person}{Carlos Segarra}, \bibinfo{person}{Ricard
  Delgado{-}Gonzalo}, \bibinfo{person}{Mathieu Lemay},
  \bibinfo{person}{Pierre{-}Louis Aublin}, \bibinfo{person}{Peter~R. Pietzuch},
  {and} \bibinfo{person}{Valerio Schiavoni}.} \bibinfo{year}{2019}\natexlab{}.
\newblock \showarticletitle{Using Trusted Execution Environments for Secure
  Stream Processing of Medical Data}. In \bibinfo{booktitle}{\emph{{DAIS}}}
  \emph{(\bibinfo{series}{LNCS}, Vol.~\bibinfo{volume}{11534})}.
  \bibinfo{publisher}{Springer}, \bibinfo{pages}{91--107}.
\newblock
\urldef\tempurl%
\url{https://doi.org/10.1007/978-3-030-22496-7_6}
\showURL{%
\tempurl}


\bibitem[\protect\citeauthoryear{{SELIS Project}}{{SELIS Project}}{2019}]%
        {PubSubGit}
\bibfield{author}{\bibinfo{person}{{SELIS Project}}.}
  \bibinfo{year}{2019}\natexlab{}.
\newblock \bibinfo{title}{{The SELIS Publish/Subscribe system}}.
\newblock
\newblock
\urldef\tempurl%
\url{https://github.com/selisproject/pubsub}
\showURL{%
\tempurl}


\bibitem[\protect\citeauthoryear{sengsgx}{sengsgx}{2020}]%
        {sengsgx}
\bibfield{author}{\bibinfo{person}{sengsgx}.} \bibinfo{year}{2020}\natexlab{}.
\newblock \bibinfo{title}{{SENG, the SGX-Enforcing Network Gateway}}.
\newblock
\newblock
\urldef\tempurl%
\url{https://github.com/sengsgx/sengsgx}
\showURL{%
\tempurl}


\bibitem[\protect\citeauthoryear{shakevsky}{shakevsky}{2020}]%
        {Keybuster}
\bibfield{author}{\bibinfo{person}{shakevsky}.}
  \bibinfo{year}{2020}\natexlab{}.
\newblock \bibinfo{title}{{Keybuster}}.
\newblock
\newblock
\urldef\tempurl%
\url{https://github.com/shakevsky/keybuster}
\showURL{%
\tempurl}


\bibitem[\protect\citeauthoryear{Shakevsky, Ronen, and Wool}{Shakevsky
  et~al\mbox{.}}{2022}]%
        {DBLP:conf/uss/ShakevskyRW22}
\bibfield{author}{\bibinfo{person}{Alon Shakevsky}, \bibinfo{person}{Eyal
  Ronen}, {and} \bibinfo{person}{Avishai Wool}.}
  \bibinfo{year}{2022}\natexlab{}.
\newblock \showarticletitle{Trust Dies in Darkness: Shedding Light on
  {Samsung's} {TrustZone} Keymaster Design}. In
  \bibinfo{booktitle}{\emph{{USENIX} Sec.}} \bibinfo{publisher}{{USENIX}
  Association}, \bibinfo{pages}{251--268}.
\newblock
\urldef\tempurl%
\url{https://www.usenix.org/conference/usenixsecurity22/presentation/shakevsky}
\showURL{%
\tempurl}


\bibitem[\protect\citeauthoryear{Shaon, Kantarcioglu, Lin, and Khan}{Shaon
  et~al\mbox{.}}{2017}]%
        {DBLP:conf/ccs/ShaonKLK17}
\bibfield{author}{\bibinfo{person}{Fahad Shaon}, \bibinfo{person}{Murat
  Kantarcioglu}, \bibinfo{person}{Zhiqiang Lin}, {and} \bibinfo{person}{Latifur
  Khan}.} \bibinfo{year}{2017}\natexlab{}.
\newblock \showarticletitle{SGX-BigMatrix: {A} Practical Encrypted Data
  Analytic Framework With Trusted Processors}. In
  \bibinfo{booktitle}{\emph{{ACM} {CCS}}}. \bibinfo{publisher}{{ACM}},
  \bibinfo{pages}{1211--1228}.
\newblock
\urldef\tempurl%
\url{https://doi.org/10.1145/3133956.3134095}
\showURL{%
\tempurl}


\bibitem[\protect\citeauthoryear{Shen, Tian, Chen, Chen, Wang, Xu, Xia, and
  Yan}{Shen et~al\mbox{.}}{2020}]%
        {DBLP:conf/asplos/ShenTCCWXXY20}
\bibfield{author}{\bibinfo{person}{Youren Shen}, \bibinfo{person}{Hongliang
  Tian}, \bibinfo{person}{Yu Chen}, \bibinfo{person}{Kang Chen},
  \bibinfo{person}{Runji Wang}, \bibinfo{person}{Yi Xu}, \bibinfo{person}{Yubin
  Xia}, {and} \bibinfo{person}{Shoumeng Yan}.} \bibinfo{year}{2020}\natexlab{}.
\newblock \showarticletitle{{Occlum}: Secure and Efficient Multitasking Inside
  a Single Enclave of {Intel} {SGX}}. In \bibinfo{booktitle}{\emph{{ASPLOS}}}.
  \bibinfo{publisher}{{ACM}}, \bibinfo{pages}{955--970}.
\newblock
\urldef\tempurl%
\url{https://doi.org/10.1145/3373376.3378469}
\showURL{%
\tempurl}


\bibitem[\protect\citeauthoryear{Shepherd, Akram, and Markantonakis}{Shepherd
  et~al\mbox{.}}{2017}]%
        {DBLP:conf/IEEEares/ShepherdAM17}
\bibfield{author}{\bibinfo{person}{Carlton Shepherd},
  \bibinfo{person}{Raja~Naeem Akram}, {and} \bibinfo{person}{Konstantinos
  Markantonakis}.} \bibinfo{year}{2017}\natexlab{}.
\newblock \showarticletitle{Establishing Mutually Trusted Channels for Remote
  Sensing Devices with Trusted Execution Environments}. In
  \bibinfo{booktitle}{\emph{{ARES}}}. \bibinfo{publisher}{{ACM}},
  \bibinfo{pages}{7:1--7:10}.
\newblock
\urldef\tempurl%
\url{https://doi.org/10.1145/3098954.3098971}
\showURL{%
\tempurl}


\bibitem[\protect\citeauthoryear{Signal}{Signal}{2017}]%
        {ContactDiscoveryService}
\bibfield{author}{\bibinfo{person}{Signal}.} \bibinfo{year}{2017}\natexlab{}.
\newblock \bibinfo{title}{{Private Contact Discovery Service}}.
\newblock
\newblock
\urldef\tempurl%
\url{https://github.com/signalapp/ContactDiscoveryService}
\showURL{%
\tempurl}
\newblock
\shownote{Latest rel.\ 2022.}


\bibitem[\protect\citeauthoryear{Silva, Mokhtar, Contiu, N{\'{e}}gru,
  R{\'{e}}veill{\`{e}}re, and Rivi{\`{e}}re}{Silva et~al\mbox{.}}{2019}]%
        {DBLP:conf/middleware/SilvaMCNRR19}
\bibfield{author}{\bibinfo{person}{Simon~Da Silva}, \bibinfo{person}{Sonia~Ben
  Mokhtar}, \bibinfo{person}{Stefan Contiu}, \bibinfo{person}{Daniel
  N{\'{e}}gru}, \bibinfo{person}{Laurent R{\'{e}}veill{\`{e}}re}, {and}
  \bibinfo{person}{Etienne Rivi{\`{e}}re}.} \bibinfo{year}{2019}\natexlab{}.
\newblock \showarticletitle{{PrivaTube}: Privacy-Preserving Edge-Assisted Video
  Streaming}. In \bibinfo{booktitle}{\emph{Middleware}}.
  \bibinfo{publisher}{{ACM}}, \bibinfo{pages}{189--201}.
\newblock
\urldef\tempurl%
\url{https://doi.org/10.1145/3361525.3361546}
\showURL{%
\tempurl}


\bibitem[\protect\citeauthoryear{Su, Yang, Luo, Zhang, Bai, and Zhu}{Su
  et~al\mbox{.}}{2020}]%
        {DBLP:conf/msn/SuYLZBZ20}
\bibfield{author}{\bibinfo{person}{Guoxiong Su}, \bibinfo{person}{Wenyuan
  Yang}, \bibinfo{person}{Zhengding Luo}, \bibinfo{person}{Yinghong Zhang},
  \bibinfo{person}{Zhiqiang Bai}, {and} \bibinfo{person}{Yuesheng Zhu}.}
  \bibinfo{year}{2020}\natexlab{}.
\newblock \showarticletitle{{BDTF}: A Blockchain-Based Data Trading Framework
  with Trusted Execution Environment}. In \bibinfo{booktitle}{\emph{{MSN}}}.
  \bibinfo{publisher}{{IEEE}}, \bibinfo{pages}{92--97}.
\newblock
\urldef\tempurl%
\url{https://doi.org/10.1109/MSN50589.2020.00030}
\showURL{%
\tempurl}


\bibitem[\protect\citeauthoryear{Subramanyan, Sinha, Lebedev, Devadas, and
  Seshia}{Subramanyan et~al\mbox{.}}{2017}]%
        {DBLP:conf/ccs/Subramanyan0LDS17}
\bibfield{author}{\bibinfo{person}{Pramod Subramanyan}, \bibinfo{person}{Rohit
  Sinha}, \bibinfo{person}{Ilia~A. Lebedev}, \bibinfo{person}{Srinivas
  Devadas}, {and} \bibinfo{person}{Sanjit~A. Seshia}.}
  \bibinfo{year}{2017}\natexlab{}.
\newblock \showarticletitle{A Formal Foundation for Secure Remote Execution of
  Enclaves}. In \bibinfo{booktitle}{\emph{{ACM} {CCS}}}.
  \bibinfo{publisher}{{ACM}}, \bibinfo{pages}{2435--2450}.
\newblock
\urldef\tempurl%
\url{https://doi.org/10.1145/3133956.3134098}
\showURL{%
\tempurl}


\bibitem[\protect\citeauthoryear{Sun, Sun, Wang, and Jing}{Sun
  et~al\mbox{.}}{2015}]%
        {DBLP:conf/ccs/SunSWJ15}
\bibfield{author}{\bibinfo{person}{He Sun}, \bibinfo{person}{Kun Sun},
  \bibinfo{person}{Yuewu Wang}, {and} \bibinfo{person}{Jiwu Jing}.}
  \bibinfo{year}{2015}\natexlab{}.
\newblock \showarticletitle{{TrustOTP}: Transforming Smartphones into Secure
  One-Time Password Tokens}. In \bibinfo{booktitle}{\emph{{ACM} {CCS}}}.
  \bibinfo{publisher}{{ACM}}, \bibinfo{pages}{976--988}.
\newblock
\urldef\tempurl%
\url{https://doi.org/10.1145/2810103.2813692}
\showURL{%
\tempurl}


\bibitem[\protect\citeauthoryear{Sun, Wang, Li, and Li}{Sun
  et~al\mbox{.}}{2021}]%
        {DBLP:journals/pvldb/SunWL021}
\bibfield{author}{\bibinfo{person}{Yuanyuan Sun}, \bibinfo{person}{Sheng Wang},
  \bibinfo{person}{Huorong Li}, {and} \bibinfo{person}{Feifei Li}.}
  \bibinfo{year}{2021}\natexlab{}.
\newblock \showarticletitle{Building Enclave-Native Storage Engines for
  Practical Encrypted Databases}.
\newblock \bibinfo{journal}{\emph{Proc. {VLDB} Endow.}} \bibinfo{volume}{14},
  \bibinfo{number}{6} (\bibinfo{year}{2021}), \bibinfo{pages}{1019--1032}.
\newblock
\urldef\tempurl%
\url{http://www.vldb.org/pvldb/vol14/p1019-sun.pdf}
\showURL{%
\tempurl}


\bibitem[\protect\citeauthoryear{Suzaki, Tsukamoto, Green, and Mannan}{Suzaki
  et~al\mbox{.}}{2020}]%
        {DBLP:conf/acsac/SuzakiTGM20}
\bibfield{author}{\bibinfo{person}{Kuniyasu Suzaki}, \bibinfo{person}{Akira
  Tsukamoto}, \bibinfo{person}{Andy Green}, {and} \bibinfo{person}{Mohammad
  Mannan}.} \bibinfo{year}{2020}\natexlab{}.
\newblock \showarticletitle{Reboot-Oriented {IoT}: Life Cycle Management in
  Trusted Execution Environment for Disposable {IoT} devices}. In
  \bibinfo{booktitle}{\emph{{ACSAC}}}. \bibinfo{publisher}{{ACM}},
  \bibinfo{pages}{428--441}.
\newblock
\urldef\tempurl%
\url{https://doi.org/10.1145/3427228.3427293}
\showURL{%
\tempurl}


\bibitem[\protect\citeauthoryear{Tamrakar}{Tamrakar}{2017}]%
        {Tamrakar2017}
\bibfield{author}{\bibinfo{person}{Sandeep Tamrakar}.}
  \bibinfo{year}{2017}\natexlab{}.
\newblock \emph{\bibinfo{title}{{Applications of Trusted Execution Environments
  (TEEs)}}}.
\newblock Doctoral thesis. \bibinfo{school}{Aalto University}.
\newblock
\showISBNx{978-952-60-7463-4 (electronic), 978-952-60-7464-1 (printed)}
\showISSN{1799-4942 (electronic), 1799-4934 (printed), 1799-4934 (ISSN-L)}
\urldef\tempurl%
\url{http://urn.fi/URN:ISBN:978-952-60-7463-4}
\showURL{%
\tempurl}


\bibitem[\protect\citeauthoryear{{The Apache Software Foundation}}{{The Apache
  Software Foundation}}{2017}]%
        {TeaclaveSGX2022}
\bibfield{author}{\bibinfo{person}{{The Apache Software Foundation}}.}
  \bibinfo{year}{2017}\natexlab{}.
\newblock \bibinfo{title}{{Teaclave SGX SDK}}.
\newblock
\newblock
\urldef\tempurl%
\url{https://github.com/apache/incubator-teaclave-sgx-sdk}
\showURL{%
\tempurl}
\newblock
\shownote{{Latest rel.\ 2022}.}


\bibitem[\protect\citeauthoryear{{The Apache Software Foundation}}{{The Apache
  Software Foundation}}{2020}]%
        {TeaclaveIncubating2022}
\bibfield{author}{\bibinfo{person}{{The Apache Software Foundation}}.}
  \bibinfo{year}{2020}\natexlab{}.
\newblock \bibinfo{title}{{Teaclave: A Universal Secure Computing Platform}}.
\newblock
\newblock
\urldef\tempurl%
\url{https://github.com/apache/incubator-teaclave}
\showURL{%
\tempurl}
\newblock
\shownote{Latest rel.\ 2022.}


\bibitem[\protect\citeauthoryear{{The Apache Software Foundation}}{{The Apache
  Software Foundation}}{2021}]%
        {TeaclaveTZ2022}
\bibfield{author}{\bibinfo{person}{{The Apache Software Foundation}}.}
  \bibinfo{year}{2021}\natexlab{}.
\newblock \bibinfo{title}{{Teaclave TrustZone SDK}}.
\newblock
\newblock
\urldef\tempurl%
\url{https://github.com/apache/incubator-teaclave-trustzone-sdk}
\showURL{%
\tempurl}
\newblock
\shownote{{Latest rel.\ 2022}.}


\bibitem[\protect\citeauthoryear{{The Gramine Project}}{{The Gramine
  Project}}{2022}]%
        {Gramine2022}
\bibfield{author}{\bibinfo{person}{{The Gramine Project}}.}
  \bibinfo{year}{2022}\natexlab{}.
\newblock \bibinfo{title}{{Gramine Library OS with Intel SGX Support}}.
\newblock
\newblock
\urldef\tempurl%
\url{https://github.com/gramineproject/gramine}
\showURL{%
\tempurl}


\bibitem[\protect\citeauthoryear{Trach, Oleksenko, Gregor, Bhatotia, and
  Fetzer}{Trach et~al\mbox{.}}{2019}]%
        {DBLP:conf/systor/TrachOGBF19}
\bibfield{author}{\bibinfo{person}{Bohdan Trach}, \bibinfo{person}{Oleksii
  Oleksenko}, \bibinfo{person}{Franz Gregor}, \bibinfo{person}{Pramod
  Bhatotia}, {and} \bibinfo{person}{Christof Fetzer}.}
  \bibinfo{year}{2019}\natexlab{}.
\newblock \showarticletitle{{Clemmys}: towards secure remote execution in
  {FaaS}}. In \bibinfo{booktitle}{\emph{{SYSTOR}}}. \bibinfo{publisher}{{ACM}},
  \bibinfo{pages}{44--54}.
\newblock
\urldef\tempurl%
\url{https://doi.org/10.1145/3319647.3325835}
\showURL{%
\tempurl}


\bibitem[\protect\citeauthoryear{Tramer}{Tramer}{2021}]%
        {SlalomGit}
\bibfield{author}{\bibinfo{person}{Florian Tramer}.}
  \bibinfo{year}{2021}\natexlab{}.
\newblock \bibinfo{title}{{SLALOM}}.
\newblock
\newblock
\urldef\tempurl%
\url{https://github.com/ftramer/slalom}
\showURL{%
\tempurl}


\bibitem[\protect\citeauthoryear{Tram{\`{e}}r and Boneh}{Tram{\`{e}}r and
  Boneh}{2019}]%
        {DBLP:conf/iclr/TramerB19}
\bibfield{author}{\bibinfo{person}{Florian Tram{\`{e}}r} {and}
  \bibinfo{person}{Dan Boneh}.} \bibinfo{year}{2019}\natexlab{}.
\newblock \showarticletitle{{Slalom}: Fast, Verifiable and Private Execution of
  Neural Networks in Trusted Hardware}. In \bibinfo{booktitle}{\emph{{ICLR}}}.
  \bibinfo{publisher}{OpenReview.net}.
\newblock
\urldef\tempurl%
\url{https://openreview.net/forum?id=rJVorjCcKQ}
\showURL{%
\tempurl}


\bibitem[\protect\citeauthoryear{Tran, Luu, Kang, Bentov, and Saxena}{Tran
  et~al\mbox{.}}{2018}]%
        {DBLP:conf/acsac/TranLKBS18}
\bibfield{author}{\bibinfo{person}{Muoi Tran}, \bibinfo{person}{Loi Luu},
  \bibinfo{person}{Min~Suk Kang}, \bibinfo{person}{Iddo Bentov}, {and}
  \bibinfo{person}{Prateek Saxena}.} \bibinfo{year}{2018}\natexlab{}.
\newblock \showarticletitle{{Obscuro}: A {Bitcoin} Mixer using Trusted
  Execution Environments}. In \bibinfo{booktitle}{\emph{{ACSAC}}}.
  \bibinfo{publisher}{{ACM}}, \bibinfo{pages}{692--701}.
\newblock
\urldef\tempurl%
\url{https://doi.org/10.1145/3274694.3274750}
\showURL{%
\tempurl}


\bibitem[\protect\citeauthoryear{Truong, Gallagher, Guo, and Walls}{Truong
  et~al\mbox{.}}{2021}]%
        {DBLP:conf/ic2e/TruongGGW21}
\bibfield{author}{\bibinfo{person}{Jean{-}Baptiste Truong},
  \bibinfo{person}{William Gallagher}, \bibinfo{person}{Tian Guo}, {and}
  \bibinfo{person}{Robert~J. Walls}.} \bibinfo{year}{2021}\natexlab{}.
\newblock \showarticletitle{Memory-Efficient Deep Learning Inference in Trusted
  Execution Environments}. In \bibinfo{booktitle}{\emph{{IC2E}}}.
  \bibinfo{publisher}{{IEEE}}, \bibinfo{pages}{161--167}.
\newblock
\urldef\tempurl%
\url{https://doi.org/10.1109/IC2E52221.2021.00031}
\showURL{%
\tempurl}


\bibitem[\protect\citeauthoryear{TrustedFirmware.org}{TrustedFirmware.org}{2014}]%
        {Optee2022}
\bibfield{author}{\bibinfo{person}{TrustedFirmware.org}.}
  \bibinfo{year}{2014}\natexlab{}.
\newblock \bibinfo{title}{{OP-TEE Documentation}}.
\newblock
\newblock
\urldef\tempurl%
\url{https://optee.readthedocs.io/en/latest/}
\showURL{%
\tempurl}
\newblock
\shownote{{Latest rel.\ 2022}.}


\bibitem[\protect\citeauthoryear{Trustonic}{Trustonic}{2015}]%
        {trustonic-tee-user-space}
\bibfield{author}{\bibinfo{person}{Trustonic}.}
  \bibinfo{year}{2015}\natexlab{}.
\newblock \bibinfo{title}{{Trustonic TEE User Space}}.
\newblock
\newblock
\urldef\tempurl%
\url{https://github.com/Trustonic/trustonic-tee-user-space/}
\showURL{%
\tempurl}


\bibitem[\protect\citeauthoryear{Trustonic}{Trustonic}{2018}]%
        {KinibiM20}
\bibfield{author}{\bibinfo{person}{Trustonic}.}
  \bibinfo{year}{2018}\natexlab{}.
\newblock \bibinfo{title}{{Secure IoT Development with Kinibi-M}}.
\newblock
  \bibinfo{howpublished}{\url{https://www.trustonic.com/technical-articles/kinibi-m/}}.
\newblock
\newblock
\shownote{{Latest rel.\ 2020}.}


\bibitem[\protect\citeauthoryear{Tsai, Porter, and Vij}{Tsai
  et~al\mbox{.}}{2017}]%
        {DBLP:conf/usenix/TsaiPV17}
\bibfield{author}{\bibinfo{person}{Chia{-}che Tsai}, \bibinfo{person}{Donald~E.
  Porter}, {and} \bibinfo{person}{Mona Vij}.} \bibinfo{year}{2017}\natexlab{}.
\newblock \showarticletitle{{Graphene-SGX}: {A} Practical Library {OS} for
  Unmodified Applications on {SGX}}. In \bibinfo{booktitle}{\emph{{USENIX}
  {ATC}}}. \bibinfo{publisher}{{USENIX} Association},
  \bibinfo{pages}{645--658}.
\newblock
\urldef\tempurl%
\url{https://www.usenix.org/conference/atc17/technical-sessions/presentation/tsai}
\showURL{%
\tempurl}


\bibitem[\protect\citeauthoryear{{USB armory}}{{USB armory}}{2022}]%
        {GoTEE2022}
\bibfield{author}{\bibinfo{person}{{USB armory}}.}
  \bibinfo{year}{2022}\natexlab{}.
\newblock \bibinfo{title}{{GoTEE - example application}}.
\newblock
\newblock
\urldef\tempurl%
\url{https://github.com/usbarmory/GoTEE-example}
\showURL{%
\tempurl}


\bibitem[\protect\citeauthoryear{{utds3lab}}{{utds3lab}}{2017}]%
        {SgxLogGit}
\bibfield{author}{\bibinfo{person}{{utds3lab}}.}
  \bibinfo{year}{2017}\natexlab{}.
\newblock \bibinfo{title}{{SGX-Log: Securing System Logs With SGX}}.
\newblock
\newblock
\urldef\tempurl%
\url{https://github.com/utds3lab/sgx-log}
\showURL{%
\tempurl}


\bibitem[\protect\citeauthoryear{Valadares, de~Carvalho C{\'{e}}sar~Sobrinho,
  Perkusich, and Gorg{\^{o}}nio}{Valadares et~al\mbox{.}}{2021}]%
        {DBLP:journals/iotj/ValadaresSPG21}
\bibfield{author}{\bibinfo{person}{Dalton C{\'{e}}zane~Gomes Valadares},
  \bibinfo{person}{{\'{A}}lvaro~Alvares de Carvalho C{\'{e}}sar~Sobrinho},
  \bibinfo{person}{Angelo Perkusich}, {and} \bibinfo{person}{Kyller~Costa
  Gorg{\^{o}}nio}.} \bibinfo{year}{2021}\natexlab{}.
\newblock \showarticletitle{Formal Verification of a Trusted Execution
  Environment-Based Architecture for {IoT} Applications}.
\newblock \bibinfo{journal}{\emph{{IEEE} Internet Things J.}}
  \bibinfo{volume}{8}, \bibinfo{number}{23} (\bibinfo{year}{2021}),
  \bibinfo{pages}{17199--17210}.
\newblock
\urldef\tempurl%
\url{https://doi.org/10.1109/JIOT.2021.3077850}
\showURL{%
\tempurl}


\bibitem[\protect\citeauthoryear{{Valve Software}}{{Valve Software}}{2019}]%
        {steamlink-sdk}
\bibfield{author}{\bibinfo{person}{{Valve Software}}.}
  \bibinfo{year}{2019}\natexlab{}.
\newblock \bibinfo{title}{{SDK for the Valve Steam Link}}.
\newblock
\newblock
\urldef\tempurl%
\url{https://github.com/ValveSoftware/steamlink-sdk}
\showURL{%
\tempurl}
\newblock
\shownote{Latest rel.\ 2021.}


\bibitem[\protect\citeauthoryear{van Rijswijk{-}Deij and Poll}{van
  Rijswijk{-}Deij and Poll}{2013}]%
        {DBLP:conf/openidentity/Rijswijk-DeijP13}
\bibfield{author}{\bibinfo{person}{Roland van Rijswijk{-}Deij} {and}
  \bibinfo{person}{Erik Poll}.} \bibinfo{year}{2013}\natexlab{}.
\newblock \showarticletitle{Using Trusted Execution Environments in Two-factor
  Authentication: comparing approaches}. In \bibinfo{booktitle}{\emph{Open
  Identity Summit}} \emph{(\bibinfo{series}{{LNI}},
  Vol.~\bibinfo{volume}{{P-223}})}. \bibinfo{publisher}{{GI}},
  \bibinfo{pages}{20--31}.
\newblock
\urldef\tempurl%
\url{https://dl.gi.de/20.500.12116/17195}
\showURL{%
\tempurl}


\bibitem[\protect\citeauthoryear{Volos, Vaswani, and Bruno}{Volos
  et~al\mbox{.}}{2018}]%
        {DBLP:conf/osdi/VolosVB18}
\bibfield{author}{\bibinfo{person}{Stavros Volos}, \bibinfo{person}{Kapil
  Vaswani}, {and} \bibinfo{person}{Rodrigo Bruno}.}
  \bibinfo{year}{2018}\natexlab{}.
\newblock \showarticletitle{{Graviton}: Trusted Execution Environments on
  {GPUs}}. In \bibinfo{booktitle}{\emph{{OSDI}}}. \bibinfo{publisher}{{USENIX}
  Association}, \bibinfo{pages}{681--696}.
\newblock
\urldef\tempurl%
\url{https://www.usenix.org/conference/osdi18/presentation/volos}
\showURL{%
\tempurl}


\bibitem[\protect\citeauthoryear{Wang, Sun, Feng, Wang, Li, and Ding}{Wang
  et~al\mbox{.}}{2020a}]%
        {DBLP:journals/corr/abs-2005-05996}
\bibfield{author}{\bibinfo{person}{Huibo Wang}, \bibinfo{person}{Mingshen Sun},
  \bibinfo{person}{Qian Feng}, \bibinfo{person}{Pei Wang},
  \bibinfo{person}{Tongxin Li}, {and} \bibinfo{person}{Yu Ding}.}
  \bibinfo{year}{2020}\natexlab{a}.
\newblock \showarticletitle{Towards Memory Safe Python Enclave for Security
  Sensitive Computation}.
\newblock \bibinfo{journal}{\emph{CoRR}}  \bibinfo{volume}{abs/2005.05996}
  (\bibinfo{year}{2020}).
\newblock
\urldef\tempurl%
\url{https://arxiv.org/abs/2005.05996}
\showURL{%
\tempurl}


\bibitem[\protect\citeauthoryear{Wang}{Wang}{2019}]%
        {LightBoxGit}
\bibfield{author}{\bibinfo{person}{Patrick Wang}.}
  \bibinfo{year}{2019}\natexlab{}.
\newblock \bibinfo{title}{{LightBox}}.
\newblock
\newblock
\urldef\tempurl%
\url{https://github.com/patrickwang96/LightBox}
\showURL{%
\tempurl}


\bibitem[\protect\citeauthoryear{Wang, Zhuang, and Yan}{Wang
  et~al\mbox{.}}{2020b}]%
        {DBLP:journals/scn/WangZY20}
\bibfield{author}{\bibinfo{person}{Ziwang Wang}, \bibinfo{person}{Yi Zhuang},
  {and} \bibinfo{person}{Zujia Yan}.} \bibinfo{year}{2020}\natexlab{b}.
\newblock \showarticletitle{{TZ-MRAS}: A Remote Attestation Scheme for the
  Mobile Terminal Based on {ARM} {TrustZone}}.
\newblock \bibinfo{journal}{\emph{Secur. Commun. Networks}}
  \bibinfo{volume}{2020} (\bibinfo{year}{2020}),
  \bibinfo{pages}{1756130:1--1756130:16}.
\newblock
\urldef\tempurl%
\url{https://doi.org/10.1155/2020/1756130}
\showURL{%
\tempurl}


\bibitem[\protect\citeauthoryear{{webinos}}{{webinos}}{2013}]%
        {Webinos2013}
\bibfield{author}{\bibinfo{person}{{webinos}}.}
  \bibinfo{year}{2013}\natexlab{}.
\newblock \bibinfo{title}{{Secure Web Operating System Application Delivery
  Environment}}.
\newblock
\newblock
\urldef\tempurl%
\url{https://github.com/webinos/Webinos-Platform}
\showURL{%
\tempurl}


\bibitem[\protect\citeauthoryear{Weiser, Werner, Brasser, Malenko, Mangard, and
  Sadeghi}{Weiser et~al\mbox{.}}{2019}]%
        {DBLP:conf/ndss/WeiserWBMMS19}
\bibfield{author}{\bibinfo{person}{Samuel Weiser}, \bibinfo{person}{Mario
  Werner}, \bibinfo{person}{Ferdinand Brasser}, \bibinfo{person}{Maja Malenko},
  \bibinfo{person}{Stefan Mangard}, {and} \bibinfo{person}{Ahmad{-}Reza
  Sadeghi}.} \bibinfo{year}{2019}\natexlab{}.
\newblock \showarticletitle{{TIMBER-V}: Tag-Isolated Memory Bringing
  Fine-grained Enclaves to {RISC-V}}. In \bibinfo{booktitle}{\emph{{NDSS}}}.
  \bibinfo{publisher}{The Internet Society}.
\newblock
\urldef\tempurl%
\url{https://www.ndss-symposium.org/ndss-paper/timber-v-tag-isolated-memory-bringing-fine-grained-enclaves-to-risc-v/}
\showURL{%
\tempurl}


\bibitem[\protect\citeauthoryear{W{\"{u}}st, Matetic, Schneider, Miers,
  Kostiainen, and Capkun}{W{\"{u}}st et~al\mbox{.}}{2019}]%
        {DBLP:conf/fc/WustMSMKC19}
\bibfield{author}{\bibinfo{person}{Karl W{\"{u}}st}, \bibinfo{person}{Sinisa
  Matetic}, \bibinfo{person}{Moritz Schneider}, \bibinfo{person}{Ian Miers},
  \bibinfo{person}{Kari Kostiainen}, {and} \bibinfo{person}{Srdjan Capkun}.}
  \bibinfo{year}{2019}\natexlab{}.
\newblock \showarticletitle{{ZLiTE}: Lightweight Clients for Shielded {Zcash}
  Transactions Using Trusted Execution}. In \bibinfo{booktitle}{\emph{Financial
  Cryptography}} \emph{(\bibinfo{series}{LNCS}, Vol.~\bibinfo{volume}{11598})}.
  \bibinfo{publisher}{Springer}, \bibinfo{pages}{179--198}.
\newblock
\urldef\tempurl%
\url{https://doi.org/10.1007/978-3-030-32101-7_12}
\showURL{%
\tempurl}


\bibitem[\protect\citeauthoryear{Xu, Zhu, Andrzejak, and Zhang}{Xu
  et~al\mbox{.}}{2021}]%
        {DBLP:conf/icnidc/XuZ0Z21}
\bibfield{author}{\bibinfo{person}{Tianxing Xu}, \bibinfo{person}{Konglin Zhu},
  \bibinfo{person}{Artur Andrzejak}, {and} \bibinfo{person}{Lin Zhang}.}
  \bibinfo{year}{2021}\natexlab{}.
\newblock \showarticletitle{Distributed Learning in Trusted Execution
  Environment: A Case Study of Federated Learning in {SGX}}. In
  \bibinfo{booktitle}{\emph{{IC-NIDC}}}. \bibinfo{publisher}{{IEEE}},
  \bibinfo{pages}{450--454}.
\newblock
\urldef\tempurl%
\url{https://doi.org/10.1109/IC-NIDC54101.2021.9660433}
\showURL{%
\tempurl}


\bibitem[\protect\citeauthoryear{Zhang}{Zhang}{2021}]%
        {TownCrierGit}
\bibfield{author}{\bibinfo{person}{Fan Zhang}.}
  \bibinfo{year}{2021}\natexlab{}.
\newblock \bibinfo{title}{{Town Crier: An Authenticated Data Feed For Smart
  Contracts}}.
\newblock
\newblock
\urldef\tempurl%
\url{https://github.com/bl4ck5un/Town-Crier}
\showURL{%
\tempurl}


\bibitem[\protect\citeauthoryear{Zhang, Cecchetti, Croman, Juels, and
  Shi}{Zhang et~al\mbox{.}}{2016}]%
        {DBLP:conf/ccs/ZhangCCJS16}
\bibfield{author}{\bibinfo{person}{Fan Zhang}, \bibinfo{person}{Ethan
  Cecchetti}, \bibinfo{person}{Kyle Croman}, \bibinfo{person}{Ari Juels}, {and}
  \bibinfo{person}{Elaine Shi}.} \bibinfo{year}{2016}\natexlab{}.
\newblock \showarticletitle{Town Crier: An Authenticated Data Feed for Smart
  Contracts}. In \bibinfo{booktitle}{\emph{{ACM} {CCS}}}.
  \bibinfo{publisher}{{ACM}}, \bibinfo{pages}{270--282}.
\newblock
\urldef\tempurl%
\url{https://doi.org/10.1145/2976749.2978326}
\showURL{%
\tempurl}


\bibitem[\protect\citeauthoryear{Zhang, Wang, Cao, Hou, and Meng}{Zhang
  et~al\mbox{.}}{2021}]%
        {DBLP:conf/cf/ZhangWCHM21}
\bibfield{author}{\bibinfo{person}{Yuhui Zhang}, \bibinfo{person}{Zhiwei Wang},
  \bibinfo{person}{Jiangfeng Cao}, \bibinfo{person}{Rui Hou}, {and}
  \bibinfo{person}{Dan Meng}.} \bibinfo{year}{2021}\natexlab{}.
\newblock \showarticletitle{{ShuffleFL}: gradient-preserving federated learning
  using trusted execution environment}. In \bibinfo{booktitle}{\emph{{CF}}}.
  \bibinfo{publisher}{{ACM}}, \bibinfo{pages}{161--168}.
\newblock
\urldef\tempurl%
\url{https://doi.org/10.1145/3457388.3458665}
\showURL{%
\tempurl}


\bibitem[\protect\citeauthoryear{Zhao, Shuang, Xu, Huang, Cui, Bettadpur, and
  Lie}{Zhao et~al\mbox{.}}{2019a}]%
        {DBLP:journals/corr/abs-1910-04957}
\bibfield{author}{\bibinfo{person}{Lianying Zhao}, \bibinfo{person}{He Shuang},
  \bibinfo{person}{Shengjie Xu}, \bibinfo{person}{Wei Huang},
  \bibinfo{person}{Rongzhen Cui}, \bibinfo{person}{Pushkar Bettadpur}, {and}
  \bibinfo{person}{David Lie}.} \bibinfo{year}{2019}\natexlab{a}.
\newblock \showarticletitle{SoK: Hardware Security Support for Trustworthy
  Execution}.
\newblock \bibinfo{journal}{\emph{CoRR}}  \bibinfo{volume}{abs/1910.04957}
  (\bibinfo{year}{2019}).
\newblock
\urldef\tempurl%
\url{http://arxiv.org/abs/1910.04957}
\showURL{%
\tempurl}


\bibitem[\protect\citeauthoryear{Zhao, Li, Zhang, and Lin}{Zhao
  et~al\mbox{.}}{2022}]%
        {DBLP:conf/sp/ZhaoLZL22}
\bibfield{author}{\bibinfo{person}{Shixuan Zhao}, \bibinfo{person}{Mengyuan
  Li}, \bibinfo{person}{Yinqian Zhang}, {and} \bibinfo{person}{Zhiqiang Lin}.}
  \bibinfo{year}{2022}\natexlab{}.
\newblock \showarticletitle{{vSGX}: Virtualizing {SGX} Enclaves on {AMD}
  {SEV}}. In \bibinfo{booktitle}{\emph{{IEEE} S{\&}P}}.
  \bibinfo{publisher}{{IEEE}}, \bibinfo{pages}{321--336}.
\newblock
\urldef\tempurl%
\url{https://doi.org/10.1109/SP46214.2022.9833694}
\showURL{%
\tempurl}


\bibitem[\protect\citeauthoryear{Zhao, Zhang, Qin, Feng, and Feng}{Zhao
  et~al\mbox{.}}{2019b}]%
        {DBLP:conf/ccs/ZhaoZQFF19}
\bibfield{author}{\bibinfo{person}{Shijun Zhao}, \bibinfo{person}{Qianying
  Zhang}, \bibinfo{person}{Yu Qin}, \bibinfo{person}{Wei Feng}, {and}
  \bibinfo{person}{Dengguo Feng}.} \bibinfo{year}{2019}\natexlab{b}.
\newblock \showarticletitle{{SecTEE}: {A} Software-based Approach to Secure
  Enclave Architecture Using {TEE}}. In \bibinfo{booktitle}{\emph{{ACM}
  {CCS}}}. \bibinfo{publisher}{{ACM}}, \bibinfo{pages}{1723--1740}.
\newblock
\urldef\tempurl%
\url{https://doi.org/10.1145/3319535.3363205}
\showURL{%
\tempurl}


\bibitem[\protect\citeauthoryear{Zhou}{Zhou}{2020}]%
        {SafeBricksGit}
\bibfield{author}{\bibinfo{person}{Yang Zhou}.}
  \bibinfo{year}{2020}\natexlab{}.
\newblock \bibinfo{title}{{SafeBricks}}.
\newblock
\newblock
\urldef\tempurl%
\url{https://github.com/YangZhou1997/SafeBricks}
\showURL{%
\tempurl}


\bibitem[\protect\citeauthoryear{Zhu, Hou, Wang, Wang, Cao, Zhao, Wang, Zhang,
  Ying, Zhang, and Meng}{Zhu et~al\mbox{.}}{2020}]%
        {DBLP:conf/sp/ZhuH0WCZWZYZM20}
\bibfield{author}{\bibinfo{person}{Jianping Zhu}, \bibinfo{person}{Rui Hou},
  \bibinfo{person}{XiaoFeng Wang}, \bibinfo{person}{Wenhao Wang},
  \bibinfo{person}{Jiangfeng Cao}, \bibinfo{person}{Boyan Zhao},
  \bibinfo{person}{Zhongpu Wang}, \bibinfo{person}{Yuhui Zhang},
  \bibinfo{person}{Jiameng Ying}, \bibinfo{person}{Lixin Zhang}, {and}
  \bibinfo{person}{Dan Meng}.} \bibinfo{year}{2020}\natexlab{}.
\newblock \showarticletitle{Enabling Rack-scale Confidential Computing using
  Heterogeneous Trusted Execution Environment}. In
  \bibinfo{booktitle}{\emph{{IEEE} S{\&}P}}. \bibinfo{publisher}{{IEEE}},
  \bibinfo{pages}{1450--1465}.
\newblock
\urldef\tempurl%
\url{https://doi.org/10.1109/SP40000.2020.00054}
\showURL{%
\tempurl}


\end{thebibliography}

\end{document}